\documentclass[12pt,12pt]{article}
\usepackage{graphicx}
\usepackage{epsfig}
\usepackage{amsmath,amssymb}
\catcode`\@=11
\newdimen\ex@
\font\dozeb=cmmib10 scaled \magstep1
\font\dozesyb=cmbsy10 scaled \magstep1
\font\dezb=cmmib10
\def\bm{\fam9}
\def\beq{\begin{equation}}
\def\eeq{\end{equation}}
\def\beqa{\begin{eqnarray}}
\def\eeqa{\end{eqnarray}}
\newcommand\BA{\begin{array}}
\newcommand\EA{\end{array}}

\def\Jb{{\bm J}}
\def\lb{{\bm l}}
\def\Lb{{\bm L}}
\def\rb{{\bm r}}
\def\kb{{\bm k}}
\def\va{{\bm \alpha }}
\def\vpi{{\bm \pi }}

\def\vnabla{{\bm  \nabla }}

\def\vc{{\bm  c}}

\begin{document}
\def\thefootnote{\fnsymbol{footnote}}

\title{\bf Exact Canonically Conjugate Momenta to\\
\vspace{0.3cm}
Quadrupole-Type Collective Coordinates\\
\vspace{0.3cm}
$\!\!$and Derivation of Nuclear \\
\vspace{0.3cm}
Quadrupole-Type Collective Hamiltonian\footnotemark[1]}
\vskip1cm
\author{Seiya NISHIYAMA\footnotemark[2]~ 
and Jo\~{a}o da PROVID\^{E}NCIA\footnotemark[3]
\\
\\
Centro de F\'\i sica Computacional,
Departamento de F\'\i sica,\\
\\
Universidade de Coimbra,
P-3004-516 Coimbra, Portugal\\[0.5cm]
{\it Dedicated to the Memory of Toshio Marumori}}

\def\bm#1{\mbox{\boldmath $#1$}}
\def\bra#1{\langle #1 |}
\def\ket#1{| #1 \rangle }

\maketitle

\footnotetext[1]
{A preliminary version of
this  work was first presented by S. Nishiyama
at the {\it Pre-Symposium of Tokyo Conference} held at
Institute for Nuclear Study, University of Tokyo,
Tokyo, JAPAN, September 4, 1978.}
\footnotetext[2]
{Corresponding author.
~E-mail address: seikoceu@khe.biglobe.ne.jp,~
nisiyama@teor.fis.uc.pt}
\footnotetext[3]
{E-mail address: providencia@teor.fis.uc.pt}


\begin{abstract}
Exact canonically conjugate momenta $\Pi_{2\mu }$ in 
quadrupole nuclear collective motions are proposed.
The basic idea lies in the introduction of 
a discrete integral equation
for the strict definition of canonically conjugate momenta
to collective variables $\phi_{2\mu }$.
A part of our collective Hamiltonian, the $\Pi_{2\mu }$-dependence
of the kinetic part of the Hamiltonian, is given exactly.
Further,
$\phi_{2\mu }$-dependence of the kinetic part of the Hamiltonian 
is also given.
\end{abstract}


\newpage

\def\thesection{\arabic{section}}

\section{Introduction}

~~~~A proper treatment of 
collective variables in nuclear physics
\cite{Maru.55,MaruYama.55,MaruL.55,NagaTamaAmaMaru.58,Maru.60}
and of
collective field formalisms in QCD
\cite{SJ.80}
and
\cite{Nishi.98}
have
been attempted in different two ways.
In Refs.
\cite{Maru.55}
and
\cite{MaruYama.55},
Marumori $et~al$. first gave a foundation of unified model
of collective motion and independent particle motion in nuclei
and further investigated the collective motion from the standpoint
of particle excitations
\cite{Maru.60}.
On the other hand,
in Ref.
\cite{SJ.80},
Jevicki and Sakita reformulated quantum collective field theory in 
terms of gauge-invariant variables.
It leads to an effective Hamiltonian 
in which the behavior of quarks and SU($N$) colored non-Abelian gluons
in the large-{\it N} limit are determined by classical solutions.
The deep understanding of such QCD properties arising from strongly non-linear 
gluon couplings requires investigations of large-{\it N} limit of two-dimensional QCD.
In Ref.$\!$
\cite{Nishi.98}
one of the present author (S.N.)
proposed an $exact$ canonically conjugate momenta approach to a SU($N$) quantum system
and derived a collective Hamiltonian in terms of the $exact$ canonical variables
up to the order of $\frac{1}{N}$.

Applying the Tomonaga's basic idea in his collective motion theory 
\cite{Tomo.55,Nakai.63},
to nuclei with the aid of the Sunakawa's integral equation method
\cite{SYN.62},
one of the present authors (S.N.) 
developed a collective description of surface oscillations of nuclei
\cite{Nishi.77}.
$\!$This description is considered to give one of possible microscopic 
foundations of nuclear collective motion derived from the famous 
Bohr-Mottelson model (BMM)$\!$ 
\cite{BM.74}$\!$
(concerning BMM see textbooks
\cite{EG.87,RoweWood.2010}).
Introducing appropriate collective 
variables, this collective description was formulated by using the first 
quantized language, contrary to the second quantized manner
in the Sunakawa method.
$\!$Preceding the previous work
\cite{Nishi.77},
extending the Tomonaga's idea 
($\!\!$ \cite{Tomo.55} $\!\!$ and $\!\!$ Nakamura-Suzuki
\cite{Suzuki.80} $\!\!$)
to three-dimensional case,
Miyazima-Tamura
\cite{MiyaTamu.56,Tamura.56}
and Goto
\cite{Goto.69}
successfully proposed a collective 
description of the surface oscillations of nuclei in relation 
to the BMM.
$\!$As they and Villars
\cite{Villars.57}
already pointed out, 
there exist difficulties in their theory:(i) The collective momenta defined by the Tomonaga's method are not exactly canonical conjugate
to collective coordinates:(ii) The collective momenta do not commute,
namely, they are not independent each other:(iii) $\!$A correction term
due to quadrupole-type phonon-phonon interaction is incorrect and
then an investigation of this problem becomes insufficient. 

The problem of finding a set of $exact$
canonical variables in collective description is a very important task.
Until very recently, it has not been considered yet in nuclear physics.
An $exact$ canonically conjugate momenta approach
was proposed by Gulshani and Rowe
\cite{GulRowe.76}.
In contrast to their approach, in this paper, 
with the help of the Sunakawa's method
\cite{SYN.62},
we propose another $exact$ canonically conjugate momenta to 
collective coordinates whose original work
was already presented
(see  Ref. \cite{Nishi.77}).
Contrary to the above mentioned approaches, 
the other attempts were made from quite different viewpoints, i.e., 
canonical transformation theory
\cite{Wata.56}
and group theoretical theory
\cite{UiBi.70}.
It should be noticed that
Corrigan $et~al$. presented the $exact$ solution of the BMM
in the body-fixed coordinates
\cite{CMW.76}
using a natural basis of Lie orthogonal group O(5) and its chain
O(5) $\!\!\supset\!\!$ O(4) $\!\cong\!$ SU(2) $\!\!\times\!\!$ SU(2).

In Section 2,
we introduce collective variables $\phi_{2\mu }$ in nuclei
and
briefly recapitulate their associated relations.
In Section 3,
the $exact$ canonically conjugate momenta $\Pi_{2\mu }$ are
defined in the form of discrete integral equation.
Section 4 is devoted to a proof on the commutativity of the 
$exact$ collective momenta $\Pi_{2\mu }$.
In Section 5,
the
$exact$ $\Pi_{2\mu }$-dependent kinetic part of Hamiltonian is derived.
The $\phi_{2\mu }$-dependent kinetic part of Hamiltonian 
is also given. 
Finally the constant term of this part is determined.
In the last section, we give some concluding remarks. 
In Appendices,
calculations of some commutators and quantities are carried out.


\newpage

\def\thesection{\arabic{section}}
\setcounter{equation}{0}
\renewcommand{\theequation}{\arabic{section}.
\arabic{equation}}

\section{Collective variables and the associated relations}

~~~~Our original Hamiltonian $H$ is given by\\[-18pt]
\begin{eqnarray}
\begin{array}{c}
H 
= 
T+V   
= 
-{\displaystyle \frac{\hbar^2}{2m}} 
\sum_{n=1}^A \sum_{\nu } 
(-1)^\nu \vnabla_\nu^n \vnabla_{-\nu }^n 
+ {\displaystyle \frac{1}{2}} 
\sum_{n,n' = 1}^A 
V(\rb_n ,\rb_{n'}) ,
\end{array}
\label{hamiltonian}
\end{eqnarray}\\[-12pt]
where $m$ is the nucleon mass. 
The $\rb_\nu^n$ and $\nabla_{\mu }^n$ mean
components of the position operator
and 
those of the gradient operator,
respectively,
of the $n$-th nucleon in the spherical tensor representation.
The term $V$ is the interaction potential between $n$-th and 
$n' $-th nucleons.

Following Miyazima and Tamura
\cite{Tamura.56},
we introduce the quadrupole collective variables $\phi_{2\mu }$ and
the associated dynamical variables $\eta_{2\mu }$
defined in the following forms:\\[-16pt]
\begin{eqnarray}
\begin{array}{c}
\phi_{2\mu } 
\equiv 
{\displaystyle \frac{4\pi }{3AR_0^2}} 
\sum_{n=1}^A  
r_n^2 Y_{2\mu } (\theta_n ,\varphi_n) ,~~
\left(
\phi_{2\mu }^* 
= 
(-1)^\mu \phi_{2-\mu } 
\right) ,
\end{array}
\label{definitionphi}
\end{eqnarray}\\[-30pt]
\begin{eqnarray}
\begin{array}{rcl}
\eta_{2\mu } 
\!\!&\equiv&\!\! 
{\displaystyle 
\frac{3AR_0^2}{4\pi } \cdot \frac{m}{2} \cdot \frac{i}{\hbar }
} 
[H,~(-1)^\mu \phi_{2-\mu }] ~~
\left(
\eta_{2\mu }^* 
= 
-(-1)^\mu \eta_{2-\mu } 
\right)\\
\\[-12pt]
&=&\!\! 
-i\hbar {\displaystyle \frac{\sqrt{2 \cdot 5}}{2}} 
\sum_{\kappa \nu } 
(-1)^\mu \langle 1\kappa 1\nu|2-\mu \rangle 
\sum_{n=1}^A  
r_n Y_{1\kappa }(\theta_n ,\varphi_n) 
\cdot \vnabla_\nu^n .
\end{array}
\label{definitioneta}
\end{eqnarray}\\[-12pt]
$R_0$ stands for the nuclear equilibrium radius.
At the first stage, we consider $\eta_{2\mu }$ as the collective 
momenta conjugate to $\phi_{2\mu }$ (velocity potential) in the
sense of Tomonaga. 
However, commutation relations among these variables become 
as follows:\\[-12pt]
\begin{equation}
[\phi_{2\mu } ,~\phi_{2\mu'}] = 0,
\label{commuphi}
\end{equation}
\begin{equation}
\begin{array}{c}
[\eta_{2\mu } ,~\phi_{2\mu'}]
=
-i\hbar R_0^{-2} R^2 \delta_{\mu \mu'}
+
i\hbar \sqrt{{\displaystyle \frac{35}{8\pi }}} 
\sum_\Lambda 
(-1)^\mu \langle 2-\mu 2\mu'|2\Lambda \rangle \phi_{2\Lambda },
\end{array}
\label{commurelation1}
\end{equation}
\begin{equation}
\begin{array}{c}
[\eta_{2\mu } ,~\eta_{2\mu'}]
=
\hbar {\displaystyle \frac{15\sqrt{5}}{8\sqrt{2}\pi }}
\sum_\Lambda 
\langle 2\mu 2\mu'|1\Lambda \rangle \lb_{1\Lambda },
\end{array}
\label{commurelation2}
\end{equation} 
where the quantities $R^2$ and $\lb_{1\Lambda }$ are defined as\\[-8pt]
\begin{equation}
\begin{array}{c}
R^2 
\equiv 
{\displaystyle \frac{5}{3A}} 
\sum_{n=1}^A 
r_n^2 ,
\end{array}
\label{definitionR2}
\end{equation}\\[-26pt]
\begin{equation}
\begin{array}{c}
\lb_{1\Lambda }
\equiv
-\hbar \sqrt{2} 
\sum_{\kappa \nu } (-1)^\Lambda
\langle 1\kappa 1\nu|1-\Lambda \rangle 
\sqrt{{\displaystyle \frac{4\pi }{3}}}
\sum_{n=1}^A 
r_n Y_{1\kappa } 
(\theta_n ,\varphi_n) 
\cdot 
\vnabla_\nu^n .
\end{array}
\end{equation} 
$\lb_{1\Lambda }$ is a component of
the total angular momentum operator $\lb$ of the system.
Here we have used the spherical tensor representation for $\rb_\nu$,
i.e.,
$
\rb_\nu
\!=\!
\sqrt{{\displaystyle \frac{4\pi }{3}}} r Y_{1\nu } (\theta ,\varphi)
$.

As a next step, to make the first term on the R.H.S. of 
(\ref{commurelation1}) 
into $-i\hbar \delta_{\mu \mu'}$,
we introduce modified variables $\pi_{2\mu }$ defined by\\[-14pt]
\begin{equation}
\pi_{2\mu } 
\equiv 
\frac{R_0^2}{2} (R^{-2} \eta_{2\mu }
+
\eta_{2\mu }R^{-2}) ,~~
\left(
\pi_{2\mu }^* 
= 
-(-1)^\mu \pi_{2-\mu }
\right) ,~~
\eta_{2\mu } 
= 
{\displaystyle \frac{1}{2R_0^2}}
(R^2 {\pi }_{2\mu } + {\pi }_{2\mu } R^2 ) ,
\label{definitionpi}
\end{equation}\\[-10pt]
the last relation of which is easily verified
by using the commutation relations
(\ref{commuphipir2})
given below.
The commutation relations 
(\ref{commurelation1}) and (\ref{commurelation2}) 
are rewritten in the
following forms:\\[-12pt]
\begin{equation}
\begin{array}{c}
[\pi_{2\mu } ,~\phi_{2\mu'}] 
= 
-i\hbar \delta_{\mu \mu'}
+
i\hbar \sqrt{{\displaystyle \frac{35}{8\pi }}} R_0^2 R^{-2} 
\sum_\Lambda 
(-1)^\mu \langle 2-\mu 2\mu' |2\Lambda \rangle \phi_{2\Lambda } ,
\end{array}
\label{commurelation1p}
\end{equation}\\[-30pt]
\begin{eqnarray}
\begin{array}{c}
[\pi_{2\mu },~\pi_{2\mu'}] 
=
\hbar {\displaystyle \frac{15\sqrt{5}}{8\sqrt{2}\pi }} R_0^4 R^{-4} 
\sum_\Lambda 
\langle 2\mu 2\mu'|1\Lambda \rangle \lb_{1\Lambda } 
+ 
i\hbar {\displaystyle \frac{2\cdot 5}{4\pi }} R_0^4 R^{-4}
(\phi_{2\mu }^* \pi_{2\mu'} -\phi_{2\mu'}^* \pi_{2\mu }),
\end{array}
\label{commurelation2p}
\end{eqnarray}
together with the additional relation
$[\phi_{2\mu },R^{\pm 2}] \!=\! 0$
and\\[-18pt]
\begin{eqnarray}
[\pi_{2\mu },~R^{+2}] 
\!=\! 
-i\hbar \frac{2\cdot 5}{4\pi }
R_0^4 R^{-2} \!
\phi_{2\mu }^*,~~~~~
[\pi_{2\mu },~R^{-2}] 
\!=\! 
i\hbar \frac{2\cdot 5}{4\pi }
R_0^4 R^{-6} \!
\phi_{2\mu }^*,
\label{commuphipir2}
\end{eqnarray}\\[-12pt]
which are written in a lump as
$
[\pi_{2\mu },~R^{\pm 2}] 
\!=\! 
-i\hbar {\displaystyle \frac{2\cdot 5}{4\pi }}
R_0^4 R^{-2} \!
\left( \!\!\!
\begin{array}{c}
1 \\
-R^{-4} \\
\end{array} \!\!\!
\right) \!
\phi_{2\mu }^*
$. 
The second equation of
(\ref{commuphipir2})
is derived through\\[-18pt]
\begin{eqnarray}
\begin{array}{rcl}
[\eta_{2\mu },~R^{\pm 2}] 
\!\!&=&\!\!
-i\hbar {\displaystyle \frac{\sqrt{2 \cdot 5}}{2}} 
\sum_{\kappa \nu } 
(-1)^\mu \langle 1\kappa 1\nu|2-\mu \rangle 
\sum_{n=1}^A  
r_n Y_{1\kappa }(\theta_n ,\varphi_n) 
\cdot 
[\vnabla_\nu^n,R^{\pm 2}] \\
\\[-10pt]
\!\!&=&\!\!
-i\hbar {\displaystyle \frac{2 \cdot 5}{4\pi }}
R_0^2 \! 
\left( \!\!
\begin{array}{c}
1 \\
-R^{-4} \\
\end{array} \!\!
\right) \!
\phi_{2\mu }^* ,
\end{array}
\label{commuphietar2}
\end{eqnarray}\\[-12pt]
where we have used the commutation relation
\begin{eqnarray}
\begin{array}{c}
[\vnabla_\nu^n,~R^{\pm 2}] 
=

{\displaystyle \frac{2 \cdot 5}{3A}}
\sqrt{{\displaystyle \frac{4\pi }{3}}} \!
\left( \!\!
\begin{array}{c}
1 \\
-R^{-4} \\
\end{array} \!\!
\right) \!
r_n Y_{1\nu }(\theta_n ,\varphi_n) .
\end{array}
\label{nablar2}
\end{eqnarray}\\[-10pt]
Neither the second term on the R.H.S. of 
(\ref{commurelation1p})
nor the R.H.S. of 
(\ref{commurelation2p}) 
is zero.
Then from this fact, it is self-evident that $\phi_{2\mu }$
and $\pi_{2\mu }$ are not canonical conjugate to each other.

The $exact$ canonically conjugate momenta
were given by Gulshani in two different ways
in 1976
\cite{GulRowe.76}.
Before going to an investigation on a prescription
how to find the $exact$ canonical conjugate momenta to $\phi_{2\mu }$,
we consider about some historical aspects of studies of collective description by the BMM
\cite{BM.74}.
In the traditional collective description of the quadrupole-type surface oscillations of nuclei by the BMM,
collective coordinates $\va_{2 \mu }$ are introduced.
These coordinates and their canonical conjugate momenta $\vpi_{2 \mu }$
are regarded as bosonic quantized operators and satisfy
the standard commutation relations
$[\vpi_{2 \mu }, \va_{2 \mu' }] \!\!=\!\! -i\hbar \delta_{\mu\mu' },~
[\vpi_{2 \mu }, \vpi_{2 \mu' }] \!\!=\!\! 0$
and
$[\va_{2 \mu }, \va_{2 \mu' }] \!\!=\!\! 0$.
In 2007
Baerdemacker $et~al.$
have calculated the matrix elements of the quadrupole operators and
canonical conjugate momenta operators of the collective model,
in an algebraic straightforward way within a SU(1,1) $\!\!\times\!\!$ O(5) basis
\cite{BHH.07}. 
For this purpose, they have introduced the ten operators
$L_M \!\!=\!\! -i \sqrt{10}/\hbar [\va \vpi^*]^{(1)}_M$
and
$O_M \!\!=\!\! -i \sqrt{10}/\hbar [\va \vpi^*]^{(3)}_M$.
Here, the short hand notation
$[\va \vpi^*]^{(J)}_M$
stands for Clebsch-Gordan coefficient in which the summation over
the hidden indices $\mu$ of $\va$ and $\vpi$ is made.
These ten operators span the Lie algebra of the O(5) group.
The quadratic Casimir operator of O(5) and the Cartan-Weyl reduction chain of O(5), 
O(5) $\!\!\supset\!\!$ O(4) $\!\cong\!$ SU(2) $\!\!\times\!\!$ SU(2),
as well as the representation of O(5),
have been intensively used to compute
various kinds of the above matrix elements. 
On the other hand,
preceding the Baerdemacker $et~al.$'s work,
Yamamura and one of the present authors (S.N.)
proposed a microscopic description of collective rotational motion in deformed nucleus
standing on the basic idea of the Bohr's rotational model
in 1972
\cite{N.Y.72}.
In this earlier and pioneering work,
they introduced the ten operators
$B_{1M} \!=\!  [\vc^\dag \vc]^{(1)}_M$ (angular momentum operator)
and
$B_{3M} \!=\!  [\vc^\dag \vc]^{(3)}_M$ (magnetic octupole-moment operator)
adding to the six operators
$B_{JM} \!=\!  [\vc^\dag \vc]^{(J)}_M~(J \!=\! 0~\mbox{and}~ 2)$
(number operator and quadrupole-moment operator)
where $\vc^\dag_{jm}$ and $\vc_{jm}$ are fermion (nucleon) operators.
The two kinds of scalar operators
$F \!=\!  \sum_{I = 0,2} \sum_M [B_{JM}^\dag B_{JM}]^{(I)}$
and
$G \!=\!  \sum_{I = 1,3}  \sum_M [B_{JM}^\dag B_{JM}]^{(I)}$
and the so-called kinematical constraints governing the operators $B_{JM}$,
composed of a usual tensor product of $\vc^\dag$ and $\vc$,
played central roles in that pioneering work
aiming an algebraic derivation of the Bohr's rotational model.
The above inquiries are the very short historical reviews of the theoretical,
particularly algebraic foundation of the BMM.

Now let us turn to our main problem
how to find the $exact$ canonical conjugate momenta to $\phi_{2\mu }$.
In the next section, we  pursue such an issue.


\newpage

\def\thesection{\arabic{section}}
\setcounter{equation}{0}
\renewcommand{\theequation}{\arabic{section}.
\arabic{equation}}

\section{Exact canonically conjugate momenta}

~~~~In order to overcome the difficulties mentioned in the preceding
section, we define the $exact$ canonically conjugate  momenta $\Pi_{2\mu }$ by
a discrete integral equation\\[-12pt] 
\begin{eqnarray}
\begin{array}{c}
\Pi_{2\mu }
\equiv
\pi_{2\mu }
+
\sqrt{{\displaystyle \frac{35}{8\pi }}} R_0^2 
\sum_{\mu_1 \mu_2} 
(-1)^{\mu_1} \langle 2-\mu_1 2\mu_2 |2\mu \rangle
\phi_{2\mu_1} 
{\displaystyle \frac{1}{2}}
\left(
R^{-2} \Pi_{2\mu_2} \!+\! \Pi_{2\mu_2} R^{-2}
\right).
\end{array}
\label{exactpi}
\end{eqnarray}
This type of the integral equation was first introduced intuitively but not logically
by Sunakawa $\!et~\!al.$
in their collective description of interacting boson systems.
The new variable $\Pi_{2\mu }$, as is clear from the structure of the definition
(\ref{exactpi}),
is no longer a one-body operator but is essentially a many-body operator.
Following the Sunakawa's manner, the proof of the $exact$ canonical commutation 
relation between $\phi_{2\mu }$ and $\Pi_{2\mu }$ can be given as follows:
Iterating the discrete integral equation 
(\ref{exactpi}) 
and using  
(\ref{commuphi}), (\ref{commurelation1p}) and 
first of (\ref{commuphipir2}), 
we have\\[-12pt] 
\begin{eqnarray}
[\Pi_{2\mu },~\phi_{2\mu'}]
& \nonumber \\
\nonumber \\[-10pt]
&\hspace{-3.0cm}
~\!=
[\pi_{2\mu },~\phi_{2\mu'}]
\!+\!
\sqrt{{\displaystyle \frac{35}{8\pi }}} R_0^2 \!
\sum_{\mu_1 \mu_2} \!
(\!-\!1)^{\mu_1} \langle 2 \!-\! \mu_1 2\mu_2 |2\mu \rangle
\phi_{2\mu_1} 
{\displaystyle \frac{1}{2}}
\left( 
R^{-2}[\Pi_{2\mu_2},~\phi_{2\mu'}] \!+\! [\Pi_{2\mu_2},~\phi_{2\mu'}]R^{-2}
\right) \nonumber \\
\nonumber \\[-4pt]
&\hspace{-3.0cm}
=
[\pi_{2\mu },~\phi_{2\mu'}]
\!+\!
\sqrt{{\displaystyle \frac{35}{8\pi }}} R_0^2 \!
\sum_{\mu_1 \mu_2} \!
(\!-\!1)^{\mu_1} \langle 2 \!-\! \mu_1 2\mu_2 |2\mu \rangle
\phi_{2\mu_1} 
{\displaystyle \frac{1}{2}}
\left( 
R^{-2}[\pi_{2\mu_2},~\phi_{2\mu'}] \!+\! [\pi_{2\mu_2},~\phi_{2\mu'}]R^{-2}
\right) \nonumber \\
\nonumber \\[-4pt]
&\hspace{-6.5cm}
\!\!\!\!+
\sqrt{{\displaystyle \frac{35}{8\pi }}} R_0^2 
\sum_{\mu_1 \mu_2} 
(-1)^{\mu_1} \langle 2 - \mu_1 2\mu_2 |2\mu \rangle
\phi_{2\mu_1} \nonumber \\
\nonumber \\[-6pt]
&\hspace{-3.0cm}
\times
{\displaystyle \frac{1}{2}} \!
\left\{ \!
R^{-2} \!
\left[ \!
\sqrt{{\displaystyle \frac{35}{8\pi }}} R_0^2 
\sum_{\mu_3 \mu_4} 
(-1)^{\mu_3} \langle 2 - \mu_3 2\mu_4 |2\mu_2 \rangle \phi_{2\mu_3}
{\displaystyle \frac{1}{2}} \!
\left(
R^{-2} \Pi_{2\mu_4} \!+\! \Pi_{2\mu_4} R^{-2}
\right) \! ,~
\phi_{2\mu'}
\right]
\right.\nonumber \\
\nonumber \\[-2pt]
&\hspace{-3.0cm}
\left.
+
\left[ \!
\sqrt{{\displaystyle \frac{35}{8\pi }}} R_0^2 \!
\sum_{\mu_3 \mu_4} 
(-1)^{\mu_3} \langle 2 - \mu_3 2\mu_4 |2\mu_2 \rangle \phi_{2\mu_3}
{\displaystyle \frac{1}{2}} \!
\left(
R^{-2} \Pi_{2\mu_4} \!+\! \Pi_{2\mu_4} R^{-2}
\right) \! ,~
\phi_{2\mu'}
\right] \!
R^{-2} \!
\right\} \nonumber \\
\nonumber \\[-8pt]
&\hspace{-3.0cm}
=
[\pi_{2\mu },~\phi_{2\mu'}]
\!+\!
\sqrt{{\displaystyle \frac{35}{8\pi }}} R_0^2 \!
\sum_{\mu_1 \mu_2} \!
(\!-\!1)^{\mu_1} \langle 2 \!-\! \mu_1 2\mu_2 |2\mu \rangle
\phi_{2\mu_1} 
{\displaystyle \frac{1}{2}} \!
\left( \!
R^{-2}[\pi_{2\mu_2},~\phi_{2\mu'}] \!+\! [\pi_{2\mu_2},~\phi_{2\mu'}]R^{-2}
\right) \nonumber \\
\nonumber \\[-6pt]
&\hspace{-3.0cm}
\!+
\left( \!
\sqrt{{\displaystyle \frac{35}{8\pi }}} R_0^2 \!
\right)^{\!\!2} \!
\sum_{\mu_1 \mu_2} \! \sum_{\mu_3 \mu_4}
(-1)^{\mu_1} \langle 2 - \mu_1 2\mu_2 |2\mu \rangle
(-1)^{\mu_3} \langle 2 - \mu_3 2\mu_4 |2\mu_2 \rangle
\phi_{2\mu_1} \phi_{2\mu_3}\nonumber \\
\nonumber \\[-6pt]
&\hspace{-3.0cm}
\times
{\displaystyle \frac{1}{2}} \!
\left\{ \!\!
R^{-2}
{\displaystyle \frac{1}{2}} \!
\left(
R^{-2} \! \left[\Pi_{2\mu_4},~\phi_{2\mu'}\right]
\!+\!
\left[\Pi_{2\mu_4},~\phi_{2\mu'}\right] \! R^{-2}
\right) 
\!+\!
{\displaystyle \frac{1}{2}} \!
\left( \!
R^{-2} \! \left[\Pi_{2\mu_4},~\phi_{2\mu'}\right]
\!+\!
\left[\Pi_{2\mu_4},~\phi_{2\mu'}\right] \! R^{-2}
\right) \!
R^{-2} \!
\right\} \nonumber \\
\nonumber
\\[-8pt]
&\hspace{-5.0cm}
=
-i\hbar\delta_{\mu \mu'}
+
i\hbar \sqrt{{\displaystyle \frac{35}{8\pi }}} R_0^2 R^{-2}
\sum_\Lambda 
(-1)^\mu \langle 2-\mu 2\mu'|2\Lambda \rangle 
\phi_{2\Lambda } \nonumber \\ 
\nonumber \\[-6pt]
&\hspace{-5.0cm}
-i\hbar \sqrt{{\displaystyle \frac{35}{8\pi }}} R_0^2 R^{-2}
\sum_\Lambda 
(-1)^\mu \langle 2-\mu 2\mu' |2\Lambda \rangle 
\phi_{2\Lambda } 
+
\cdots ,
\label{commurelpiphi}
\end{eqnarray}
All terms which involve the higher powers concerning $\phi_{2\mu }$
cancel each other except the first term on the R.H.S. of 
(\ref{commurelpiphi})
and thus we get the $exact$ canonical commutation relation
\begin{equation}
[\Pi_{2\mu },~\phi_{2\mu'}]
=-
i\hbar\delta_{\mu \mu'} .
\label{exactcommurelpiphi}
\end{equation}
The relation
$\Pi_{2\mu }^* \!=\! - (\!-\! 1)^\mu \Pi_{2-\mu }$
holds and
the hermiticity condition for 
$
\Pi_{2\mu } \!
\left( \!
\Pi_{2\mu }^\dagger
\!=\!
(- 1)^\mu \Pi_{2-\mu } \!
\right)
$
is satisfied with the aid of 
(\ref{exactcommurelpiphi}) 
and the trivial relation
$
\sum_{\mu'} (\!-\! 1)^{\mu'} 
\langle 2\mu' 2 \!-\! \mu' |2\mu \rangle 
\!=\! 0
$.


\newpage

\def\thesection{\arabic{section}}
\setcounter{equation}{0}
\renewcommand{\theequation}{\arabic{section}.
\arabic{equation}}

\section{$\!\!\!\!\!$Commtativity of exact canonically conjugate momenta}

~~~~In order to assert that the operators $\Pi_{2\mu }$ are exact canonically
conjugate to $\phi_{2\mu }$, we must give a proof on commutativity
of the $exact$ canonically conjugate momenta $\Pi_{2\mu }$ among different $\mu$.
The commutation relations among $\Pi_{2\mu }$ lead to the following
form through a tedious but straightforward calculation
(for detailed calculation see Appendix A):\\[-14pt]
\begin{eqnarray}
\!\!\!\!\!\!\!\!\!\!\!\!
\begin{array}{lll}
&&
[\Pi_{2\mu },~\Pi_{2\mu'}] \\
\\[-10pt]
&=&\!\!\!\!
\hbar {\displaystyle \frac{15\sqrt{5}}{8\sqrt{2}\pi }} R_0^4 \!
\sum_\Lambda
\langle 2\mu 2\mu'|1\Lambda \rangle \!
\left(
R^{-2}\lb_{1\Lambda }R^{-2} \!-\! (-1)^\Lambda \Lb_{1-\Lambda }^* 
\right) \\
\\[-10pt]
&&
-
\sqrt{{\displaystyle \frac{35}{8\pi}}} R_0^2 \!
\sum_{\mu_1 \mu_2}
(-1)^{\mu_1} \langle 2-\mu_1 2\mu_2 |2\mu \rangle 
\phi_{2\mu_1}
{\displaystyle \frac{1}{2}} \!
\left(
R^{-2}[\Pi_{2\mu },~\Pi_{2\mu_2}] \!+\! [\Pi_{2\mu },~\Pi_{2\mu_2}]R^{-2}
\right) \\
\\[-10pt]
&&
+ (\mu \leftrightarrow \mu') \\
\\[-10pt]
&&-
{\displaystyle \frac{1}{2}} \!
\left( \!
\sqrt{{\displaystyle \frac{35}{8\pi }}} R_0^2 \!
\right)^{\!\!2} \!\!
\sum_{\mu_1 \mu_2} \! \sum_{\mu'_1 \mu'_2} 
(-1)^{\mu_1} \langle 2-\mu_1 2\mu_2 |2\mu \rangle 
(-1)^{\mu'_1} \langle 2-\mu'_1 2\mu'_2 |2\mu' \rangle \\
\\[-20pt]
&&
\hspace{100pt}
\times 
\phi_{2\mu_1} \phi_{2\mu'_1}
{\displaystyle \frac{1}{2}} \!
\left\{ \!
R^{-2}
{\displaystyle \frac{1}{2}} \!
\left(
R^{-2}[\Pi_{2\mu_2 },~\Pi_{2\mu' _2}]
\!+\!
[\Pi_{2\mu_2 },~\Pi_{2\mu' _2}]R^{-2}
\right)
\right. \\
\\[-14pt]
&&
\left.
\hspace{180pt}
\!\!+
{\displaystyle \frac{1}{2}} \!
\left(
R^{-2}[\Pi_{2\mu_2 },~\Pi_{2\mu' _2}]
\!+\!
[\Pi_{2\mu_2 },~\Pi_{2\mu' _2}]R^{-2}
\right) \!
R^{-2} \!
\right\}\\
\\[-18pt]
&&
+ (\mu \leftrightarrow \mu') .
\end{array}
\label{commurelpipi}
\end{eqnarray}
$(\mu \!\leftrightarrow\! \mu')$ denotes a term obtained
by the exchange of $\mu$ and $\mu'$.
In the above, we notice that the determinant of
coefficients of all the terms which involve commutators 
$[\Pi_{2\mu },\Pi_{2\mu' }]$ does not generally vanish
under the convection to conceive $R^{2}$ as a constant $c$-number.
As for the constant, we will discuss in the end of the next section.
Thus we have the conclusion\\[-10pt]
\begin{equation}
[\Pi_{2\mu },~\Pi_{2\mu'}] = 0 ,
\label{commurelpipi0}
\end{equation}
if an inhomogeneous term 
$\!
R^{-2} \lb_{\!1\Lambda } R^{-2}
\!-\!
(\!-\!1)^\Lambda \! \Lb_{\!1\!-\!\Lambda }^* \!
$
becomes zero.
Thus, we could give a proof on the commutativity of 
collective momenta $\Pi_{2\mu }$.
From this proof, it turns out that $\Pi_{2\mu }$ are 
exactly canonical conjugate to $\phi_{2\mu }$. 
Due to this fact, 
the $\!(\!-\!1)^\Lambda \! R^{2} \! \Lb_{\!1-\Lambda }^* \! R^{2}\!$
can be understood as the collective angular momentum 
$\Lb_{1\Lambda }^{\mbox{coll}}$.
Therefore, the set
$\{ \phi_{2\mu } ,\Pi_{2\mu } \}$ 
is regarded as a set of $exact$ canonical variables, 
if we restrict the Hilbert 
space to the collective sub-space 
$|\Psi^{\mbox{coll}} \rangle$ 
which satisfies the following subsidiary condition:\\[-8pt]
\begin{equation}
\left(
\lb_{1\Lambda } - \Lb_{1\Lambda }^{\mbox{coll}}
\right)
|\Psi^{\mbox{coll}} \rangle = 0 .
\label{Lsubsidiarycondition}
\end{equation}
This condition is very reasonable and also implies that 
our choice of
collective variables is suitable for our aim.
Finally, owing to the relations
(\ref{exactpi})
and
(\ref{commurelpipi0}), 
the commutation relation
$[\pi_{2\mu },~\Pi_{2\mu'}]$
is rewritten as\\[-14pt]
\begin{eqnarray}
\begin{array}{rcl}
& &[\pi_{2\mu },~\Pi_{2\mu'}] 
=
-i\hbar \sqrt{{\displaystyle \frac{35}{8\pi }}} R_0^2 \!
\sum_\Lambda
\langle 2\mu 2\mu'|2\Lambda \rangle 
{\displaystyle \frac{1}{2}} \!
\left(
R^{-2} \Pi_{2\Lambda } \!+\! \Pi_{2\Lambda } R^{-2}
\right)\\
\\[-12pt]
&+&\!\!\!\!
\sqrt{{\displaystyle \frac{35}{8\pi }}} R_0^2 \!
\sum_{\mu_1 \mu_2} 
(-1)^{\mu_1} \langle 2-\mu_1 2\mu_2 |2\mu \rangle
\phi_{2\mu_1} 
{\displaystyle \frac{1}{2}} \!
\left(
[\Pi_{2\mu'},R^{-2}] \Pi_{2\mu_2} \!+\! \Pi_{2\mu_2} [\Pi_{2\mu'},R^{-2}] 
\right) ,
\end{array}
\label{commurelpi2Pi2}
\end{eqnarray}
which is approximated as\\[-16pt]
\begin{eqnarray}
\approx 
-i\hbar \sqrt{{\displaystyle \frac{35}{8\pi }}} R_0^2 R^{-2} 
\langle 2\mu 2\mu' |2 \mu \!+\! \mu' \rangle \Pi_{2\mu \!+\! \mu' } .
\label{commurelpiPi2}
\end{eqnarray}


\newpage

\def\thesection{\arabic{section}}
\setcounter{equation}{0}
\renewcommand{\theequation}{\arabic{section}.
\arabic{equation}}

\section{$\!\!\!\!\!\!\Pi_{2\mu}$- and $\phi_{2\mu}$- dependence of  kinetic part of Hamiltonian}

~~~~A remaining task is to express a kinetic part $T$ of Hamiltonian 
in terms of $\phi_{2\mu }$ and $\Pi_{2\mu }$.
Following Sunakawa's method, we first investigate 
$\Pi_{2\mu }$-dependence of $T$.
For this purpose, 
we expand it in a power series of the exact canonical conjugate momenta 
$\Pi_{2\mu }$ 
as follows:\\[-20pt]
\begin{eqnarray}
\begin{array}{rcl}
T
\!\!&=&\!\! 
T^{(0)} (\phi ;R^2)
+
\sum_\mu 
{\displaystyle \frac{1}{2}}
\left\{
T_{2\mu }^{(1)} (\phi;R^2) \Pi_{2\mu }
+
\Pi_{2\mu } T_{2\mu }^{(1)} (\phi;R^2)
\right\} \\
\\[-6pt]
& &\!\!\!\!
+ 
\sum_{\mu \mu'} 
\langle 2\mu 2\mu' |00 \rangle 
{\displaystyle \frac{1}{2}}
\left\{
T_{00 }^{(2)} (\phi:R^2) 
\Pi_{2\mu } \Pi_{2\mu'}
+
\Pi_{2\mu } \Pi_{2\mu'} 
T_{00 }^{(2)} (\phi:R^2) 
\right\} \\
\\[-6pt]
& &\!\!\!\!
+ 
\sum_{L(\neq 0)\Lambda } \sum_{\mu \mu'} 
\langle 2\mu 2\mu' |L\Lambda \rangle 
{\displaystyle \frac{1}{2}}
\left\{
T_{L\Lambda }^{(2)} (\phi:R^2) 
\widehat{\Pi }_{2\mu } \widehat{\Pi }_{2\mu'}
\!+\!
\widehat{\Pi }_{2\mu } \widehat{\Pi }_{2\mu'} 
T_{L\Lambda }^{(2)} (\phi:R^2) 
\right\}  
\!+\!
\cdots ,
\end{array}
\label{expansionT}
\end{eqnarray}\\[-12pt]
where $\widehat{\Pi }_{2\mu }$ is defined as
$\widehat{\Pi }_{2\mu }
\!\equiv\!
{\displaystyle \frac{1}{2}}
\left(
R^{-2}\Pi_{2\mu } \!+\! \Pi_{2\mu }R^{-2}
\right)
$.
The
$T_{L\Lambda }^{(n)}(\phi;R^2)$ are unknown expansion coefficients
and satisfy the relations\\[-16pt]
\begin{eqnarray}
T_{L\mu }^{(1)}(\phi;R^2)
=
-(-1)^{\mu }T_{L-\mu }^{(1)*}(\phi;R^2),~~
T_{L\Lambda }^{(2)}(\phi;R^2)
=
(-1)^{L+\Lambda}T_{L-\Lambda }^{(2)*}(\phi;R^2) .
\end{eqnarray}
In order to get the explicit expression for $T^{(n)}~(n\neq 0)$,
using the commutation relation
$
[\widehat{\Pi }_{2\mu },~\phi_{2\mu' } ]
\!=\!
-
i\hbar R^{-2} \delta_{\mu \mu'}
$, 
we take the commutators with $\phi_{2\mu }$ in the following way:\\[-16pt]
\begin{eqnarray}
\begin{array}{rcl}
&\!\!\!\!&\!\!\!\!\!\!\!\!\!\!\!\!
[T,~\phi_{2\mu }]
\!=\! 
-
i\hbar T_{2\mu }^{(1)} (\phi;R^2)
\!-\!
i\hbar 
\sum_{\mu'} 
\langle 2\mu 2\mu'|00 \rangle 
\left\{ \!
T_{00 }^{(2)} (\phi;R^2) \Pi_{2\mu'} 
\!+\!
\Pi_{2\mu'} 
T_{00 }^{(2)} (\phi;R^2) \!
\right\} \\
\\[-6pt]
& &\!\!\!\!\!\!\!\!\!\!\!\!
-i\hbar 
\sum_{L(\neq 0)\Lambda }\sum_{\mu'} 
\langle 2\mu 2\mu'|L\Lambda \rangle 
{\displaystyle \frac{1}{2}}
\left\{ \!
T_{L\Lambda }^{(2)} (\phi;R^2) R^{-2} \widehat{\Pi }_{2\mu'} 
\!+\!
R^{-2} \widehat{\Pi }_{2\mu'} 
T_{L\Lambda }^{(2)} (\phi;R^2) \!
\right\} \\
\\[-6pt]
& &\!\!\!\!\!\!\!\!\!\!\!\!
-i\hbar
\sum_{L(\neq 0)\Lambda } (-1)^L \sum_{\mu'}
\langle 2\mu 2\mu' |L\Lambda \rangle 
{\displaystyle \frac{1}{2}}
\left\{ \!
T_{L\Lambda }^{(2)} (\phi;R^2) \widehat{\Pi }_{2\mu'} R^{-2} 
\!+\!
\widehat{\Pi }_{2\mu'} 
R^{-2} T_{L\Lambda }^{(2)} (\phi;R^2) \!
\right\}
+
\cdots ,
\end{array}
\label{commutatorTphi}
\end{eqnarray}\\[-24pt]
\begin{eqnarray}
\begin{array}{rcl}
[[T,~\phi_{2\mu }],~\phi_{2\mu'}] 
=&&\!\!\!\!\!\!
(-i\hbar)^2 2
\langle 2\mu 2\mu' |00 \rangle
T_{00 }^{(2)} (\phi;R^2) \\
\\[-10pt]
&&\!\!\!\!\!\!\!\!\!\!
+
(-i\hbar)^2 
\sum_{L(\neq 0)\Lambda } 
\left(1 + (-1)^L \right)
\langle 2\mu 2\mu' |L\Lambda \rangle 
T_{L\Lambda }^{(2)} (\phi;R^2) R^{-4}
+
\cdots .
\end{array}
\label{commutatorTphiphi}
\end{eqnarray}
We can easily calculate the L.H.S's of 
(\ref{commutatorTphi}) and (\ref{commutatorTphiphi})
by making explicit use of the definitions 
(\ref{definitioneta}), (\ref{definitionpi}) 
and 
(\ref{exactpi})
and by taking commutators with $\phi_{2\mu}$ successively:\\[-16pt]
\begin{eqnarray}
\begin{array}{rcl}
& &
[T,~\phi_{2\mu }] 
= 
{\displaystyle \frac{4\pi }{3AR_0^2} \frac{\hbar }{i} \frac{2}{m}} 
(\!-\!1)^\mu \eta_{2-\mu } 
= 
-i\hbar 
{\displaystyle \frac{4\pi }{3AR_0^2} \frac{2}{m} \frac{1}{2R_0^2}} \\
\\[-10pt]
& &
\times 
\left\{ \!
R^2 
\!\cdot\!
\left[ 
(-1)^\mu \Pi_{2-\mu } 
\!-\! 
\sqrt{{\displaystyle \frac{35}{8\pi }}} R_0^2 
\sum_{\mu',\Lambda } 
\langle 2\mu 2\mu'|2\Lambda \rangle \phi_{2\Lambda } 
\widehat{\Pi }_{2\mu'}
\right]
\right. \\ 
\\[-10pt] 
& &
\hspace{31pt}
\!+\!
\left. 
\left[ 
(-1)^\mu \Pi_{2-\mu } 
\!-\!
\sqrt{{\displaystyle \frac{35}{8\pi }}} R_0^2 
\sum_{\mu',\Lambda } 
\langle 2\mu 2\mu'|2\Lambda \rangle \phi_{2\Lambda } 
\widehat{\Pi }_{2\mu'}
\right]
\!\cdot\! 
R^2 \!
\right\} ,
\end{array}
\label{commutatorTphi2}
\end{eqnarray}\\[-28pt]
\begin{eqnarray}
\begin{array}{c}
[[T,~\phi_{2\mu }],~\phi_{2\mu'}] 
=
(-i\hbar)^2 
{\displaystyle \frac{4\pi }{3AR_0^4} \frac{2}{m}} 
\left\{ \!
\sqrt{5}R^2
\langle 2\mu 2\mu'|00 \rangle 
\!-\!
\sqrt{{\displaystyle \frac{35}{8\pi }}}R_0^2
\sum_\Lambda
\langle 2\mu 2\mu'|2\Lambda \rangle \phi_{2\Lambda } \!
\right\} ,
\end{array}
\label{commutatorTphiphi2}
\end{eqnarray}
\begin{equation}
[[[T,~\phi_{2\mu }],~\phi_{2\mu'}],~\phi_{2\mu''}] = 0 .
\label{commutatorTphiphiphi}
\end{equation}
The commutation relations
$[\phi_{2\mu },~R^{-2}]$
and
$[\pi_{2\mu },~R^{-2}]$
are already given in
(\ref{commuphipir2}).
On the other hand
the commutation relation
$[\Pi_{2\mu },~R^{-2}]$
is computed as

\begin{eqnarray}
\begin{array}{rcl}
& &\!\!\!\!\!\!\!\!\!\!\!\!
[\Pi_{2\mu },~R^{-2}]
=
[\pi_{2\mu },~R^{-2}] \\
\\[-14pt]
& &\!\!\!\!\!\!\!\!\!\!\!\!
+
\sqrt{{\displaystyle \frac{35}{8\pi }}}
R_0^2  
\sum_{\mu_1 \mu_2} 
(-1)^{\mu_1} \!
\langle 2-\mu_1 2\mu_2 |2\mu \rangle 
\phi_{2\mu_1}
{\displaystyle \frac{1}{2}} \!
\left(
R^{-2}[\Pi_{2\mu_2},~R^{-2}] \!+\! [\Pi_{2\mu_2},~R^{-2}]R^{-2}
\right) \\
\\[-14pt]
& &\!\!\!\!\!\!\!\!\!\!\!\!
\hspace{55pt}
=
i\hbar {\displaystyle \frac{2 \cdot 5}{4\pi }}
R_0 ^4 R^{-6}
\phi_{2\mu }^* \\
\\[-16pt]
& &\!\!\!\!\!\!\!\!\!\!\!\!
+
\sqrt{{\displaystyle \frac{35}{8\pi }}}
R_0^2  
\sum_{\mu_1 \mu_2} 
(-1)^{\mu_1} \!
\langle 2-\mu_1 2\mu_2 |2\mu \rangle 
\phi_{2\mu_1}
{\displaystyle \frac{1}{2}} \!
\left(
R^{-2}[\pi_{2\mu_2},~R^{-2}] \!+\! [\pi_{2\mu_2},~R^{-2}]R^{-2}
\right) 
+
\cdots ,\\
\\[-18pt]
& &
\!\!\!\!\!\!\!\!\!\!\!\!
\hspace{55pt}
=
i\hbar {\displaystyle \frac{2 \cdot 5}{4\pi }}
R_0 ^4 R^{-6} \!
\left\{ \!
\phi_{2\mu }^*
\!+\!
\sqrt{{\displaystyle \frac{35}{8\pi }}}
R_0^2 R^{-2} \!
\sum_{\mu_1 \mu_2} 
\langle 2\mu_1 2\mu_2 |2\mu \rangle
\phi_{2\mu_1}^* \phi_{2\mu_2}^*
\right. \\
\\[-24pt]
& &
\left.
\!\!\!\!\!\!\!\!\!\!\!\!\!\!\!\!\!\!\!\!\!\!\!\!\!\!\!\!\!\!\!\!\!
+\!
\left( \!\!
\sqrt{{\displaystyle \frac{35}{8\pi }}} \!\!
R_0^2 R^{-2} \!\!
\right)^{\!\!-2} \!\!\!\!
\sum_{L \Lambda } \!\!
\sqrt{5(2L \!\!+\!\! 1)}W(2222;2L) \!
\sum_{\mu_1 \mu_3 \mu_5} \!
\langle 2\mu_1 2\mu_3 |L\Lambda \rangle \!
\langle L\Lambda 2\mu_4 |2\mu \rangle \!
\phi_{2\mu_1}^* \phi_{2\mu_3}^* \phi_{2\mu_4}^* 
\!\!+
\!\cdots\! \!\!\!\!
{}^{^{^{^{^{^{^{.}}}}}}}
\right\} \! , 
\end{array}
\label{commutatorpiR2}
\end{eqnarray}\\[-16pt]
which is  approximated as\\[-22pt]
\beq
[\Pi_{2\mu },~R^{-2}]
\approx
i\hbar {\displaystyle \frac{2 \cdot 5}{4\pi }}
R_0 ^4 R^{-6}
\phi_{2\mu }^* .
\label{apprcommuPaiR}
\eeq
Due to
(\ref{commuphipir2}) and
(\ref{commutatorpiR2}) or (\ref{apprcommuPaiR}),
we have the commutation relations\\[-16pt]
\begin{eqnarray}
[[\Pi_{2\mu },~R^{-2}],~R^{\pm 2}] = 0, ~~
[[\Pi_{2\mu },~R^{-2}],~T_{L\Lambda }^{(n)} (\phi;R^2)] = 0 .
\label{commutatorpiR2R2}
\end{eqnarray}
Using 
(\ref{commutatorpiR2R2})
and
comparing 
(\ref{commutatorTphi2})
$\sim$
(\ref{commutatorTphiphiphi}) 
with 
(\ref{commutatorTphi})
and
(\ref{commutatorTphiphi}),  
$T^{(n)}$ are determined as\\[-16pt]
\begin{eqnarray}
\!\!\!\!\!\!\!\!\!\!\!\!\!\!\!\!
\left.
\begin{array}{rcl}
&& T_{2\mu }^{(1)} (\phi;R^2) \!=\! 0 ,~
T_{00}^{(2)} (\phi;R^2)
\!=\!
{\displaystyle \frac{4\pi }{3AR_0^4} \frac{1}{m}}
\sqrt{5} R^2 ,~
T_{2\Lambda }^{(2)} (\phi;R^2)
\!=\!
-{\displaystyle \frac{4\pi }{3AR_0^2} \frac{1}{m}} 
\sqrt{{\displaystyle \frac{35}{8\pi }}} R^4
\phi_{2\Lambda } ,\\
\\[-10pt]
&& 
T_{L\Lambda }^{(2)} (\phi;R^2) \!=\! 0 
~\mbox{for $L=1,3,4$},~~ 
T_{L\Lambda }^{(n)} (\phi;R^2) \!=\! 0
~\mbox{for $n\geq 3$} .
\end{array} \!\!
\right\}
\label{Tcoefficients}
\end{eqnarray}
Substituting 
(\ref{Tcoefficients}) into 
(\ref{expansionT}), 
we can get the exact 
$\Pi_{2\mu }$-dependence of the kinetic part $T$ of Hamiltonian as follows:\\[-18pt]
\begin{eqnarray}
\begin{array}{rcl}
T 
\!\!&=&\!\! 
T^{(0)} (\phi;R^2) 
\!+\!
{\displaystyle \frac{4\pi }{3AR_0^4} \frac{1}{m} } \!
\sum_\mu \!
{\displaystyle \frac{1}{2}} \!
\left\{ 
R^2\Pi_{2\mu }(-1)^\mu \Pi_{2-\mu }
\!+\!
\Pi_{2\mu }(-1)^\mu \Pi_{2-\mu }R^2
\right\} \\
\\[-12pt]
& &\!\! 
-{\displaystyle \frac{4\pi }{3AR_0^2} \frac{1}{m}} 
\sqrt{{\displaystyle \frac{35}{8\pi }}} \!
\sum_{\mu \mu'} 
\langle 2\mu 2\mu'|2\Lambda \rangle \!
\sum_\Lambda \!
{\displaystyle \frac{1}{2}} \!
\left\{ \!
R^4 \phi_{2\Lambda }\widehat{\Pi }_{2\mu } \widehat{\Pi }_{2\mu'} 
\!+\! 
\widehat{\Pi }_{2\mu } \widehat{\Pi }_{2\mu'}R^4 \phi_{2\Lambda } \!
\right\} \! .
\end{array}
\label{expansionT2}
\end{eqnarray}\\[-9pt]
In 
(\ref{expansionT2}), 
the second term is a kinetic energy part of a quadrupole-type nuclear collective motion. 
Indeed it has the kinetic form of five-dimensional harmonic oscillators.
It is remarkable that
the mass parameter is in full accordance with the irrotational flow assumption
of the BMM
which is but inconsistent with the microscopic dynamics of the nucleons.
$\!\!$To correct for the irrotational assumption,
Gneuss and Greiner
proposed a deformation-dependent mass parameter
using the Bohr-Mottelson-Frankfurt collective model
\cite{GG.71}
(see Chap.8 of textbook
\cite{EG.87}
and Chap.4 of textbook
\cite{RoweWood.2010}).
This deformation mass parameter is in exact correspondence with the third term in
(\ref{expansionT2}) 
which also coincides with the mass parameter given by Miyazima and Tamura
\cite{Tamura.56} 
if we neglect an effect induced by the non-commutativity
(\ref{apprcommuPaiR}).
Thus we derive a part of collective Hamiltonian,
the
$\Pi_{2\mu }$-dependence of the Hamiltonian.
In the recent topical review on the Bohr Hamiltonian
\cite{PR.09},
with the use of a mean-field theory,
Pr\'{o}chniak and Rohozi\'{n}ski
have obtained the generalized form of the Bohr Hamiltonian
which contains a mass tensor in the kinetic energy.
On the contrary,
we could get naturally such a term
only using the exact canonically conjugate momenta defined by
(\ref{exactpi})
without making
any mean-field approximation.
A potential energy part of the collective 
harmonic oscillation and another part of the anharmonic effects
(quadrupole-type phonon-phonon interaction) is derived from the two-body
potential $V$ and also from the term $T^{(0)} (\phi;R^2)$.

Finally, we should stress the fact that up to this stage,
all the expressions are exact.

\newpage

Our next task is to determine the term $T^{(0)}(\phi;R^2)$.
For this purpose, we expand it in a power series of the collective
coordinates $\phi_{2\mu }$ in the form
\begin{eqnarray}
\begin{array}{c}
T^{(0)} (\phi ;R^2)
\!=\!
C_0 (R^2)
\!+\!
\sum_\mu \!
C_{1 \mu } (R^2) \phi_{2\mu }^*
\!+\! 
\sum_{\mu \mu'} \!
C_{2 \mu  \mu' } (R^2) \phi_{2\mu }^* \phi_{2\mu'}^* 
\!+\!
\cdots ,
\end{array}
\label{expansionT0}
\end{eqnarray}
where
$
C_{2 \mu  \mu' } (R^2)
\!=\!
C_{2 \mu'  \mu } (R^2)
$.
In the above,
the expansion coefficients $C_n  (R^2)$ are determined
in a manner quite parallel to the manner used before.

First, from the definition 
(\ref{exactpi}), 
we have the following discrete integral equation:\\[-14pt]
\begin{eqnarray}
\!\!\!\!\!\!\!\!
\begin{array}{rcl}
& &\!\!\!\!
[\Pi_{2\mu },~T^{(0)}(\phi;R^2)]
=
f_{2\mu }\\
\\[-8pt]
& &
\!\!\!\!
+
\sqrt{{\displaystyle \frac{35}{8\pi }}}
R_0^2  
\sum_{\mu_1 \mu_2} 
(-1)^{\mu_1} 
\langle 2-\mu_1 2\mu_2 |2\mu \rangle 
\phi_{2\mu_1} \!\!
\left[
{\displaystyle \frac{1}{2}}
\left(
R^{-2}\Pi_{2\mu_2} + \Pi_{2\mu_2}R^{-2}
\right),
~T^{(0)}(\phi;R^2)
\right] ,
\end{array}
\label{commutatorpiT0}
\end{eqnarray}\\[-4pt]
and the inhomogeneous term $f_{2\mu }$ becomes\\[-14pt]
\begin{eqnarray}
\begin{array}{cc}
&
f_{2\mu } 
\equiv
[\pi_{2\mu },~T^{(0)}(\phi;R^2)] 
\\
\\[-8pt]
&= 
[\pi_{2\mu },~T]
 - {\displaystyle \frac{4\pi }{3AR_0^4} \frac{1}{m}}
\sum_{\mu_1} \!\!
\left[
\pi_{2\mu },~
{\displaystyle \frac{1}{2}} 
\left\{ 
R^2\Pi_{2\mu_1 }(-1)^{\mu_1} \Pi_{2-\mu_1 } 
+  \Pi_{2\mu_1 }(-1)^{\mu_1} \Pi_{2-\mu_1 }R^2
\right\} \!
\right] 
\\ 
\\[-10pt]
&
+
{\displaystyle \frac{4\pi }{3AR_0^2} \frac{1}{m}} 
\sqrt{{\displaystyle \frac{35}{8\pi }}}
\sum_{\mu_1 \mu_2} 
\sum_\Lambda 
\langle 2\mu_1 2\mu_2|2\Lambda \rangle \!\!
\left[
\pi_{2\mu },~
{\displaystyle \frac{1}{2}}
\left\{ \!
R^4 \phi_{2\Lambda }\widehat{\Pi }_{2\mu_1 } \widehat{\Pi }_{2\mu_2} 
+ 
\widehat{\Pi }_{2\mu_1 } \widehat{\Pi }_{2\mu_2 }R^4 \phi_{2\Lambda } \!
\right\} 
\right]
\\ 
\\[-12pt]
&\!\!\!\!
= 
[\pi_{2\mu },~T]
-
{\displaystyle \frac{4\pi }{3AR_0^4} \frac{\sqrt{5}}{m}} 
\sum_{\mu_1 \mu_2} \langle 2\mu_1 2\mu_2|00 \rangle \!
\cdot 
{\displaystyle \frac{1}{2}} 
\left\{
[\pi_{2\mu },~R^2] \Pi_{2\mu_1} \Pi_{2\mu_2}
\!+\!
\Pi_{2\mu_1} \Pi_{2\mu_2} [\pi_{2\mu },~R^2] 
\right. 
\\ 
\\[-8pt]
&
\left.
+
R^2 [\pi_{2\mu },~\Pi_{2\mu_1} ] \Pi_{2\mu_2}
\!+\!
R^2 \Pi_{2\mu_1} [\pi_{2\mu },~\Pi_{2\mu_2} ] 
\!+\!
[\pi_{2\mu },~\Pi_{2\mu_1} ] \Pi_{2\mu_2} R^2 
\!+\!
\Pi_{2\mu_1} [\pi_{2\mu },~\Pi_{2\mu_2} ] R^2
\right\} 
\\
\\[-8pt]
& 
+
{\displaystyle \frac{4\pi }{3AR_0^2} \frac{1}{m}} 
\sqrt{{\displaystyle \frac{35}{8\pi }}}
\sum_{\mu_1 \mu_2} 
\sum_\Lambda 
\langle 2\mu_1 2\mu_2|2\Lambda \rangle
\cdot 
{\displaystyle \frac{1}{2}} 
\\ 
\\[-8pt]
&\!\!\!\!\!\!\!\!\!\!\!\!\!\!\!
\!\times\!
\left\{ \!
[\pi_{2\mu },\!~R^4] \phi_{2\Lambda } 
\widehat{\Pi }_{2\mu_1} \! \widehat{\Pi }_{2\mu_2}
\!\!+\!\!
\widehat{\Pi }_{2\mu_1} \! \widehat{\Pi }_{2\mu_2} 
[\pi_{2\mu },\!~R^4] \phi_{2\Lambda } 
\!\!+\!\!
R^4 [\pi_{2\mu },\!~\phi_{2\Lambda }]  
\widehat{\Pi }_{2\mu_1} \! \widehat{\Pi }_{2\mu_2}
\!\!+\!\!
\widehat{\Pi }_{2\mu_1} \! \widehat{\Pi }_{2\mu_2} 
R^4 [\pi_{2\mu },\!~\phi_{2\Lambda }]
\right. 
\\ 
\\[-6pt]
&\!\!\!\!\!\!\!\!\!\!\!\!\!\!\!
\left.
+
R^4 \! \phi_{2\Lambda } [\pi_{2\mu },\!~\widehat{\Pi }_{2\mu_1} \!] 
\widehat{\Pi }_{2\mu_2}
\!\!+\!\!
R^4 \! \phi_{2\Lambda } \widehat{\Pi }_{2\mu_1} \!
[\pi_{2\mu },\!~\widehat{\Pi }_{2\mu_2} \!] 
\!\!+\!\!
[\pi_{2\mu },\!~\widehat{\Pi }_{2\mu_1} \!] 
\widehat{\Pi }_{2\mu_2} R^4 \! \phi_{2\Lambda }
\!\!+\!\!
\widehat{\Pi }_{2\mu_1} \!
[\pi_{2\mu },\!~\widehat{\Pi }_{2\mu_2} \!] 
R^4 \! \phi_{2\Lambda } \!
\right\} \!\! , 
\end{array}
\label{fcommutatorpiT0}
\end{eqnarray}\\[-4pt]
in which the commutators
$[\pi_{2\mu },~R^4]$ 
and 
$[\pi_{2\mu },~\widehat{\Pi }_{2\mu'} ]$
are given as\\[-12pt]
\begin{eqnarray}
[\phi_{2\mu },~R^{\pm 4}] = 0 ,~~
[\pi_{2\mu },~R^{\pm 4}] 
= 
-i\hbar \frac{4 \cdot 5}{4\pi }
R_0^4 R^{-2} \!
\left( \!\!
\begin{array}{c}
R^2 \\
-R^{-6} \\
\end{array} \!\!
\right) \! 
\phi_{2\mu }^* ,
\label{commuphipir4}
\end{eqnarray}
\vspace{-0.5cm}
\begin{eqnarray}
\!\!\!\!
\BA{rl}
&[\pi_{2\mu },~\widehat{\Pi }_{2\mu'} ]
\!=\!
i\hbar {\displaystyle \frac{5}{4\pi }}
R_0^4 \!
\left( \!
R^{-6} \phi_{2\mu }^* {\Pi }_{2\mu'}
\!+\!
{\Pi }_{2\mu'} R^{-6} \phi_{2\mu }^* 
\right)
\!+\!
{\displaystyle \frac{1}{2}} \!
\left( \!
R^{-2} [\pi_{2\mu },~{\Pi }_{2\mu'} ]
\!+\!
[\pi_{2\mu },~{\Pi }_{2\mu'} ] R^{-2} 
\right) \\
\\[-8pt]
&
\!=\!
-
(i\hbar )^2 {\displaystyle \frac{5}{4\pi }}
R_0^4 R^{-6}
(-1)^\mu \delta_{\mu', -\mu }
\!+\!
i\hbar {\displaystyle \frac{2 \cdot 5}{4\pi }}
R_0^4 R^{-6} \phi_{2\mu }^* {\Pi }_{2\mu'}
\!+\!
i\hbar {\displaystyle \frac{3 \cdot 5}{4\pi }}
R_0^4 R^{-4} \phi_{2\mu }^* 
[{\Pi }_{2\mu'},~R^{-2}] \\
\\[-8pt]
&
~~\!+
R^{-2} [\pi_{2\mu },~{\Pi }_{2\mu'} ]
+
{\displaystyle \frac{1}{2}}
[[\pi_{2\mu },~{\Pi }_{2\mu'} ], ~R^{-2}] \\
\\[-12pt]
&
\!=\!
-
(i\hbar )^2 {\displaystyle \frac{5}{4\pi }}
R_0^4 R^{-6}
(-1)^\mu \delta_{\mu', -\mu }
\!+\!
i\hbar {\displaystyle \frac{2 \cdot 5}{4\pi }}
R_0^4 R^{-6} \phi_{2\mu }^* {\Pi }_{2\mu'}
+
(i\hbar )^2 {\displaystyle \frac{3}{2 }}
{\displaystyle \frac{(2 \cdot 5)^2}{(4\pi )^2}}
R_0^8 R^{-10} \phi_{2\mu }^* \phi_{2\mu' }^* \\
\\[-12pt]
&
~-
i\hbar
\sqrt{{\displaystyle \frac{35}{8\pi }}}
R_0^2 R^{-4} 
\langle 2\mu 2\mu'|2\mu \!+\!\mu' \rangle {\Pi }_{2\mu \!+\!\mu'}
-
(i\hbar)^2
{\displaystyle \frac{1}{2}}
{\displaystyle \frac{2 \cdot 5}{4\pi }}
 \sqrt{{\displaystyle \frac{35}{8\pi }}}
R_0^6 R^{-8} \langle 2\mu 2\mu'|2\mu \!+\!\mu' \rangle 
{\phi }_{2\mu \!+\!\mu'}^*,
\EA
\label{commurelpihatPi2}
\end{eqnarray}
in the last two lines of 
(\ref{commurelpihatPi2})
we have used the previous approximate relations
(\ref{commurelpiPi2}) and (\ref{apprcommuPaiR}).
Using
(\ref{expansionT2}),
after troublesome calculation, 
the commutation relation between
the functional operators $f_{2\mu }$
and
$\phi_{2\mu'}$
reads

\newpage

\begin{eqnarray}
&
\!\!\!\!\!\!\!\!\!\!\!\!\!\!\!\!\!\!\!\!\!\!\!\!\!\!\!\!\!\!\!\!\!\!\!\!\!\!\!\!\!\!\!\!\!\!\!\!\!\!\!\!\!\!\!\!\!\!\!\!\!\!\!\!
\!\!\!\!\!\!\!\!\!\!\!\!\!\!\!\!\!\!\!\!\!\!\!\!\!\!\!\!\!\!\!\!\!\!\!\!\!\!\!\!\!\!\!\!\!\!\!\!\!\!\!\!\!\!\!\!\!\!\!\!\!\!\!\!
\!\!\!\!\!\!\!\!\!\!\!\!\!\!\!\!\!\!\!\!\!\!\!\!\!\!\!\!\!\!\!\!\!\!\!
[f_{2\mu },~ \phi_{2\mu' }] 
= 
[[\pi_{2\mu },~T],~ \phi_{2\mu' }] \nonumber \\
\nonumber \\[-12pt]
&
-
i\hbar
{\displaystyle \frac{4\pi }{3A R_0^4} \frac{1}{m}} 
\left\{
i\hbar
{\displaystyle \frac{2 \!\cdot\! 5}{4\pi }} R_0^4 
\left[
R^{-2} \phi_{2\mu }^*,~(-1)^{\mu'} \Pi_{2-\mu'}
\right]_+
-
[[\pi_{2\mu },~(-1)^{\mu'} \Pi_{2-\mu'} ],~R^2]_+
\right\} \nonumber \\
\nonumber \\[-8pt]
&\!\!\!\!\!\!\!\!
-
{\displaystyle \frac{4\pi }{3A R_0^4} \frac{1}{m}} \!
\sum_{\mu_1} \!\! 
\left[ 
{\displaystyle \frac{1}{2}}
\left\{
(-1)^{\mu_1} \Pi_{2-\mu_1} 
[[\pi_{2\mu },~\Pi_{2\mu_1} ],~\phi_{2\mu' } ]
\!+\!
[[\pi_{2\mu },~\Pi_{2\mu_1} ],~\phi_{2\mu' } ]
(-1)^{\mu_1} \Pi_{2-\mu_1}
\right\} \! ,\!~R^2 
\right]_{\!\!+} \nonumber \\
\nonumber \\[-12pt]
&
+
(i\hbar )^2
{\displaystyle \frac{4\pi }{3A R_0^2} \frac{1}{m}} 
\sqrt{{\displaystyle \frac{35}{8\pi }}}
\sum_{\mu_1} \!
\langle 2\mu' 2\mu_1|2\mu \rangle \! 
\left[
{\displaystyle \frac{1}{2}} \!
\left(
R^{-2} \widehat{\Pi }_{2\mu_1} + \widehat{\Pi }_{2\mu_1} R^{-2}
\right) \! ,~R^4
\right]_{\!\!+} \nonumber \\
\nonumber \\[-8pt]
&
+
(i\hbar )^2
{\displaystyle \frac{4 \!\cdot\! 5}{3A} \frac{1}{m}} 
\sqrt{{\displaystyle \frac{35}{8\pi }}} R_0^2
\sum_{\mu_1} \sum_\Lambda \!
\langle 2\mu' 2\mu_1|2\Lambda \rangle \! 
\left[
{\displaystyle \frac{1}{2}} \!
\left(
R^{-2} \widehat{\Pi }_{2\mu_1} + \widehat{\Pi }_{2\mu_1} R^{-2}
\right) \! ,~\phi_{2\mu }^* \phi_{2\Lambda }
\right]_{\!\!+} \nonumber \\
\nonumber \\[-10pt]
&\!\!\!\!\!\!\!\!
-
(i\hbar )^2
{\displaystyle \frac{35}{6A} \frac{1}{m}} \!
\sum_{\mu_1} \! \sum_{\Lambda \Lambda'} \!
\langle 2\mu' 2\mu_1|2\Lambda \rangle \! 
(-1)^{\mu }
\langle 2-\mu 2\Lambda|2\Lambda' \rangle \!
\left[
{\displaystyle \frac{1}{2}} \!
\left( \!
R^{-2} \widehat{\Pi }_{2\mu_1} 
\!+\! 
\widehat{\Pi }_{2\mu_1} R^{-2} \!
\right) \! ,~R^2 \phi_{2\Lambda' } \!
\right]_{\!\!+} \nonumber \\
\nonumber \\[-10pt]
& 
-
i\hbar {\displaystyle \frac{4\pi }{3A R_0^2} \frac{1}{m}} 
\sqrt{{\displaystyle \frac{35}{8\pi }}}
\sum_{\mu_1} 
\sum_\Lambda \!
\langle 2\mu' 2\mu_1|2\Lambda \rangle \! 
\left[
{\displaystyle \frac{1}{2}} \!
\left( \!
R^{-2} [\pi_{2\mu },~\widehat{\Pi }_{2\mu_1} \!] 
\!+\! 
[\pi_{2\mu },~\widehat{\Pi }_{2\mu_1} \!] R^{-2} \!
\right) \! ,~R^4 \phi_{2\Lambda }
\right]_{\!\!+} \nonumber \\
\nonumber \\[-12pt]
&
+
{\displaystyle \frac{4\pi }{3A R_0^2} \frac{1}{m}} 
\sqrt{{\displaystyle \frac{35}{8\pi }}} 
\sum_{\mu_1 \mu_2} 
\sum_\Lambda 
\langle 2\mu_1 2\mu_2|2\Lambda \rangle \nonumber \\
\nonumber \\[-12pt]
&
\times 
\left[ 
{\displaystyle \frac{1}{2}} \! 
\left( 
\widehat{\Pi }_{2\mu_1} 
[[\pi_{2\mu },~\widehat{\Pi }_{2\mu_2} \!],~\phi_{2\mu'} ]
+
[[\pi_{2\mu },~\widehat{\Pi }_{2\mu_1} \!],~\phi_{2\mu'} ]
\widehat{\Pi }_{2\mu_2} 
\right) \! , ~R^4 \phi_{2\Lambda } 
\right]_{\!\!+}  , 
\label{fcommutatorphi2}
\end{eqnarray}\\[-16pt]
in which the commutators
$
[[\pi_{2\mu },~\Pi_{2\mu_1} ],~\phi_{2\mu' } ]
$
and
$
[[\pi_{2\mu },~\widehat{\Pi }_{2\mu_1} \!],~\phi_{2\mu'} ]
$
are given, respectively as\\[-16pt]
\begin{eqnarray}
\begin{array}{c}
\!\!\!\!\!\!\!\!
[[\pi_{2\mu }, \Pi_{2\mu_1} ], \phi_{2\mu' } ]
\!=\!
(i\hbar )^2 \!
\sqrt{{\displaystyle \frac{35}{8\pi }}} \!
R_0^2 R^{-2} \!
\left\{ \!\!
\langle 2\mu 2\mu_1|2\mu' \rangle
\!-\!
{\displaystyle \frac{2 \!\cdot\! 5}{4\pi }} \!
R_0^4 R^{-4} \!
\sum_{\mu'_1} \!
\langle 2\mu'_1 2\mu'|2\mu \rangle 
\phi_{2\mu'_1 }^* \phi_{2\mu_1}^* \!\!
\right\} \! ,
\end{array}
\label{commurelpiPiphi}
\end{eqnarray}
\vspace{-0.6cm}
\begin{eqnarray}
\begin{array}{ll}
[[\pi_{2\mu }, \widehat{\Pi }_{2\mu_1} \!], \phi_{2\mu'} ]
\!=\!&\!\!\!
-
(i\hbar )^2  
{\displaystyle \frac{2 \!\cdot\! 5}{4\pi }} 
R_0^4 R^{-6} \phi_{2\mu }^*
\!\cdot\!
\delta_{\mu_1 \mu'}
\!+\!
(i\hbar )^2 
\sqrt{{\displaystyle \frac{35}{8\pi }}}
R_0^2 R^{-4}
\langle 2\mu 2\mu_1|2\mu' \rangle \\
\\[-14pt]
&\!\!\!
-
(i\hbar )^2  
{\displaystyle \frac{2 \!\cdot\! 5}{4\pi }}
\sqrt{{\displaystyle \frac{35}{8\pi }}}
R_0^6 R^{-8} \!
\sum_{\mu'_1} 
\langle 2\mu'_1 2\mu'|2\mu \rangle 
\phi_{2\mu'_1 }^* \phi_{2\mu_1}^* \! ,
\end{array}
\label{commurelpihatPiphi}
\end{eqnarray}\\[-10pt]
and the commutator
$[[\pi_{2\mu },~T],~ \phi_{2\mu' }]$
is given exactly by
(\ref{commutatorpiTphi})
in Appendix B.
In
(\ref{fcommutatorphi2})
the symbol
$[A,~B]_{\!+}$
denotes the anticommutator between the operators $A$ and $B$.

Substituting 
(\ref{commurelpiPiphi}) and (\ref{commurelpihatPiphi})
into
(\ref{fcommutatorphi2}),
we calculate the commutator
$[f_{2\mu },~ \phi_{2\mu' }]$
as\\[-18pt]
\begin{eqnarray}
\!\!\!\!\!\!\!\!
&
\!\!\!\!\!\!\!\!\!\!\!\!\!\!\!\!\!\!\!\!\!\!\!\!\!\!\!\!\!\!\!\!\!\!\!\!\!\!\!\!\!\!\!\!\!\!\!\!\!\!\!\!\!\!\!\!\!\!\!\!\!\!\!\!
\!\!\!\!\!\!\!\!\!\!\!\!\!\!\!\!\!\!\!\!\!\!\!\!\!\!\!\!\!\!\!\!\!\!\!\!\!\!\!\!\!\!\!\!\!\!\!\!\!\!\!\!\!\!
\left[f_{2\mu },~ \phi_{2\mu' }\right]
= 
\left[
\left[
{\pi }_{2\mu },~T^{(0)}(\phi ;R^2 )
\right],~ \phi_{2\mu' }
\right] \nonumber \\
\nonumber \\[-16pt]
&
\!\!\!\!\!\!\!\!\!\!\!\!\!\!\!\!\!\!\!\!\!\!\!\!\!\!\!\!\!\!\!\!\!\!\!\!\!\!\!\!\!\!\!\!\!\!\!\!\!\!\!\!\!\!\!\!\!\!\!\!\!\!\!\!
\!\!\!\!\!\!\!\!\!\!\!\!\!\!\!\!\!\!\!\!\!\!\!\!\!\!\!\!\!\!\!\!\!\!\!\!\!\!\!\!\!\!\!\!\!\!\!\!\!\!\!\!\!\!\!\!\!\!\!\!\!\!\!\!
\!\!\!\!\!\!\!\!\!\!\!\!\!\!\!\!\!\!\!\!\!\!\!\!\!\!\!\!\!\!\!\!\!\!\!\!\!\!\!\!
=
\left[
\left[\pi_{2\mu },~T
\right],~ \phi_{2\mu' }
\right] \nonumber \\
\nonumber \\[-14pt]
&\!\!\!\!\!\!\!\!
-
{\displaystyle \frac{\hbar^2 R_0^2}{2m}} \!
\left[
i\hbar
{\displaystyle \frac{20}{3A R_0^2 }} R^{-2}
\delta_{\mu\mu'}
\!-\!
{\displaystyle \frac{40}{3A R_0^2 }} R^{-2}
\phi_{2\mu }^* (\!-\! 1)^{\mu'} \Pi_{2-\mu'}
\!-\!
i\hbar
{\displaystyle \frac{50}{3 \pi A }} R_0^2 R^{-6}
\phi_{2\mu }^* 
(\!-\! 1)^{\mu'} \phi_{2-\mu'}^* 
\right. \nonumber \\
\nonumber \\[-10pt]
&\!\!\!\!\!\!\!\!
\left.
-
{\displaystyle \frac{4 \!\cdot\! 4\pi }{3A R_0^4}} \!
\sqrt{{\displaystyle \frac{35}{8\pi }}} \!
\sum_{\Lambda } \!
(\!-\! 1)^{\mu'}
\langle 2 \mu 2 \!-\! \mu'|2 \Lambda \rangle 
\Pi_{2\Lambda } 
\!-\!
i\hbar
{\displaystyle \frac{20}{3A}} \!
\sqrt{{\displaystyle \frac{35}{8\pi }}} R^{-4} \!
\sum_{\Lambda } \!
(\!-\! 1)^{\mu'} \!
\langle 2 \mu 2 \!-\! \mu'|2 \Lambda \rangle  
\phi_{2\Lambda }^*  
\right. \nonumber \\
\nonumber \\[-14pt]
&
\left.
-
{\displaystyle \frac{i}{\hbar } \frac{2 \!\cdot\! 4\pi }{3A R_0^4}} \!
\sqrt{{\displaystyle \frac{35}{8\pi }}} R^2 \!
\sum_{\mu_1\mu_2} 
\langle 2 \mu_1 2 \mu_2|2 \mu \rangle 
\phi_{2\mu_1 }^*
\right. 
\nonumber \\
\nonumber \\[-12pt]
&\!\!\!\!\!\!\!\!
\left.
\times\!
\left( \!
2 \!
\left[(\!-\! 1)^{\mu'} \! \Pi_{2-\mu'}, R^{-2}\right] \! \Pi_{2\mu_2}
\!\!-\!\!
R^2 \!
\left[ 
\Pi_{2\mu_2}, R^{-2} 
\right] \!
\left[(\!-\! 1)^{\mu'} \! \Pi_{2-\mu'}, R^{-2}\right]
\!\!+\!\!
\left[ 
\Pi_{2\mu_2},
\left[(\!-\! 1)^{\mu'} \! \Pi_{2-\mu'}, R^{-2}\right] \!
\right] \!
\right)
\right. \nonumber \\
\nonumber \\[-14pt]
&
\left.
-
{\displaystyle \frac{i}{\hbar } \frac{2 \!\cdot\! 4\pi }{3A R_0^4}} \!
\sqrt{{\displaystyle \frac{35}{8\pi }}} R^2 \!
\sum_{\mu_1} 
(\!-\! 1)^{\mu'} \!
\langle 2 \mu 2 \!-\! \mu'|2 \mu_1 \rangle 
\phi_{2\mu_1 }^* 
\right. \nonumber \\
\nonumber \\[-12pt]
&\!\!\!\!\!\!\!\!\!\!\!\!
\left.
\times\!
\sum_{\mu_2} \!
\left( \!
2
\left[ 
\Pi_{2\mu_2},R^{-2} 
\right] \!
(\!-\! 1)^{\mu_2} \!
\Pi_{2-\mu_2} 
\!\!-\!
R^2 \!
\left[ 
\Pi_{2\mu_2},R^{-2} 
\right] \!
\left[ 
(\!-\! 1)^{\mu_2} \! \Pi_{2-\mu_2}, R^{-2}
\right]
\!\!+\!
\left[ 
\Pi_{2\mu_2},
\left[(\!-\! 1)^{\mu_2} \! \Pi_{2-\mu_2}, R^{-2}\right]
\right] \!
\right)
\right. \nonumber \\
\nonumber \\[-14pt]
&
\left.
-
i\hbar 
{\displaystyle \frac{40}{3A}} \!
\sqrt{{\displaystyle \frac{35}{8\pi }}} R^{-4} \!
\sum_\Lambda 
(\!-\! 1)^{\mu'} \!
\langle 2 \mu 2 \!-\! \mu'|2 \Lambda \rangle 
\phi_{2\Lambda }^*
\!+\!
i\hbar 
{\displaystyle \frac{35}{3A R_0^2} } R^{-2} 
\delta_{\mu \mu'}
\right. \nonumber \\
\nonumber \\[-12pt]
&\!\!\!\!\!\!\!\!
\left.
+
{\displaystyle \frac{70}{3A R_0^2} } R^{2} \!
\sum_{L \Lambda } \!
\left\{
1
\!-\! 
(\!-\! 1)^L
\right\} \!
5W(2222;2L)
(\!-\! 1)^{\mu' } \!
\langle 2 \mu 2 \!-\! \mu'|L \Lambda \rangle 
\widehat{\Jb}_{L \Lambda }
\right. \nonumber 
\end{eqnarray}
\begin{eqnarray}
&\!\!\!\!\!\!\!\!\!\!\!\!\!\!\!\!
\left.
-
{\displaystyle \frac{35}{3A R_0^2} }  \!
\sum_{L \Lambda } \!
\left\{
1
\!-\! 
(\!-\! 1)^L
\right\} \!
5W(2222;2L)
(\!-\! 1)^{\mu' } \!
\langle 2 \mu 2 \!-\! \mu'|L \Lambda \rangle \!\!
\sum_{\Lambda' \Lambda'' } \!
\langle 2 \Lambda' 2 \Lambda'' |L \Lambda \rangle 
\phi_{2\Lambda' }^* \!
\left[{\Pi }_{2\Lambda''},R^{-2}
\right]
\right. \nonumber \\
\nonumber \\[-10pt]
&\!\!\!\!\!\!\!\!\!\!\!\!\!\!
\left.
+
i\hbar 
{\displaystyle \frac{20}{3A}} \!
\sqrt{{\displaystyle \frac{35}{8\pi }}} R^{-4} \!
\sum_\Lambda \!
(\!-\! 1)^{\mu'} \!
\langle 2 \mu 2 \!-\! \mu'|2 \Lambda \rangle 
\phi_{2\Lambda }^* \!
\!-\!
{\displaystyle \frac{350}{3 \pi A }} R_0^2 R^{-2} \!
\sum_{\Lambda' \Lambda'' }
(\!-\! 1)^{\mu' } \!
\langle 2 \Lambda' 2 \Lambda'' |2 \!-\! \mu' \rangle 
\phi_{2 \Lambda'  }^*
\phi_{2 \Lambda'' }^*
\widehat{\Jb}_{2 \mu }  
\right. \nonumber \\
\nonumber \\[-8pt]
&\!\!\!\!\!\!\!\!\!
\left.
+
i\hbar
{\displaystyle \frac{175}{3 \pi A } } R_0^2 R^{-6} \!
\sum_{L \Lambda } \!
\left\{
1
\!+\! 
(\!-\! 1)^L
\right\} \!
5W(2222;2L)
(\!-\! 1)^{\mu' } \!
\langle 2\mu 2 \!-\! \mu'|L \Lambda \rangle \!
\sum_{\Lambda' \Lambda'' } \!\!
\langle 2 \Lambda' 2 \Lambda''|L \Lambda \rangle 
\phi_{2\Lambda' }^*
\phi_{2\Lambda'' }^*
\right. \nonumber \\
\nonumber \\[-8pt]
&\!\!\!\!\!\!\!\!\!\!
\left.
-
{\displaystyle \frac{i}{\hbar } \frac{175}{3A R_0^2} } R^2 \!
\sum_{\mu_1\mu_2} 
\langle 2 \mu_1 2 \mu_2|2 \mu \rangle 
\phi_{2\mu_1 }^* \!
\left[
{\Pi }_{2\mu_2}, R^{-2}
\right] 
\sum_{\mu'_1\mu'_2}
(\!-\! 1)^{\mu' } \!
\langle 2 \mu'_1 2 \mu'_2|2 \!-\! \mu' \rangle 
\phi_{2\mu'_1 }^* \!
\left[
{\Pi }_{2\mu'_2}, R^{-2}
\right]  
\right. \nonumber \\
\nonumber \\[-12pt]
&
\left.
+
{\displaystyle \frac{70}{3A R_0^2}} R^2 \!
\sum_{\Lambda } 
(\!-\! 1)^{\mu' } \!
\langle 2 \mu 2 \!-\! \mu'|2 \Lambda \rangle 
\widehat{\Jb}_{2 \Lambda } 
\!+\!
i\hbar 
{\displaystyle \frac{10}{3A}} \!
\sqrt{{\displaystyle \frac{35}{8\pi }}} R^{-4} \!
\sum_{\Lambda } 
(\!-\! 1)^{\mu' } \!
\langle 2 \mu 2 \!-\! \mu'|2 \Lambda \rangle 
\phi_{2\Lambda }^* 
\right. \nonumber \\
\nonumber \\[-8pt]
&\!\!\!\!\!\!\!\!\!\!\!\!
\left. 
-
{\displaystyle \frac{175}{3 \pi A }} R_0^2 R^{-6} \!
\sum_{\Lambda } 
(\!-\! 1)^{\mu' } \!
\langle 2 \mu 2 \!-\! \mu'|2 \Lambda \rangle 
\phi_{2\Lambda }^* \!
\sum_{\mu_1 \mu_2 \Lambda'}
(\!-\! 1)^{\Lambda' }  \!
\langle 2\mu_1 2\mu_2|2 \!-\! \Lambda' \rangle 
\phi_{2\Lambda' }^* 
{\phi }_{2\mu_1}^*  
{\Pi }_{2\mu_2} \!
\right] \! . 
\label{fcommutatorphi5}
\end{eqnarray}\\[-14pt]
In the above equation,
using
$
\widehat{\Pi }_{2\mu }
\!\!=\!\!
{\displaystyle \frac{1}{2}} \!
\left( \!
R^{-2} {\Pi }_{2\mu }
\!\!+\!\!
{\Pi }_{2\mu } R^{-2} \!
\right)
$,
we have introduced the new quantity
$\widehat{\Jb}_{\!L\Lambda }$
together with the associated relation\\[-20pt]
\begin{eqnarray}
\left.
\begin{array}{rcl}
&&
\sum_{\mu \mu'}
\left< 2 \mu 2 \mu'|L \Lambda \right>
\phi_{2\mu }^* \widehat{\Pi }_{2\mu'}
\!=\!
{\displaystyle \frac{1}{2}} \!
\left( \!
R^2 \widehat{\Jb}_{\!L\Lambda }
\!+\!
\widehat{\Jb}_{\!L\Lambda } R^2 \!
\right) , \\
\\[-12pt]
&&
{\displaystyle \frac{1}{2}} \!
\left( \!
R^2 \widehat{\Jb}_{\!L\Lambda }
\!+\!
\widehat{\Jb}_{\!L\Lambda } R^2 \!
\right)
\!=\!
R^2 
\widehat{\Jb}_{\!L\Lambda }
\!-\!
{\displaystyle \frac{1}{2}} \!
\sum_{\mu, \mu'} \!
\left< 2 \mu 2 \mu' | \right. \!\!
\left. L \Lambda \right> \!
\phi_{2\mu }^* \!
\left[
\Pi_{2\mu'}, R^{-2}
\right] ,
\end{array}
\right\}
\label{jllambda1}
\end{eqnarray}\\[-12pt]
and used the commutation relations\\[-18pt]
\begin{eqnarray}
\left.
\begin{array}{rcl}
&&\left[ 
\widehat{\Jb}_{\!L\Lambda }, \phi_{2\mu }^* 
\right] 
\!=\!
-i\hbar R^{-4} \!
\sum_{\mu'}
(-1)^{\mu }
\left< 2 \mu'2 -\mu |L \Lambda \right> \!
\phi_{2\mu'}^* \\
\\[-10pt]
&&\left[ 
\widehat{\Jb}_{\!L\Lambda }, R^2 
\right] 
\!=\!
- \!
\sum_{\mu,\mu'} \!
\left< 2 \mu 2 \mu'|L \Lambda \right> \!
\phi_{2\mu }^* \!
\left[
\Pi_{2\mu'}, R^{-2}
\right]  .
\end{array}
\right\}
\label{jllambda2} 
\end{eqnarray}\\[-12pt]
Substituting the approximate relation
(\ref{apprcommuPaiR})
into remaining terms same as
$[\Pi_{2\mu },R^{-2}]$
and
using the relations
(\ref{sigmaphipi2}) and (\ref{defjandhatj})
in Appendices C and D, 
a calculation of
(\ref{fcommutatorphi5})
is carried out.
Further, 
with the aid of
(\ref{R4sigmaphijhat}), (\ref{sigmaphiphiphi}) and (\ref{R4sigmaphijhat2}),
(\ref{fcommutatorphi5})
is cast into
(\ref{fcommutatorphi10})
in Appendix E.
As mentioned in the end of this Appendix,
we obtain the approximate expression for
$\left[f_{2\mu },\phi_{2\mu' }\right]$
in the following form:\\[-20pt]
\begin{eqnarray}
&
\left[f_{2\mu },\phi_{2\mu' }\right]
\approx
-
{\displaystyle \frac{\hbar^2 R_0^2}{2m}} \!
\left[ {}^{^{^{^{^{^{^{}}}}}}} \!\!
i\hbar \!
\left( \! 
{\displaystyle \frac{35}{3A} }
+
{\displaystyle \frac{56}{3 A } } \!\!
\right) \!\!
R_0^{-2} \! R^{-2}
\delta_{\mu\mu'}
-
i\hbar 
{\displaystyle \frac{50}{3 \pi A }} \! 
R_0^2 R^{-6}
\phi_{2\mu }^* 
\phi_{2 \mu'}
\right. \nonumber \\
\nonumber \\[-10pt]
&
\left.
+
i\hbar
10
\left( \!
1
\!-\!
{\displaystyle \frac{7}{3A}} \!
\right) \!
\sqrt{{\displaystyle \frac{35}{8\pi }}} R^{-4} \!
\sum_{\Lambda } \!
(\!-\! 1)^{\mu'} \!
\langle 2 \mu 2 \!-\! \mu'|2 \Lambda \rangle  
\phi_{2\Lambda }^*  
\right. \nonumber \\
\nonumber \\[-10pt]
&
\left.
-
i\hbar
{\displaystyle \frac{440}{7} }
{\displaystyle \frac{1}{3 A } } 
\sqrt{{\displaystyle \frac{35}{8 \pi }}} \! 
R^{-4} \!
\sum_{\Lambda } 
(\!-\! 1)^{\mu' } \!
\langle 2\mu 2 \!-\! \mu'|2 \Lambda \rangle \!
\phi_{2\Lambda }^*
-
i\hbar
{\displaystyle \frac{1250}{7} }
{\displaystyle \frac{1}{3 \pi A } }
R_0^2 R^{-6}
\phi_{2\mu }^*
\phi_{2 \mu' }
\right. \nonumber \\
\nonumber \\[-8pt]
&
\left.
+
i{\displaystyle \frac{15}{3A R_0^{2}}} 
\sqrt{10} R^2 \!
\sum_{\Lambda } 
(\!-\! 1)^{\mu'} \!
\left< 2\mu 2 \!-\! \mu'|1 \Lambda \right> \!
\left\{
R^{-2} 
\lb_{1 \Lambda }
R^{-2}
-
(\!-\! 1)^{\Lambda } 
\Lb_{1 \!-\! \Lambda }^*
\right\} 
\right. \nonumber \\
\nonumber \\[-8pt]
&
\left.
\!-\!
i\hbar
10 \!
\left( \!
5
\!+\!
{\displaystyle \frac{7 \sqrt{21}}{3}} \!
\right) \!
{\displaystyle \frac{1}{3A }} \! 
\sqrt{{\displaystyle \frac{35}{8\pi }}} \!
R^{-4} \!
\sum_{\Lambda } 
(\!-\! 1)^{\mu'} \!\!
\langle 2 \mu 2 \!-\! \mu'|2 \Lambda \rangle 
\phi_{2\Lambda }^* \!
\right. \nonumber \\
\nonumber \\[-10pt]
&
\left.
\times
{\displaystyle \sqrt{\frac{4\pi }{3}}} \!
\sum_{\nu } \!\!
\sum_{n=1}^A \!
r_{\! n} 
Y_{\! 1 \nu }(\!\theta_{n} ,\!\varphi_{n}) 
\!\cdot\!
(\!-\! 1)^{\nu }
\vnabla_{\!-\nu }^n \!
\right] \! .
\label{fcommutatorphi13}
\end{eqnarray}\\[-16pt]
Here we have discarded the term 
$(\!-\! 1)^{\mu'} \! \phi_{2\mu' }^* \widehat{\Jb}_{2 \mu }$
and used the approximate relations given as\\[-18pt]
\begin{eqnarray}
\begin{array}{c}
\sum_{\mu } \!
\phi_{2\mu }^*
\phi_{2\mu }
\!\approx\!
{\displaystyle \frac{4 \pi }{5}}
R_0^{-4}R^{4} ,~
\sum_{\mu \mu' }
\langle 2 \mu 2 \mu'|2 \Lambda \rangle 
\phi_{2\mu }
\phi_{2\mu' }
\!\approx\!
-\sqrt{{\displaystyle \frac{8 \pi }{35}}}
R_0^{-2}R^{2}
\phi_{2\Lambda } ,
\label{approxrelations}
\end{array} 
\end{eqnarray}\\[-14pt]
the second relation of which has no effects on the exactness of the operator
$T^{(0)}(\phi;R^2)$
since it has a non-scalar form.
\newpage
Thus we obtain the following final result:\\[-20pt]
\begin{eqnarray}
\!\!\!\!
\begin{array}{rcl}
& & 
\left[f_{2\mu },\phi_{2\mu' }\right]
=
[[{\pi }_{2\mu },T^{(0)}(\phi ;R^2 )],\phi_{2\mu'}]
= 
-
{\displaystyle \frac{\hbar^2 R_0^2}{2m}} \!
\left[ 
i\hbar \! 
{\displaystyle \frac{91}{3A} }
R_0^{-2}R^{-2}
\delta_{\mu\mu'} 
\right. \\ 
\\[-14pt]
&+&
\left.
\!\!\!\!\!\!
i\hbar
10 \!
\left( \!\!
1
\!-\!
{\displaystyle \frac{93}{7} }
{\displaystyle \frac{1}{3A}} \!\!
\right) \!\!
\sqrt{{\displaystyle \frac{35}{8\pi }}} R^{-4} \!
\sum_{\Lambda } \!
(\!-\! 1)^{\mu'} \!
\langle 2 \mu 2 \!-\! \mu'|2 \Lambda \rangle  
\phi_{2\Lambda }^*
\!-\!
 i\hbar 
{\displaystyle \frac{1600}{7} } 
{\displaystyle \frac{1}{3 \pi A }}  
R_0^2 R^{-6}
\phi_{2\mu }^* 
\phi_{2 \mu'}
\right.  \\
\\[-12pt]
& &
\left.
\!-\!
i\hbar
10 \!
\left( \!
5
\!+\!
{\displaystyle \frac{7 \sqrt{21}}{3}} \!
\right) \!\!
{\displaystyle \frac{1}{3A }} \! 
\sqrt{{\displaystyle \frac{35}{8\pi }}} \!
R^{-4} \!
\sum_{\Lambda } 
(\!-\! 1)^{\mu'} \!
\langle 2 \mu 2 \!-\! \mu'|2 \Lambda \rangle 
\phi_{2\Lambda }^* 
\right. \\
\\[-22pt]
& &
~~~~~~~~~~~~~~~~~~~~~~~~~~~~~~~~~~~~~~~~~~~~~
\times\!
{\displaystyle \sqrt{\frac{4\pi }{3}}} \!
\sum_{\nu } \!
\sum_{n=1}^A \!
r_{n} 
Y_{1 \nu }(\theta_{n} ,\varphi_{n}) 
\!\cdot\!
(\!-\! 1)^{\nu }
\vnabla_{-\nu }^n \! \\
\\[-16pt]
& &
\left.
+
i{\displaystyle \frac{15}{3A R_0^{2}}} 
\sqrt{10} R^2 \!
\sum_{\Lambda } 
(\!-\! 1)^{\mu'} \!\!
\left< 2\mu 2 \!-\! \mu'|1 \Lambda \right> \!
\left\{ \!
R^{-2} 
\lb_{\!1 \Lambda }
R^{-2}
\!-\!
(\!-\! 1)^{\Lambda } \!
\Lb_{\!1 \!-\! \Lambda }^* \!
\right\} \!
\right] ,
\end{array}
\label{fcommutatorphi14}
\end{eqnarray}\\[-12pt]
in which
a bilinear term $\phi_{2\mu }^* \phi_{2 \mu'}$ is neglected
compared to linear terms $\phi_{2\Lambda }^*$.
$\!\!$The last term in
(\ref{fcommutatorphi14})
becomes zero on the collective sub-space 
$|\Psi^{\mbox{coll}} \rangle$
due to the first subsidiary condition
(\ref{Lsubsidiarycondition}).
To get the commutativity 
$\left[f_{2\mu },\phi_{2\mu' }\right] \!=\! 0$,
further we strongly demand that the equation
(\ref{fcommutatorphi14})
must vanish
on the collective sub-space 
$|\Psi^{\mbox{coll}} \rangle$. 
It leads us to  the following second subsidiary condition:\\[-20pt]
\begin{eqnarray}
\!\!\!\!\!\!\!\!\!\!\!\!\!\!\!\!
\begin{array}{rcl}
& &
\left\{
i\hbar
{\displaystyle \frac{91}{3A} }
R_0^{-2}R^{-2}
\delta_{\mu\mu'}
\!+\!
i\hbar
10 \!
\left( \!\!
1
\!-\!
{\displaystyle \frac{93}{7} }
{\displaystyle \frac{1}{3A}} \!\!
\right) \!
\sqrt{{\displaystyle \frac{35}{8\pi }}} R^{-4} \!
\sum_{\Lambda } \!
(\!-\! 1)^{\mu'}
\langle 2 \mu 2 \!-\! \mu'|2 \Lambda \rangle  
\phi_{2\Lambda }^*
\right.  \\
\\[-16pt]
& &
-
i\hbar
10 \!
\left( \!
5
\!+\!
{\displaystyle \frac{7 \sqrt{21}}{3}} \!
\right) \!\!
{\displaystyle \frac{1}{3A }} \! 
\sqrt{{\displaystyle \frac{35}{8\pi }}} \!
R^{-4} \!
\sum_{\Lambda } 
(\!-\! 1)^{\mu'} \!
\langle 2 \mu 2 \!-\! \mu'|2 \Lambda \rangle 
\phi_{2\Lambda }^* \\
\\[-24pt]
& &
\left.
~~~~~~~~~~~~~~~~~~~~~~~~~~~~
\!\times\!
{\displaystyle \sqrt{\frac{4\pi }{3}}} \!
\sum_{\nu } \!
\sum_{n=1}^A \!
r_{n} 
Y_{1 \nu }(\theta_{n} ,\varphi_{n}) 
\!\cdot\!
(\!-\! 1)^{\nu }
\vnabla_{-\nu }^n 
\right\} \!
|\Psi^{\mbox{coll}} \rangle 
= 0 . 
\end{array}
\label{vanishcondition}
\end{eqnarray}\\[-12pt]
These conditions mean that $f_{\!2\mu }\!$ depends only on $\phi_\mu\!$ and $\!R^2\!$
on the collective sub-space 
$\!|\Psi^{\mbox{coll}} \rangle$.
Both the first and the second subsidiary conditions
are implementable to investigate what is a structure of
the collective sub-space satisfying such conditions.
This is a very important and interesting problem
which can be solved
in the case of simpler two-dimensional nuclei.
 
For $\mu \neq \mu'$,
the second subsidiary condition
(\ref{vanishcondition})
is simply rewritten as\\[-18pt]
\begin{eqnarray}
\!\!\!\!\!\!\!\!\!\!\!\!\!\!\!\!
\begin{array}{rcl}
& &
\left\{ \!
i\hbar
10 \!
\left( \!\!
1
\!-\!
{\displaystyle \frac{93}{7} }
{\displaystyle \frac{1}{3A}} \!\!
\right) \!
\!-\!
i\hbar \!\!
\left( \!\!
5
\!+\!
{\displaystyle \frac{7 \sqrt{21}}{3}} \!\!
\right) \!\!
{\displaystyle \frac{1}{3A }} \!
{\displaystyle \sqrt{\frac{4\pi }{3}}} \!
\sum_{\nu } \!
\sum_{n=1}^A \!
r_{n} 
Y_{1 \nu }(\theta_{n} ,\varphi_{n}) 
\!\cdot\!
(\!-\! 1)^{\nu }
\vnabla_{-\nu }^n \!\!
\right\} \!\!
|\Psi^{\mbox{coll}} \rangle
\!=\! 0 , 
\end{array}
\label{vanishcondition2}
\end{eqnarray}\\[-12pt]
from which the scalar operator
$
{\displaystyle \sqrt{\frac{4\pi }{3}}} \!
\sum_{\nu } \!
\sum_{n=1}^A \!
r_{n} 
Y_{1 \nu }(\theta_{n} ,\varphi_{n}) 
\!\cdot\!
(\!-\! 1)^{\nu }
\vnabla_{-\nu }^n
$
takes a simple form on the collective sub-space 
$|\Psi^{\mbox{coll}} \rangle$
as\\[-28pt]
\begin{eqnarray}
\begin{array}{c}
{\displaystyle \sqrt{\frac{4\pi }{3}}} \!
\sum_{\nu } \!
\sum_{n=1}^A \!
r_{n} 
Y_{1 \nu }(\theta_{n} ,\varphi_{n}) 
\!\cdot\!
(\!-\! 1)^{\nu }
\vnabla_{-\nu }^n
=
30 A
\cdot
{\displaystyle
\frac{
1
\!-\!
{\displaystyle \frac{93}{7} }
{\displaystyle \frac{1}{3A}}
}
{
5
\!+\!
{\displaystyle \frac{7 \sqrt{21}}{3}}
}
} .
\end{array}
\label{scalaroperator}
\end{eqnarray}\\[-12pt]
For $\mu = \mu'$,
using
(\ref{scalaroperator})
it is shown that the second subsidiary condition
(\ref{vanishcondition})
is satisfied since $A$ is sufficiently large.
Substituting
(\ref{scalaroperator})
into
(\ref{compiT3}),
we can get the approximate commutation relation for
$[{\pi }_{2\mu },T]$
as\\[-16pt]
\begin{eqnarray}
\!\!\!\!\!\!\!\!\!\!\!\!\!\!
\begin{array}{rcl}
& &
[{\pi }_{2\mu },~T]
\!\approx\! 
-
{\displaystyle \frac{\hbar^2 R_0^2}{2m}}
\!\cdot\!
i\hbar {\displaystyle \frac{25}{2 \pi }} \!
\left( \!
1 \!-\! 2 \cdot {\displaystyle \frac{1}{3A}}
\!+\!
{\displaystyle \frac{60}{7}} \cdot
{\displaystyle \frac{1}{3A}}
\!\cdot\!
{\displaystyle
\frac{21A \!-\! 93}
{15 \!+\! 7 \sqrt{21}}
} \!
\right) \!
R_0^2 R^{-6} \phi_{2\mu }^* ,
\end{array}
\label{compiT3} 
\end{eqnarray}\\[-28pt]

Using the definition of $f_{2\mu }$
(\ref{fcommutatorpiT0}),
now we compute the $\phi_{2\mu }$-dependence of
$f_{2\mu }$ up to the third order $\phi^3$.
For this aim, we take an operator ordering in which all $\Pi$ operators
are put into the right and all $\phi$ and $R^2$ operators
to the left of each other, i.e.,
the {\it normal product} form.
Thus we can find the final approximate expression for
the operator-valued function
$f_{2\mu }(\phi; R^2)$ up to the third order $\phi^3$
in the following form:
\newpage
\begin{eqnarray}
\begin{array}{cc}
&\!\!\!\!\!\!\!\!\!\!
f_{2\mu }(\phi; R^2)
\!\approx\! 
-
{\displaystyle \frac{\hbar^2 R_0^2}{2m}}
\!\cdot\!
i\hbar {\displaystyle \frac{25}{2 \pi }} \!
\left( \!
1 \!-\! 2
\!\cdot\! {\displaystyle \frac{1}{3A}}
\!+\!
{\displaystyle \frac{60}{7}} 
\!\cdot\!
{\displaystyle \frac{1}{3A}}
\!\cdot\!
{\displaystyle
\frac{21A \!-\! 93}
{15 \!+\! 7 \sqrt{21}} \!
}
\right) \!\!
R_0^2 R^{-6}
\phi_{2\mu }^*  \\
\\[-14pt]
& 
-
{\displaystyle \frac{2\pi }{3AR_0^4} \frac{1}{m}}
(i\hbar)^3 \!
\left\{ \!
7 \!
\left( \! {\displaystyle \frac{2 \!\cdot\! 5 }{4\pi}} \! \right)^{\! 2} \!\!
R_0^8 R^{-6}
\phi_{2\mu}^*
\!-\!
3 \!
\left( \! {\displaystyle \frac{2 \!\cdot\! 5 }{4\pi}} \! \right)^{\! 3} \!\!
R_0^{12} R^{-10}
\sum_{\mu'}
\phi_{2\mu'}^* \phi_{2\mu'} \phi_{2\mu }^* \!
\right\} \\
\\[-14pt]
& 
+
{\displaystyle \frac{2\pi }{3AR_0^2} \frac{1}{m}} 
\sqrt{{\displaystyle \frac{35}{8\pi }}}
\!\cdot\!
(i\hbar)^3 \!
\left\{ \!
-
{\displaystyle \frac{1}{4}} \!
\sum_{\mu' \mu'' }
\langle 2\mu' 2\mu'' |2\mu \rangle \!
\left( \! {\displaystyle \frac{2 \!\cdot\! 5 }{4\pi}} \! \right)^{\! 2} \!\!
R_0^8 R^{-8}
\phi_{2\mu'}^* \phi_{2\mu''}^* 
\right. \\ 
\\[-14pt]
&
+
2 \!
\left( \! {\displaystyle \frac{2 \!\cdot\! 5 }{4\pi}} \! \right)^{\! 2} \!\!
R_0^{10} R^{-10} \!
\sqrt{{\displaystyle \frac{35}{8\pi }}} \!
\sum_{LK} \!
\sqrt{ 5 (2L \!+\! 1) } W(2222;2L) \\ 
\\[-12pt]
&
\!\times\!
\sum_{\mu' \mu'' }
\sum_{\Lambda}
(\!-\! 1)^{K }
\langle 2\mu' L-K |2\Lambda \rangle
\langle 2\mu'' LK |2\mu \rangle
\phi_{2\mu'}^* \phi_{2\mu''}^* \phi_{2\Lambda} \\ 
\\[-12pt]
&
+
2 \!
\left( \! {\displaystyle \frac{2 \!\cdot\! 5 }{4\pi}} \! \right)^{\! 2} \!\!
R_0^{10} R^{-10} \!
\sqrt{{\displaystyle \frac{35}{8\pi }}} \!
\sum_{LK} \!
\sqrt{ 5 (2L \!+\! 1) } W(2222;2L) \\ 
\\[-12pt]
&
\!\times\!
\sum_{\mu' \mu'' }
\sum_{\Lambda}
(\!-\! 1)^{K }
\langle 2 \mu \!+\! \mu'  2\mu'' |LK \rangle
\langle 2\Lambda L-K |2-\mu \rangle
\phi_{2 \mu \!+\! \mu'} \phi_{2\mu''} \phi_{2\Lambda}^* \\ 
\\[-12pt]
&
-
2 \!
\left( \! {\displaystyle \frac{2 \!\cdot\! 5 }{4\pi}} \! \right)^{\! 2} \!\!
R_0^{6} R^{-6} \!
\sqrt{{\displaystyle \frac{35}{8\pi }}} \!
\sum_{LK} \!
\sqrt{ 5 (2L \!+\! 1) } W(2222;2L) \\ 
\\[-12pt]
&
\!\times\!
\sum_{\mu' \mu'' }
\sum_{\Lambda}
(\!-\! 1)^{K }
\langle 2 \mu \!+\! \mu''  2\mu' |LK \rangle
\langle 2\Lambda L-K |2-\mu \rangle
\phi_{2 \mu'} \phi_{2 \mu \!+\! \mu''} \phi_{2\Lambda}^*  \\ 
\\[-12pt]
&
+
{\displaystyle \frac{15}{4}} \!
\left( \! {\displaystyle \frac{2 \!\cdot\! 5 }{4\pi}} \! \right)^{\! 2} \!\!
R_0^{10} R^{-10} \!
\sqrt{{\displaystyle \frac{35}{8\pi }}} \!
\sum_{LK} \!
\sqrt{ 5 (2L \!+\! 1) } W(2222;2L) \\ 
\\[-18pt]
&
\left.
\!\times\!
\sum_{\mu' \mu'' }
\sum_{\Lambda}
(\!-\! 1)^{L }
(\!-\! 1)^{K }
\langle 2 \mu'  2 \mu \!+\! \mu'' |LK \rangle
\langle 2\Lambda L-K |2-\mu \rangle
\phi_{2 \mu'} \phi_{2 \mu \!+\! \mu''} \phi_{2\Lambda}^*
{}{^{^{^{^{^{^{^{.}}}}}}}}
\!\!\!\!
\right\} \! .
\end{array}
\label{fcommutator}
\end{eqnarray}\\[-16pt]
Substituting
(\ref{fcommutator})
into
(\ref{commutatorpiT0}),
we get the R.H.S. of the discrete integral equation
(\ref{commutatorpiT0})
as\\[-14pt]
\begin{eqnarray}
\!\!\!\!\!\!\!\!
\begin{array}{rcl}
& &\!\!\!\!
[\Pi_{2\mu }, T^{(0)}(\phi;R^2)]
=
-
{\displaystyle \frac{\hbar^2 R_0^2}{2m}}
\!\cdot\!
i\hbar {\displaystyle \frac{25}{2 \pi }} \!
\left( \!
1 \!-\! 4 \!\cdot\! {\displaystyle \frac{1}{3A}}
\!+\!
{\displaystyle \frac{60}{7}} \!\cdot\!
{\displaystyle \frac{1}{3A}}
\!\cdot\!
{\displaystyle
\frac{21A \!-\! 93}
{15 \!+\! 7 \sqrt{21}} \!
}
\right) \! 
R_0^2 R^{-6} \\
\\[-14pt]
& &
~~~~~~~~~~~~~~~~~~~~~~
\times\!
\left( \!
\phi_{2\mu }^*
\!+\!
\sqrt{{\displaystyle \frac{35}{8\pi }}}
R_0^2  R^{-2} \!
\sum_{\mu' \mu''}
\langle 2  \mu' 2\mu'' |2\mu \rangle 
\phi_{2\mu'}^* \phi_{2\mu'' }^*  \!
\right) \\
\\[-14pt]
& &
-
{\displaystyle \frac{\hbar^2 R_0^2}{2m}}
\!\cdot\!
i\hbar 
{\displaystyle \frac{25}{2 \pi }}
{\displaystyle \frac{1}{2}}
{\displaystyle \frac{1}{3A}}
\sqrt{{\displaystyle \frac{35}{8\pi }}} \!
R_0^2 R^{-8} \!
\sum_{\mu' \mu'' }
\langle 2\mu' 2\mu'' |2\mu \rangle \!
\phi_{2\mu'}^* \phi_{2\mu''}^* \\
\\[-14pt]
& &
-
{\displaystyle \frac{\hbar^2 R_0^2}{2m}}
\!\cdot\!
i\hbar 
{\displaystyle \frac{25}{2 \pi }}
{\displaystyle \frac{4}{3A}}
\sqrt{{\displaystyle \frac{35}{8\pi }}} \!
R_0^{6} R^{-10} \!
\sum_{LK} \!
\sqrt{ 5 (2L \!+\! 1) } W(2222;2L) \\ 
\\[-14pt]
& &
~~~~~~~~~~~~~~~~~~~~~~
\!\times\!
\sum_{\mu' \mu'' } \!
\sum_{\Lambda}
(\!-\! 1)^{K } \!
\langle 2\mu' L \!\!-\!\! K |2\Lambda \rangle
\langle 2\mu'' LK |2\mu \rangle
\phi_{2\mu'}^* \phi_{2\mu''}^* \phi_{2\Lambda} \\
\\[-14pt]
& &
 =
-
{\displaystyle \frac{\hbar^2 R_0^2}{2m}}
\!\cdot\!
i\hbar {\displaystyle \frac{25}{2 \pi }} \!
\left( \!
1 \!-\! 4 \!\cdot\! {\displaystyle \frac{1}{3A}}
\!+\!
{\displaystyle \frac{60}{7}} \!\cdot\!
{\displaystyle \frac{1}{3A}}
\!\cdot\!
{\displaystyle
\frac{21A \!-\! 93}
{15 \!+\! 7 \sqrt{21}} \!
}
\right) \! 
R_0^2 R^{-6} \\
\\[-14pt]
& &
~~~~~~~~~~~~~~~~~~~~~~
\times\!
\left( \!
\phi_{2\mu }^*
\!+\!
\sqrt{{\displaystyle \frac{35}{8\pi }}}
R_0^2  R^{-2} \!
\sum_{\mu' \mu''} \! 
\langle 2 \mu' 2\mu'' |2\mu \rangle 
\phi_{2\mu'}^* \phi_{2\mu'' }^*  \!
\right) \\
\\[-12pt]
& &
-
{\displaystyle \frac{\hbar^2 R_0^2}{2m}}
\!\cdot\!
i\hbar 
{\displaystyle \frac{25}{2 \pi }}
{\displaystyle \frac{1}{2}}
{\displaystyle \frac{1}{3A}}
\sqrt{{\displaystyle \frac{35}{8\pi }}} \!
R_0^2 R^{-8} \!
\sum_{\mu' \mu'' }
\langle 2\mu' 2\mu'' |2\mu \rangle \!
\phi_{2\mu'}^* \phi_{2\mu''}^* \\
\\[-12pt]
& &
-
{\displaystyle \frac{\hbar^2 R_0^2}{2m}}
\!\cdot\!
i\hbar 
{\displaystyle \frac{25}{2 \pi }}
2
{\displaystyle \frac{4}{3A}}
\sqrt{{\displaystyle \frac{35}{8\pi }}} \!
R_0^{6} R^{-10}  \\
\\[-10pt]
& &
\!\times
\sum_{L} \!
\sqrt{ 5 (2L \!+\! 1) } W(2222;2L) \!
\sum_{L'K'} \!
\sqrt{ 5 (2L' \!+\! 1) } W(2222;LL') \\ 
\\[-8pt]
& &
~~~~~~~~~~~~~~~~~~~~~~
\!\times\!
\sum_{\mu' \mu'' } \!
\sum_{\Lambda} \!
\langle 2 \!\!-\!\! \Lambda 2 \mu'' | L'K' \rangle
\langle 2 \!\!-\!\! \mu' L'K' |2\mu \rangle
\phi_{2\mu'} \phi_{2\mu''}^* \phi_{2\Lambda} ,
\end{array}
\label{commutatorPaiT0}
\end{eqnarray}\\[-4pt]
where we have used the first approximate relation of
(\ref{approxrelations})
and the last two long terms of
(\ref{fcommutator}).
Using
(\ref{commutatorPaiT0})
and the commutation relations
(\ref{exactcommurelpiphi})
and
(\ref{commurelpipi0})
and for the sake of simplicity
discarding the contribution from the effect by the term
$[\Pi_{2\mu}, R^{-2}]$,
we obtain\\[-20pt]
\begin{eqnarray}
\!\!\!\!\!\!\!\!
\begin{array}{rcl}
& &
\!\!\!\!
[\Pi_{2\mu'}, [\Pi_{2\mu }, T^{(0)}(\phi;R^2)]]
\!=\!
{\displaystyle \frac{\hbar^2 R_0^2}{2m}}
\!\cdot\!
(i\hbar)^2 {\displaystyle \frac{25}{2 \pi }} \!
\left( \!
1 \!-\! 4 \!\cdot\! {\displaystyle \frac{1}{3A}}
\!+\!
{\displaystyle \frac{60}{7}} \!\cdot\!
{\displaystyle \frac{1}{3A}}
\!\cdot\!
{\displaystyle
\frac{21A \!-\! 93}
{15 \!+\! 7 \sqrt{21}} \!
}
\right) \! 
R_0^2R^{-6} \\
\\[-14pt]
& &
~~~~~~~~~~~~~~
\times\!
\left( \!\!
(\!-\!1)^{\mu'} \!
\delta_{\mu',-\mu}
\!+\!
\sqrt{{\displaystyle \frac{35}{8\pi }}}
R_0^2R^{-2} \!
\sum_{\mu'' }
\left\{ \!
(\!-\!1)^{\mu'} \!
\langle 2 \!\!-\!\! \mu' 2\mu'' |2\mu \rangle \!
+\!
\langle 2\mu' 2\mu'' |2\mu \rangle \!
\right\} \! 
\phi_{2\mu''}^* \!\!
\right) \\
\\[-14pt]
& &
\!\!\!\!\!
+
{\displaystyle \frac{\hbar^2 R_0^2}{2m}}
\!\cdot\!
(i\hbar)^2
{\displaystyle \frac{25}{2 \pi }}
{\displaystyle \frac{1}{2}}
{\displaystyle \frac{1}{3A}}
\sqrt{{\displaystyle \frac{35}{8\pi }}} \!
R_0^2 R^{-8} \!
\sum_{\mu''} \!
\left\{ \!
(\!-\!1)^{\mu'} \!
\langle 2 \!\!-\!\! \mu' 2\mu'' |2\mu \rangle \!
+\!
\langle 2\mu' 2\mu'' |2\mu \rangle \!
\right\} \!
\phi_{2\mu''}^* \\
\\[-10pt]
& &
\!\!\!\!\!
+
{\displaystyle \frac{\hbar^2 R_0^2}{2m}}
\!\cdot\!
(i\hbar)^2 
{\displaystyle \frac{25}{2 \pi }}
2
{\displaystyle \frac{4}{3A}}
\sqrt{{\displaystyle \frac{35}{8\pi }}} \!
R_0^{6} R^{-10} \\
\\[-10pt]
& &
~~~~~~~~~~~~~~~
\!\times
\sum_{L}\!
\sqrt{ 5 (2L \!+\! 1) } W(2222;2L) \!
\sum_{L'K'} \!
\sqrt{ 5 (2L' \!+\! 1) } W(2222;LL') \\ 
\\[-8pt]
& &
~~~~~~~~~~~~~~~~~~~~~~
\!\times\!
\left(
\sum_{\mu''}  \!\!
\sum_{\Lambda} \!
 (\!-\!1)^{ \Lambda} \!
\langle 2 \Lambda 2 \mu'' | L'K' \rangle
\langle 2 \!\!-\!\! \mu' L'K' |2\mu \rangle
\phi_{2\mu''}^* \phi_{2\Lambda}^*
\right. \\
\\[-8pt]
& &
~~~~~~~~~~~~~~~~~~~~~~~~
\!+\!
\sum_{\mu'' } \!\!
\sum_{\Lambda} \!
(\!-\!1)^{\mu'} \! (\!-\!1)^{ \Lambda} \!
\langle 2 \Lambda 2 \!\!-\!\! \mu' | L'K' \rangle
\langle 2 \!\!-\!\! \mu'' L'K' |2\mu \rangle
\phi_{2\mu''}^* 
\phi_{2\Lambda}^* \\
\\[-10pt]
& &
\left.
~~~~~~~~~~~~~~~~~~~~~~~~
\!+\!
\sum_{\mu'' \mu''' } \!
(\!-\!1)^{\mu''} \!
\langle 2 \!\!-\!\! \mu' 2 \mu''' | L'K' \rangle
\langle 2 \mu'' L'K' |2\mu \rangle
\phi_{2\mu''}^* \phi_{2\mu'''}^* \!
\right) ,
\end{array}
\label{commutatorPaiPaiT0}
\end{eqnarray}
\vspace{-0.6cm}
\begin{eqnarray}
\!\!\!\!\!\!\!\!\!\!\!\!\!\!\!\!
\begin{array}{rcl}
& &
\!\!\!\!\!\!\!\!
[\Pi_{2\mu''}, [\Pi_{2\mu'}, [\Pi_{2\mu }, T^{(0)}(\phi;R^2)]]]
\!=\!
- {\displaystyle \frac{\hbar^2 R_0^2}{2m}}
\!\cdot\!
(i\hbar)^3 {\displaystyle \frac{25}{2 \pi }} \!
\left( \!
1 \!-\! 4 \!\cdot\! {\displaystyle \frac{1}{3A}}
\!+\!
{\displaystyle \frac{60}{7}} \!\cdot\!
{\displaystyle \frac{1}{3A}}
\!\cdot\!
{\displaystyle
\frac{21A \!-\! 93}
{15 \!+\! 7 \sqrt{21}} \!
}
\right) \! 
R_0^2R^{-6} \\
\\[-16pt]
& &
~~~~~~~~~~~~~~~~~~~~~
\times\!
\sqrt{{\displaystyle \frac{35}{8\pi }}}
R_0^2R^{-2} \!
\left\{ \!
(\!-\!1)^{\mu'} \!
(\!-\!1)^{\mu''} \!
\langle 2 \!\!-\!\! \mu' 2 \!\!-\!\! \mu'' |2\mu \rangle \!
+\!
(\!-\!1)^{\mu''} \!
\langle 2\mu' 2 \!\!-\!\! \mu'' |2\mu \rangle \!
\right\} \\
\\[-10pt]
& &
\!\!\!\!\!
-
{\displaystyle \frac{\hbar^2 R_0^2}{2m}}
\!\cdot\!
(i\hbar)^3
{\displaystyle \frac{25}{2 \pi }}
{\displaystyle \frac{1}{2}}
{\displaystyle \frac{1}{3A}}
\sqrt{{\displaystyle \frac{35}{8\pi }}} \!
R_0^2 R^{-8} \!
\left\{ \!
(\!-\!1)^{\mu'} \!
(\!-\!1)^{\mu''} \!
\langle 2 \!\!-\!\! \mu' 2 \!\!-\!\! \mu'' |2\mu \rangle \!
+\!
(\!-\!1)^{\mu''} \!
\langle 2\mu' 2 \!\!-\!\! \mu'' |2\mu \rangle \!
\right\} \\
\\[-10pt]
& &
\!\!\!\!\!
-
{\displaystyle \frac{\hbar^2 R_0^2}{2m}}
\!\cdot\!
(i\hbar)^3 
{\displaystyle \frac{25}{2 \pi }}
2
{\displaystyle \frac{4}{3A}}
\sqrt{{\displaystyle \frac{35}{8\pi }}} \!
R_0^{6} R^{-10} \\
\\[-10pt]
& &
~~~
\!\times\!
\sum_{L}\!
\sqrt{ 5 (2L \!+\! 1) } W(2222;2L) \!
\sum_{L'K'} \!
\sqrt{ 5 (2L' \!+\! 1) } W(2222;LL') \\ 
\\[-8pt]
& &
~~~
\!\times\!
\left\{ {}^{^{^{^{.}}}} \!\!\!\!
\sum_{\Lambda} \!
(\!-\!1)^{\mu'''} \!\!
\left( \! 
\langle 2 \mu''' 2 \mu'' | L'K' \rangle
\langle 2 \!\!-\!\! \mu' L'K' |2\mu \rangle \!
\!+\!
(\!-\!1)^{\mu'} \!
\langle 2 \mu''' 2 \!\!-\!\! \mu' | L'K' \rangle
\langle 2 \!\!-\!\! \mu'' L'K' |2\mu \rangle \!
\right)  
\right. \\
\\[-8pt]
& &
~~~
\!+\!
\sum_{\mu''' } \!\!
(\!-\!1)^{\mu'''} \!\!
\left( \!
\langle 2 \!\!-\!\! \mu'' 2 \!\!-\!\!  \mu''' | L'K' \rangle \!
\langle 2 \!\!-\!\! \mu' L'K' |2\mu \rangle \!
\!+\!
(\!-\!1)^{\mu'} \!
\langle 2 \!\!-\!\! \mu'' 2 \!\!-\!\! \mu' | L'K' \rangle \!
\langle 2 \mu''' L'K' |2\mu \rangle \! 
\right) \\
\\[-8pt]
& &
\!\!\!\!\!
\left.
\!+\!
\sum_{\mu'''} \!\!
\left( \!
\langle 2 \!\!-\!\! \mu' 2 \mu''' | L'K' \rangle \!
\langle 2 \!\!-\!\! \mu'' L'K' |2\mu \rangle \!
\!\!+\!\!
(\!-\!1)^{\mu''} \! (\!-\!1)^{\mu'''} \!\!
\langle 2 \!\!-\!\! \mu' 2 \!\!-\!\! \mu'' | L'K' \rangle \!
\langle 2 \mu''' L'K' |2\mu \rangle \! 
\right) \!
\right\} \! 
\phi_{2\mu'''}^* ,
\end{array}
\label{commutatorPaiPaiPaiT0}
\end{eqnarray}
\vspace{-0.7cm}
\begin{eqnarray}
\!\!\!\!\!\!\!\!
\begin{array}{rcl}
& &\!\!\!\!
[\Pi_{2\mu'''}, [\Pi_{2\mu''}, [\Pi_{2\mu'}, [\Pi_{2\mu }, T^{(0)}(\phi;R^2)]]]] 
\!=\!
{\displaystyle \frac{\hbar^2 R_0^2}{2m}}
\!\cdot\!
(i\hbar)^4 
{\displaystyle \frac{25}{2 \pi }}
{\displaystyle \frac{4}{3A}}
2
\sqrt{{\displaystyle \frac{35}{8\pi }}} \!
R_0^{6} R^{-10} \\
\\[-10pt]
& &
~~~~~~
\!\times\!
\sum_{L}\!
\sqrt{ 5 (2L \!+\! 1) } W(2222;2L) \!
\sum_{L'K'} \!
\sqrt{ 5 (2L' \!+\! 1) } W(2222;LL') \\ 
\\[-6pt]
& &
~~~~~~
\!\times\!
\left\{
\langle 2 \!\!-\!\! \mu''' 2 \mu'' | L'K' \rangle
\langle 2 \!\!-\!\! \mu' L'K' |2\mu \rangle \!
\!+\!
(\!-\!1)^{\mu'} \!
\langle 2 \!\!-\!\! \mu''' 2 \!\!-\!\! \mu' | L'K' \rangle
\langle 2 \!\!-\!\! \mu'' L'K' |2\mu \rangle \!
\right. \\
\\[-6pt]
& &
~~~~~~
\!+\!
\langle 2 \!\!-\!\! \mu'' 2 \mu''' | L'K' \rangle
\langle 2 \!\!-\!\! \mu' L'K' |2\mu \rangle \!
\!+\!
(\!-\!1)^{\mu'} \!
\langle 2 \!\!-\!\! \mu'' 2 \!\!-\!\! \mu' | L'K' \rangle
\langle 2 \!\!-\!\! \mu''' L'K' |2\mu \rangle \\
\\[-6pt]
& &
\left.
\!+
(\!-\!1)^{\mu'''} \!
\langle 2 \!\!-\!\! \mu' 2 \!\!-\!\!  \mu''' | L'K' \rangle
\langle 2 \!\!-\!\! \mu'' L'K' |2\mu \rangle \!
\!+\!
(\!-\!1)^{\mu''} \!
\langle 2 \!\!-\!\! \mu' 2 \!\!-\!\! \mu'' | L'K' \rangle
\langle 2 \!\!-\!\! \mu''' L'K' |2\mu \rangle
\right\} ,
\end{array}
\label{commutatorPaiPaiPaiPaiT0}
\end{eqnarray}
\vspace{-0.5cm}
\begin{eqnarray}
\!\!\!\!\!\!\!\!
\begin{array}{c}
[\Pi_{2\mu''''}, [\Pi_{2\mu'''}, [\Pi_{2\mu''}, [\Pi_{2\mu'}, [\Pi_{2\mu }, T^{(0)}(\phi;R^2)]]]] 
\!=\! 0.
\end{array}
\label{commutatorPaiPaiPaiPaiPaiT0}
\end{eqnarray}

We can determine $C_n (R^2) (n \!\neq\! 0)$ in
(\ref{expansionT0}).
In order to get their expressions,
we take the commutators with the canonical conjugate variable
$\Pi_{2 \mu}$
in the following way:\\[-16pt]
\begin{eqnarray}
\left.
\!\!\!\!\!\!\!\!\!\!\!\!\!\!\!\!\!\!\!\!
\begin{array}{rcl}
& &
[\Pi_{2\mu }, T^{(0)}(\phi;R^2)]
\!=\!
- i\hbar 
(\!-\!1)^{\mu}
C_{1 \!-\! \mu} (R^2)
- 2 i\hbar \!
\sum_{ \mu'} 
(\!-\!1)^{\mu}
C_{2 \!-\! \mu \mu'} (R^2) 
\phi_{2 \mu'}^*
+
\cdots , \\
\\[-8pt]
& &
[\Pi_{2\mu'}, ~[\Pi_{2\mu }, T^{(0)}(\phi;R^2)]]
\!=\!
2 (i\hbar)^2 
(\!-\!1)^{\mu}(\!-\!1)^{\mu'} \!
C_{2 \!-\! \mu \!-\! \mu'} (R^2) \\
\\[-8pt]
& &
~~~~~~~~~~~~~~~~~~~~~~~~~~~~~~~~
+
6 (i\hbar)^2  \!
\sum_{ \mu'' }
(\!-\!1)^{\mu}(\!-\!1)^{\mu'} \!
C_{3 \!-\! \mu \!-\! \mu'\mu''} (R^2)
\phi_{2 \mu''}^*
+
\cdots , \\
\\[-8pt]
& &
[\Pi_{2\mu''},~ [\Pi_{2\mu'},~ [\Pi_{2\mu }, T^{(0)}(\phi;R^2)]]] 
\!=
-~\!
6 (i\hbar)^3 
(\!-\!1)^{\mu} \! (\!-\!1)^{\mu'} \! (\!-\!1)^{\mu''} \!
C_{3 \!-\! \mu \!-\! \mu' \!-\! \mu''} (R^2) \\
\\[-8pt]
& &
~~~~~~~~~~~~~~~~~~~~~~~~~~~~~~~~~
-\!
16 (i\hbar)^3 \!
\sum_{ \mu''' }
(\!-\!1)^{\mu} \! (\!-\!1)^{\mu'} \! (\!-\!1)^{\mu''} \!
C_{4 \!-\! \mu \!-\! \mu' \!-\! \mu'' \mu'''} (R^2)
\phi_{2 \mu'''}^*
+
\cdots ,  \\
\\[-6pt]
& &
[\Pi_{2\mu'''}, [\Pi_{2\mu''}, [\Pi_{2\mu'}, [\Pi_{2\mu }, T^{(0)}(\phi;R^2)]]]] 
\!\!=\!\!
16 (i\hbar)^4
(\!-\!1)^{\mu} \! (\!-\!1)^{\mu'} \! (\!-\!1)^{\mu''} \!
C_{4 \!-\! \mu \!-\! \mu' \!-\! \mu'' \mu'''} (R^2)
\!+\!
\cdots \! ,  \\
\\[-16pt]
& &
\vdots  
\end{array} \!\!\!
\right\}
\label{expansioncoefficients}
\end{eqnarray}\\[-12pt]
Comparing from
(\ref{commutatorPaiT0})
to
(\ref{commutatorPaiPaiPaiPaiPaiT0})
with
(\ref{expansioncoefficients}),
then 
we obtain the explicit expressions for
$C_n (R^2) (n \!\neq\! 0)$
as follows:\\[-16pt]
\begin{eqnarray}
\left.
\!\!\!\!\!\!\!\!\!\!\!\!\!\!
\begin{array}{rcl}
& &
C_{1 \mu} (R^2)
\!=\!
0 , \\
\\[-12pt]
& &
C_{2 \mu \mu'} (R^2)
\!=\!
{\displaystyle \frac{\hbar^2 R_0^2}{2m}}
\!\cdot\!
{\displaystyle \frac{25}{4 \pi }} \!
\left( \!
1 \!-\! 4 \!\cdot\! {\displaystyle \frac{1}{3A}}
\!+\!
{\displaystyle \frac{60}{7}} \!\cdot\!
{\displaystyle \frac{1}{3A}}
\!\cdot\!
{\displaystyle
\frac{21A \!-\! 93}
{15 \!+\! 7 \sqrt{21}} \!
}
\right) \! 
R_0^2 R^{-6}
(\!-\!1)^{\mu} \!
\delta_{\mu', \!-\! \mu} ,  \\
\\[-14pt]
& &
C_{3 \mu \mu' \mu''} (R^2) 
\!=\!
{\displaystyle \frac{\hbar^2 R_0^2}{2m}}
\!\cdot\! 
{\displaystyle \frac{25}{12 \pi }} \!
\left\{ \!
{\displaystyle \frac{1}{2}}
\!\cdot\!
{\displaystyle \frac{1}{3A}}
\!+\!
\left( \!
1 \!-\! 4 \!\cdot\! {\displaystyle \frac{1}{3A}}
\!+\!
{\displaystyle \frac{60}{7}} \!\cdot\!
{\displaystyle \frac{1}{3A}}
\!\cdot\!
{\displaystyle
\frac{21A \!-\! 93}
{15 \!+\! 7 \sqrt{21}} \!
}
\right) \! 
R_0^2  \!
\right\} \\
\\[-14pt]
& &
~~~~~~~~~~~~~~~~~~
\!\times\! 
\sqrt{{\displaystyle \frac{35}{8\pi }}}
R_0^2R^{-8} \!
\left\{ \!
(\!-\!1)^{\mu'} \!
\langle 2 \!\!-\!\! \mu' 2\mu'' |2\mu \rangle \!
+\!
\langle 2\mu' 2\mu'' |2\mu \rangle \!
\right\} , \\
\\[-12pt]
& &
C_{4 \mu \mu' \mu'' \mu'''} (R^2) 
\!=\!
{\displaystyle \frac{\hbar^2 R_0^2}{2m}}
\!\cdot\!
(i\hbar)^4 
{\displaystyle \frac{25}{32 \pi }}
2
{\displaystyle \frac{4}{3A}}
\sqrt{{\displaystyle \frac{35}{8\pi }}} \!
R_0^{6} R^{-10} \\
\\[-10pt]
& &
~~~~~~
\!\times\!
\sum_{L}\!
\sqrt{ 5 (2L \!+\! 1) } W(2222;2L) \!
\sum_{L'K'} \!
\sqrt{ 5 (2L' \!+\! 1) } W(2222;LL') \\ 
\\[-8pt]
& &
~~~~~~
\!\times\!
\left\{
\langle 2 \!\!-\!\! \mu''' 2 \mu'' | L'K' \rangle
\langle 2 \!\!-\!\! \mu' L'K' |2\mu \rangle \!
\!+\!
(\!-\!1)^{\mu'} \!
\langle 2 \!\!-\!\! \mu''' 2 \!\!-\!\! \mu' | L'K' \rangle
\langle 2 \!\!-\!\! \mu'' L'K' |2\mu \rangle \!
\right. \\
\\[-8pt]
& &
~~~~~~
\!+\!
\langle 2 \!\!-\!\! \mu'' 2 \mu''' | L'K' \rangle
\langle 2 \!\!-\!\! \mu' L'K' |2\mu \rangle \!
\!+\!
(\!-\!1)^{\mu'} \!
\langle 2 \!\!-\!\! \mu'' 2 \!\!-\!\! \mu' | L'K' \rangle
\langle 2 \!\!-\!\! \mu''' L'K' |2\mu \rangle \\
\\[-8pt]
& &
\left.
\!+
(\!-\!1)^{\mu'''} \!
\langle 2 \!\!-\!\! \mu' 2 \!\!-\!\!  \mu''' | L'K' \rangle
\langle 2 \!\!-\!\! \mu'' L'K' |2\mu \rangle \!
\!+\!
(\!-\!1)^{\mu''} \!
\langle 2 \!\!-\!\! \mu' 2 \!\!-\!\! \mu'' | L'K' \rangle
\langle 2 \!\!-\!\! \mu''' L'K' |2\mu \rangle
\right\} , \\
\\[-8pt]
& &
C_{5 \mu \mu' \mu'' \mu''' \mu''''} (R^2) 
\!=\!
0 .  
\end{array} \!\!
\right\}
\label{expressionforexpansioncoefficients}
\end{eqnarray}\\[-6pt]
Substituting
(\ref{expressionforexpansioncoefficients})
into
$C_n (R^2) (n \neq 0)$,
in
(\ref{expansionT0}),
we have the result\\[-14pt]
\begin{eqnarray}
\!\!\!\!\!
\begin{array}{rcl}
T^{(0)} (\phi ;R^2)
& &
\!\!\!\!\!\!\!\!\!
=
C_0 (R^2)
\!+\!
{\displaystyle \frac{\hbar^2 R_0^2}{2m}}
\!\cdot\!
{\displaystyle \frac{25}{4 \pi }} \!
\left( \!
1 \!-\! 4 \!\cdot\! {\displaystyle \frac{1}{3A}}
\!+\!
{\displaystyle \frac{60}{7}}  \!\cdot\!
{\displaystyle \frac{1}{3A}}
\!\cdot\!
{\displaystyle
\frac{21A \!-\! 93}
{15 \!+\! 7 \sqrt{21}} \!
}
\right) \!\! 
R_0^2 R^{-6} \!
\sum_{\mu} \!
\phi_{2\mu }^* \phi_{2\mu} \\
\\[-14pt]
& &
~~~~~~~
+
{\displaystyle \frac{\hbar^2 R_0^2}{2m}}
\!\cdot\! 
{\displaystyle \frac{25}{12 \pi }} \!
\left\{ \!
{\displaystyle \frac{1}{2}}
\!\cdot\!
{\displaystyle \frac{1}{3A}}
\!+\!\!
\left( \!
1 \!-\! 4 \!\cdot\! {\displaystyle \frac{1}{3A}}
\!+\!
{\displaystyle \frac{60}{7}}
\!\cdot\!
{\displaystyle \frac{1}{3A}}
\!\cdot\!
{\displaystyle
\frac{21A \!-\! 93}
{15 \!+\! 7 \sqrt{21}} \!
}
\right) \!\! 
R_0^2  \!
\right\} \\
\\[-16pt]
& &
~~~
\!\times\! 
\sqrt{{\displaystyle \frac{35}{8\pi }}}
R_0^2R^{-8} \!
\sum_{\mu \mu' \mu''} \!
\left\{ \!
(\!-\!1)^{\mu'} \!
\langle 2 \!\!-\!\! \mu' 2\mu'' |2\mu \rangle \!
+\!
\langle 2\mu' 2\mu'' |2\mu \rangle \!
\right\} \!
\phi_{2\mu }^* \phi_{2\mu'}^* \phi_{2\mu''}^* \\
\\[-12pt]
& &
\!\!\!\!\!\!\!\!\!\!\!\!\!\!\!\!\!\!\!\!\!\!\!\!\!\!\!\!\!\!\!\!\!\!\!\!\!\!\!\!\!\!\!\!\!\!\!\!
+
{\displaystyle \frac{\hbar^2 R_0^2}{2m}}
\!\cdot\!
(i\hbar)^4 \!
{\displaystyle \frac{25}{16 \pi }}
{\displaystyle \frac{4}{3A}} \!
\sqrt{{\displaystyle \frac{35}{8\pi }}} \!
R_0^{6} R^{-10} \!
\sum_{L}\!\!
\sqrt{ 5 (2L \!\!+\!\! 1) } W(2222;2L) \!\!
\sum_{L'K'} \!\!
\sqrt{ 5 (2L' \!\!+\!\! 1) } W(2222;LL') \\ 
\\[-12pt]
& &
\!\!\!\!\!\!\!\!\!\!\!\!\!\!\!\!\!\!\!\!\!\!\!\!\!\!\!\!\!\!\!\!\!\!
\!\times\!
\sum_{\mu \mu' \mu'' \mu'''} \!
\left\{
\langle 2 \!\!-\!\! \mu''' 2 \mu'' | L'K' \rangle
\langle 2 \!\!-\!\! \mu' L'K' |2\mu \rangle \!
\!+\!
(\!-\!1)^{\mu'} \!
\langle 2 \!\!-\!\! \mu''' 2 \!\!-\!\! \mu' | L'K' \rangle
\langle 2 \!\!-\!\! \mu'' L'K' |2\mu \rangle \!
\right. \\
\\[-6pt]
& &
\!\!\!\!\!\!\!\!\!\!\!\!\!\!\!\!\!\!\!\!\!\!
~~~~~~~
\!+\!
\langle 2 \!\!-\!\! \mu'' 2 \mu''' | L'K' \rangle
\langle 2 \!\!-\!\! \mu' L'K' |2\mu \rangle \!
\!+\!
(\!-\!1)^{\mu'} \!
\langle 2 \!\!-\!\! \mu'' 2 \!\!-\!\! \mu' | L'K' \rangle
\langle 2 \!\!-\!\! \mu''' L'K' |2\mu \rangle \\
\\[-6pt]
& &
\!\!\!\!\!\!\!\!\!\!\!\!\!\!\!\!\!\!\!\!\!\!\!\!\!\!\!\!\!\!\!\!\!\!
\left.
\!+
(\!-\!1)^{\mu'''} \!
\langle 2 \!\!-\!\! \mu' 2 \!\!-\!\!  \mu''' | L'K' \rangle \!
\langle 2 \!\!-\!\! \mu'' L'K' |2\mu \rangle \!
\!+\!
(\!-\!1)^{\mu''} \!
\langle 2 \!\!-\!\! \mu' 2 \!\!-\!\! \mu'' | L'K' \rangle \!
\langle 2 \!\!-\!\! \mu''' L'K' |2\mu \rangle \!
\right\} \\
\\[-10pt]
& &
~~~~~~~~~~~~~~~~~~~~~~~~~~~~~~~~~~~~~~~~~~~~~~~~~~~~~~~~~~
\times
\phi_{2\mu }^* \phi_{2\mu'}^* \phi_{2\mu''}^* \phi_{2\mu'''}^*
\!+\!
\cdots .
\end{array}
\label{expressionforexpansionT0}
\end{eqnarray}
We calculate the first term
$C_0 (R^2)$
in the R.H.S. of
(\ref{expressionforexpansionT0}).
First we approximate
$\Pi_{2\mu }$ as\\[-20pt] 
\begin{eqnarray}
\!\!\!\!\!\!\!\!\!\!\!\!
\begin{array}{rcl}
& &
\Pi_{2\mu }
\!\approx\!
\pi_{2\mu }
\!+\!
\sqrt{{\displaystyle \frac{35}{8\pi }}} R_0^2 R^{-2} \!
\sum_{\mu_1 \mu_2} \!
(\!-\!1)^{\mu_1} \!
\langle 2 \!-\! \mu_1 2\mu_2 |2\mu \rangle
\phi_{2\mu_1} \!
\pi_{2 \mu_2 } \\
\\[-14pt]
& &
\!+\!
\left( \! 
\sqrt{{\displaystyle \frac{35}{8\pi }}} R_0^2 R^{-2} \!\!
\right)^{\!\!2} \!\!
\sum_{\mu_1 \mu_2} \!
\sum_{\mu_3 \mu_4}
(\!-\!1)^{\mu_1} \!
\langle 2 \!-\! \mu_1 2\mu_2 |2\mu \rangle
(\!-\!1)^{\mu_3} \!
\langle 2 \!-\! \mu_3 2\mu_4 |2 \mu_2 \rangle
\phi_{2\mu_1}
\phi_{2\mu_3}
\pi_{\mu_4}  .
\end{array}
\label{approxPi}
\end{eqnarray}\\[-14pt]
Substituting
(\ref{expressionforexpansionT0})
and
(\ref{approxPi})
into
(\ref{expansionT2})
with
$
\widehat{\Pi }_{2\mu }
\!=\!
R^{-2}
\Pi_{2\mu }
\!+\!
i\hbar 
{\displaystyle \frac{5}{4 \pi}} R_0^4 R^{-6}
(\!-\!1)^{\mu}
\phi_{2 \!-\! \mu } 
$,
the constant term
$C_0 (R^2)$
is computed up to the third order of
$\phi_{2\mu}$:\\[-18pt]
\begin{eqnarray}
\begin{array}{rcl}
C_0 (R^2) 
&\!\!\!\!=\!\!\!\!& 
T
\!-
{ \displaystyle  \frac{\hbar^2 R_0^2}{2m}\frac{10}{3AR_0^4} }
R_0^2 R^{-2} \!
\sum_{\mu} \! 1 \\
\\[-12pt]
& &
~~  
\!-
{\displaystyle \frac{\hbar^2 R_0^2}{2m}}
\!\cdot\!
{\displaystyle \frac{25}{4 \pi }} \!
\left( \!
1 \!-\! 4 \!\cdot\! {\displaystyle \frac{1}{3A}}
\!+\!
{\displaystyle \frac{60}{7}} \!\cdot\!
{\displaystyle \frac{1}{3A}}
\!\cdot\!
{\displaystyle
\frac{21A \!-\! 93}
{15 \!+\! 7 \sqrt{21}} \!
}
\right) \!\! 
R_0^2 R^{-6}
\sum_{\mu} \! 
\phi_{2\mu }^* \phi_{2\mu} \\
\\[-16pt]
& &
~~\!
-
{\displaystyle \frac{4\pi }{3AR_0^4} \frac{1}{m} }
\sum_\mu \!
\left\{ \!
R^2 \!\!
\left( \!\!
\pi_{2\mu }
\!+\!
\sqrt{{\displaystyle \frac{35}{8\pi }}} R_0^2 R^{-2} \!
\sum_{\mu_1 \mu_2} \!
\langle 2 \mu_1 2\mu_2 |2\mu \rangle
\phi_{2\mu_1}^* \!
\pi_{2 \mu_2 } \!\!
\right)
\right. \\
\\[-16pt]
& &
\left.
~~~~~~~~~~~~~~
\!\times\!
(\!-\!1)^\mu \!\!
\left( \!\!
\pi_{2 \!-\! \mu }
\!+\!
\sqrt{{\displaystyle \frac{35}{8\pi }}} R_0^2 R^{-2} \!
\sum_{\mu_1 \mu_2} \!
\langle 2 \mu_1 2\mu_2 |2 \!-\! \mu \rangle
\phi_{2\mu_1}^* \!
\pi_{2 \mu_2 } \!\!
\right)  
\right. \\
\\[-14pt]
& &\!\!
\!\!\!\!\!\!\!\!\!\!\!\!
\left.
\!+
i\hbar
{\displaystyle \frac{5}{4 \pi}} R_0^4 R^{-2}
\phi_{2\mu } \!\!
\left( \!\!
\pi_{2\mu }
\!+\!
\sqrt{{\displaystyle \frac{35}{8\pi }}} R_0^2 R^{-2} \!
\sum_{\mu_1 \mu_2} \!
\langle 2 \mu_1 2\mu_2 |2\mu \rangle
\phi_{2\mu_1}^* \!
\pi_{2 \mu_2 } \!\!
\right)
\right. \\
\\[-16pt]
& &\!\!
\!\!\!\!\!\!\!\!\!\!\!\!
\left.
\!+
i\hbar
{\displaystyle \frac{5}{4 \pi}} R_0^4 R^{-2}
\phi_{2\mu }^* 
(\!-\!1)^\mu \!\!
\left( \!\!
\pi_{2 \!-\! \mu }
\!+\!
\sqrt{{\displaystyle \frac{35}{8\pi }}} R_0^2 R^{-2} \!
\sum_{\mu_1 \mu_2} \!
\langle 2 \mu_1 2\mu_2 |2 \!-\! \mu \rangle
\phi_{2\mu_1}^* \!
\pi_{2 \mu_2 } \!\! 
\right) \!\!
\right\} \\
\\[-14pt]
& &
~
+{\displaystyle \frac{4\pi }{3AR_0^2} \frac{1}{m}} 
\sqrt{{\displaystyle \frac{35}{8\pi }}} 
\sum_{\mu \mu' \mu''}
(\!-\!1)^{\mu }
\langle 2 \mu' 2 \mu'' |2 \!-\! \mu \rangle \\
\\[-14pt]
& &\!\!
\!\!\!\!\!\!\!\!\!\!\!\!\!\!\!\!\!\!\!\!\!\!\!\!\!\!
\times\!
\left\{ \!\!
R^4 \phi_{2 \mu'' }^* \!\!
\left( \!\!
R^{-2}
\pi_{2\mu }
\!\!+\!\!
\sqrt{{\displaystyle \frac{35}{8\pi }}} R_0^2 R^{-4} \!
\sum_{\mu_1 \mu_2} \!
\langle 2 \mu_1 2\mu_2 |2\mu \rangle
\phi_{2\mu_1}^* \!
\pi_{2 \mu_2 } \!
\!+\!
i\hbar 
{\displaystyle \frac{5}{4 \pi}} R_0^4 R^{-6}
\phi_{2 \mu }^*  \!\!
\right) 
\right.  \\
\\[-14pt]
& &\!\! 
\!\!\!\!\!\!\!\!\!\!\!
\left.
\!\times\!
\left( \!\!
R^{-2}
\pi_{2\mu' }
\!\!+\!\!
\sqrt{{\displaystyle \frac{35}{8\pi }}} R_0^2 R^{-4} \!
\sum_{\mu_1 \mu_2} \!
\langle 2 \mu'_1 2 \mu''_2 |2 \mu' \rangle
\phi_{2\mu'_1}^* \!
\pi_{2 \mu''_2 } \!
\!+\!
i\hbar 
{\displaystyle \frac{5}{4 \pi}} R_0^4 R^{-6}
\phi_{2 \mu' }^*  \!\!
\right) \!\!
\right\}  \\
\\[-14pt]
& &
~
-
{\displaystyle \frac{\hbar^2 R_0^2}{2m}}
\!\cdot\! 
{\displaystyle \frac{25}{12 \pi }} \!
\left\{ \!
{\displaystyle \frac{1}{2}}
\!\cdot\!
{\displaystyle \frac{1}{3A}}
\!+\!\!
\left( \!
1 \!-\! 4 \!\cdot\! {\displaystyle \frac{1}{3A}}
\!+\!
{\displaystyle \frac{60}{7}} \!\cdot\!
{\displaystyle \frac{1}{3A}}
\!\cdot\!
{\displaystyle
\frac{21A \!-\! 93}
{15 \!+\! 7 \sqrt{21}} \!
}
\right) \!\! 
R_0^2  \!
\right\} \\
\\[-14pt]
& &
~~~~
\!\times\!
\sqrt{{\displaystyle \frac{35}{8\pi }}}
R_0^2 R^{-8} \!
\sum_{\mu \mu' \mu''} \!
\left\{ \!
(\!-\!1)^{\mu'} \!
\langle 2 \!\!-\!\! \mu' 2\mu'' |2\mu \rangle \!
+\!
\langle 2\mu' 2\mu'' |2\mu \rangle \!
\right\} \!
\phi_{2\mu }^* \phi_{2\mu'}^* \phi_{2\mu''}^* \\
\\[-14pt]
& &
~
+
{\displaystyle \frac{\hbar^2 R_0^2}{2m}} \!
\left( \! 
{\displaystyle \frac{2 \!\cdot\! 5}{4 \pi}} \!
 \right)^{\!2} \!\!
{\displaystyle \frac{4\pi }{3A R_0^2}} 
\sqrt{{\displaystyle \frac{35}{8\pi }}}
R_0^2 R^{-8} \!
\sum_{\mu \mu' \mu''} \!
(\!-\!1)^{\mu} \!  
\langle 2\mu' 2 \mu'' |2 \!-\! \mu \rangle \!
\phi_{2 \mu }^*  \phi_{2 \mu' }^* 
\phi_{2 \mu'' }^*.
\end{array}
\label{expansionC0}
\end{eqnarray}\\[-4pt]
The product of operators  $\phi_{2 \mu }$ and  $\pi_{2 \mu }$ in
(\ref{expansionC0})
is changed to the {\em normal product} form with
all operators $\phi_{2 \mu }$ standing to the left of
all operators $\pi_{2 \mu }$,
as shown in
(\ref{expansionC02})  in Appendix F.

Next we derive an approximate relation between
the $T$ and the term
$
\sum_{\mu }
\eta_{2 \mu }
(\!-\!1)^{\mu} \!
\eta_{2 \!-\!  \mu }
$.
This is made with the use of the definition of the variables
$\eta_{2 \mu }$
(\ref{definitioneta})
as follows:\\[-16pt]
\begin{eqnarray}
\!\!\!\!\!\!\!\!\!\!\!\!\!\!\!
\begin{array}{rcl}
& &
\sum_{\mu } \!
\eta_{2 \mu }
(\!-\!1)^{\mu}
\eta_{2 \!-\!  \mu }
\!=\!
(-i\hbar)^2 
{\displaystyle \frac{2 \!\cdot\! 5}{4}} \!
\sum_{\mu} \!
\left\{ \!
\sum_{\kappa \nu } 
(\!-\!1)^\mu \langle 1\kappa 1\nu|2 \!-\! \mu \rangle \!\!
\sum_{n=1}^A \! 
r_n Y_{1\kappa }(\theta_n ,\varphi_n) 
\!\cdot\! 
\vnabla_\nu^n 
\right. \\
\\[-18pt]
& &
\left.
~~~~~~~~~~~~~~~~~~~~~~~~~~~~~~~~~~~~~~~~\!\!\!
\!\times~\!\!
(\!-\!1)^{\mu} \!
\sum_{\kappa' \nu' } 
(\!-\!1)^{-\mu}
\langle 1 \kappa' 1 \nu' | 2 \mu \rangle \!\!
\sum_{n'=1}^A \! 
r_{n'} Y_{1 \kappa' }(\theta_{n'} , \varphi_{n'}) 
\!\cdot\!
\vnabla_{\nu'}^{n'} \!
\right\} \\
\\[-16pt]
& &
\!\approx\!
\hbar^2
{\displaystyle \frac{2 \!\cdot\! 5}{4}} \!
\sqrt{{\displaystyle \frac{5}{3}}} \!
\sum_{\kappa \nu \kappa' \nu' } 
(\!-\!1)^\kappa \!
\sum_{\mu} \!
\langle 1 \kappa' 1 \nu' | 2 \mu \rangle \!
\langle 2 \mu 1 \nu |1 \!-\!  \kappa \rangle \!\!
\sum_{n\!=\!1}^A \! 
r_n^2
Y_{1\kappa }(\theta, \varphi)
Y_{1\kappa' }(\theta, \varphi) \!\!
\sum_{n\!=\!1}^A \!\!
\vnabla_\nu^n
\vnabla_{\nu'}^n \\
\\[-16pt]
& &
\!\approx\!
\hbar^2 
{\displaystyle \frac{2 \!\cdot\! 5}{4}} \!
\sqrt{{\displaystyle \frac{5}{3}}}
{\displaystyle \frac{3A}{5}}
R^2
{\displaystyle \frac{1}{4 \pi}}
\sqrt{5}W(2111:20) \!
\sum_{\kappa} \! 1 \!
\sum_{\nu } \!
\langle 1 \nu 1\!-\! \nu | 0 0 \rangle \!
\sum_{n=1}^A \!\!
\vnabla_\nu^n
\vnabla_{-\nu}^n ,
\end{array}
\label{etaetaT}
\end{eqnarray}

$\!\!\!\!\!\!\!\!$from which we get an important relation\\[-16pt]
\begin{eqnarray}
\begin{array}{c}
T
\!=\!
{\displaystyle \frac{4 \pi}{5 A}}
{\displaystyle \frac{1}{m}}
R^{-2} \!
\sum_{\mu } \!
\eta_{2 \mu }
(\!-\!1)^{\mu}
\eta_{2 - \mu } .
\label{Tetaeta}
\end{array}
\end{eqnarray}\\[-16pt]
Substituting the relation and the approximate relation\\[-18pt]
\begin{eqnarray}
\!\!\!\!
\begin{array}{c}
\pi_{2\mu } 
\!=\! 
R_0^2 \!
\left( \!\!
R^{-2} \eta_{2\mu }
\!+\!
{\displaystyle \frac{1}{2 }}
[\eta_{2\mu }, R^{-2})] \!\!
\right)
\!=\!
R_0^2 R^{-2}
\eta_{2\mu }
\!+\!
i\hbar {\displaystyle \frac{5}{4\pi }}
R_0^4 R^{-4}
\phi_{2\mu }^* , 
\label{paiapproxaeta}
\end{array}
\end{eqnarray}
\vspace{-0.7cm}
\begin{eqnarray}
\!\!\!\!\!\!\!\!\!\!\!\!\!\!
\begin{array}{c}
R^2 \!
\sum_{\mu } \!
\pi_{2 \mu }
(\!-\!1)^{\mu}
\pi_{2 \!-\!  \mu } 
\!\approx\!
R_0^4 R^{-2} \!
\sum_{\mu } \!
\eta_{2\mu }
(\!-\!1)^{\mu}
\eta_{2 \!-\! \mu }
\!-\!
\left( \! i\hbar \! \right)^2 \!
{\displaystyle \frac{5}{4\pi }} \!
R_0^4 R^{-2} \!
\sum_{\mu } \! 1 \\
\\[-14pt]
+
i\hbar
{\displaystyle \frac{5}{4\pi }} \!
R_0^6 R^{-4} \!
\sum_{\mu } \!
\left( \!
\phi_{2 \mu } 
\eta_{2 \mu }
\!+\!
\phi_{2 \mu }^*
(\!-\!1)^{\mu}
\eta_{2\!-\! \mu } \!
\right)
\!+\!
5 \!
\left( \!
i\hbar {\displaystyle \frac{5}{4\pi }} \!
\right)^{\!\!2} \!\!
R_0^8 R^{-6} \!
\sum_{\mu } \!
\phi_{2 \mu }^*
\phi_{2 \mu } ,
\end{array}
\label{pieta}
\end{eqnarray}\\[-14pt]
and
(\ref{Tetaeta})
into
(\ref{expansionC02})
and
discarding third order terms of
$\phi$
and triple products of operator 
$\phi_{2 \mu } \phi_{2 \mu' } \pi_{2 \mu'' }$
etc.
except
$\phi_{2 \mu }^* \pi_{2 \mu_2 } (\!-\!1)^\mu \pi_{2 \!-\! \mu }$
and
$\phi_{2 \mu } \pi_{2 \mu' } \pi_{2 \mu'' }$,
(\ref{expansionC0})
is cast into a form as
(\ref{expansionC02}).
\\[-12pt]

Finally,
only the first constant term in the R.H.S. of
(\ref{expansionC03})
survives and
another terms may be negligibly small 
if they are operated onto the collective sub-space 
$|\Psi^{\mbox{coll}} \rangle$.
The second term may vanish
if we take account of effects by
the operators
$\phi_{2 \mu} \pi_{2 \mu} \pi_{2 \mu'}$,
etc.,
which are approximated as
$\phi_{20} \pi_{2 \mu} (\!-\!1)^{\mu}\pi_{2 \!-\! \mu}$
as did in
(\ref{expansionC02}).
It is, however,  a very difficult problem to prove vanishing of,
in particular, the third term which includes the first order differentials,
on the collective sub-space 
$|\Psi^{\mbox{coll}} \rangle$.
As a result, the constant term
$C_0 (R^2) $
leads to\\[-18pt]
\begin{eqnarray}
\begin{array}{c}
C_0 (R^2) 
\approx 
-
{ \displaystyle  \frac{\hbar^2 R_0^2}{2m}
\frac{10}{3AR_0^4} } \! 
\left\{ \!
1
\!+\!
\left( \!
1
\!+\!
{\displaystyle \frac{4}{5 }}
\sqrt{{\displaystyle \frac{7}{2 }}} \!
\right) \!
\right\} \!
R_0^2 R^{-2} \!
\sum_{\mu} \! 1 ,~
( \sum_{\mu} \! 1 \!=\! 5 ) ,
\end{array}
\label{approxC0}
\end{eqnarray}\\[-12pt]
where we have used the approximate relation for
$\phi_{2 0}$
given by\\[-20pt]
\begin{eqnarray}
\begin{array}{c}
\phi_{2 0}
\!=\!
{\displaystyle \frac{4\pi }{3AR_0^2}} \!
\sum_{n = 1}^A \!
r_n ^2
\sqrt{{\displaystyle \frac{5}{4 \pi }}} \!
\left( \!
{\displaystyle \frac{3}{2}}
\cos ^2 \! \theta_n
\!-\!
{\displaystyle \frac{1}{2}} \!
\right)
\!\approx\!
{\displaystyle \frac{4\pi }{3AR_0^2}}
\sqrt{{\displaystyle \frac{5}{4 \pi }}} \!
\sum_{n = 1}^A \!
r_n ^2
\!=\!
\sqrt{{\displaystyle \frac{4 \pi }{5}}}  
{\displaystyle \frac{R^2}{R_0^2}} .
\end{array}
\label{approxphi20}
\end{eqnarray}\\[-14pt]
The collective Hamiltonian should be expressed in terms of collective
variables only.
However, $R^2$ is not a collective variable, so it must be
replaced by the average value $< \!\!\!R^2 \!\! >$.
Following the convection to conceive $\!< \!\!\!R^2 \!\!\! >\!$
as a constant $\!R_0 ^2\!$
\cite{MiyaTamu.56},
then we have 
$< \!\!\!R^2 \!\! > =\!\! R_0 ^2$.
After that,
substituting
(\ref{expressionforexpansionT0})
and
(\ref{approxC0})
into
(\ref{expansionT2}),
we reach our final goal of expressing the kinetic part
$T$ of the original Hamiltonian $H$ in terms of the $exact$ canonical variables 
$\phi_{2 \mu }$ and $\Pi_{2 \mu }$
up to the forth order of
$\phi_{2 \mu }$.
But it is so lengthy that we avoid to write down here
its explicit full expression.
Thus we can give one of possible foundations of
the nuclear collective motion.

\vspace{-0.2cm}


\def\thesection{\arabic{section}}
\setcounter{equation}{0}
\renewcommand{\theequation}{\arabic{section}.
\arabic{equation}}

\section{Concluding remarks}

\vspace{-0.1cm}

~~~
First we give the basic spirit of the Marumori's essential idea to approach
the collective motions in nuclei which is stated in his original paper of 1955
on the foundation of the unified nuclear model as follows:
``On the basis such a description of the nucleus,
A. Bohr and his collaborators analyzed the available empirical evidences 
and gave a reasonable interpretation of nuclear properties.
According to their model, the nuclear motion is treated as the dynamics
of a coupled system of individual particle motion and collective oscillations.
Therefore, the states of this system are described in terms of individual particle-coordinates and collective-coordinates,
that is, the collective coordinates which describe the nuclear collective properties are recognized as independent dynamical variables
as well as individual particle coordinates.
However, these collective coordinates characterizing
the spatial distribution of the nucleon density should be,
strictly speaking, symmetric functions of the individual nucleon coordinates
and we cannot expect the total number of degree of freedom
to become larger than the number of particle coordinates originally given.$\cdots$",
which is cited from Ref.
\cite{Maru.55}.
Prior to the Marumori's original paper,
A. Bohr and B. Mottelson proposed the famous particle-core model 
in 1952 and 1953
\cite{BM.52.53}. 

To extend the Tomonaga's method  to the three-dimensional case,
Miyazima and Tamura introduced the collective coordinates
$\xi_{2 \mu}$
but met the difficulty of finding out appropriate (mutually independent)
momentum operators
$\pi_{2 \mu}$
which are canonically conjugate to the
$\xi_{2 \mu}$.
To overcome this difficulty,
they further introduced some new redundant coordinates
$\alpha_{2 \mu}$
and their conjugate momenta
$\beta_{2 \mu}$
but at the same time imposed an appropriate number of subsidiary conditions
so as to cancel out the increase of the degrees of freedom.
Next they performed appropriate canonical transformations,
so that the newly introduced coordinates have the meaning of
the collective motions.
Finally they rewrote the Hamiltonian in a form of a sum of three parts,
one describing the collective surface oscillation, another the internal motion
and the remaining one the interaction between them
\cite{MiyaTamu.56}.
While
Nakai derived the correct canonical momentum by using the theory of
point transformations and made clear the relation between Tomonaga's method
and the exact method of canonical transformations.
Applying his method, he investigated the problem of surface oscillations
of atomic nuclei and treated the correspondence with the BMM and
the Miyazima-Tamura's earlier work
\cite{Nakai.63}.

In this paper, on the contrary,
the $exact$ canonically conjugate momenta
$\Pi_{2 \mu}$
to
$\phi_{2 \mu}$
is derived by modifying the approximate momenta
$\pi_{2 \mu}$
with the use of the discrete version of the Sunakawa's integral equation
\cite{SYN.62}.
We have shown that
the $exact$ canonical commutation relations between
the collective variables
$\phi_{2 \mu}$
and
$\Pi_{2 \mu}$
and the commutativity of the momenta
$\Pi_{2 \mu}$ and $\Pi_{2 \mu'}$.
By making use of
the $exact$ canonical variables
$\phi_{2 \mu}$
and
$\Pi_{2 \mu}$,
we found the collective Hamiltonian
which includes the so-called surface phonon-phonon interaction
corresponding to the Hamiltonian of Bohr-Mottelson model.
We concentrate our concerns only on the collective motion in nuclei.
The discussion of the couplings between the individual particle motion and
the collective motion is beyond the scope of the present theory. 
In the case of simpler two-dimensional nuclei,
such couplings are able to be studied with the use of the basic idea of the present theory.
Particularly the structure of the collective sub-space
satisfying the subsidiary condition is investigated in detail
relating to the individual particle motion through the variable $R^2$.
This work will appear elsewhere.

Following Miyazima-Tamura and Nakai
\cite{MiyaTamu.56,Nakai.63},
we take the interaction potential function in the form of  Fourier integral
as\\[-16pt]
\begin{eqnarray}
\begin{array}{c}
V   
\!=\!
{\displaystyle \frac{1}{2}} 
\sum_{n,n' = 1}^A \!
V(\rb_n ,\rb_{n'})
\!=\!
{\displaystyle \frac{1}{2}} 
\sum_{n,n' = 1}^A \!
{\displaystyle \int} \!\! f(k) e^{i\kb \cdot (\rb_n - \rb_{n'})}d\kb, ~~
f(k)
\!=\!
{\displaystyle \frac{V_0}{2 \pi^2 \kappa}}
{\displaystyle \frac{1}{\kb^2 \!+\! k^2}} ,
\end{array}
\label{potential}
\end{eqnarray}\\[-14pt]
from which we have for the first term of
$
{\displaystyle \frac{\partial^2 V}{\partial\phi_{2 \mu}\partial\phi_{2 \mu'}} }
$:
This term plays an important role to evaluate the surface tension energy
as seen in
\cite{MiyaTamu.56,Nakai.63}.


\vspace{0.4cm}

\centerline{\bf Acknowledgements}
\vspace{0.2cm}
S. N. would like to
express his sincere thanks to 
Professor Manuel Fiolhais for kind and
warm hospitality extended to
him at the Centro de F\'\i sica Computacional,
Universidade de Coimbra.
This work was supported by FCT (Portugal) under the project
CERN/FP/83505/\\2008.
The authors thank the Yukawa Institute for Theoretical Physics
at Kyoto University.

\newpage


\leftline{\large{\bf Appendix}}
\appendix

\vspace{0.1cm}


\def\thesection{\Alph{section}}
\setcounter{equation}{0}
\renewcommand{\theequation}{\Alph{section}.\arabic{equation}}
\section{$\!\!$Calculations of commutators
$[\pi_{2\mu },~\Pi_{2\mu'}]$ and $[\Pi_{2\mu },~\Pi_{2\mu'}]$}


\vskip0.2cm

~~~First, using
(\ref{exactpi})
and 
(\ref{exactcommurelpiphi}), 
we can calculate the commutation relation for 
$[\pi_{2\mu },~\Pi_{2\mu'}]$
as
\begin{eqnarray}
\begin{array}{lll}
&&[\pi_{2\mu },~\Pi_{2\mu'}] 
=
-i\hbar \sqrt{{\displaystyle \frac{35}{8\pi }}} R_0^2  
\sum_{\mu_1 }
\langle 2\mu 2\mu'|2\mu_1 \rangle 
{\displaystyle \frac{1}{2}}
\left(
R^{-2}\Pi_{2\mu_1 } + R^{-2}\Pi_{2\mu_1 }
\right) \\
\\[-10pt]
\!\!\!\!&+&\!\!\!\!
\sqrt{{\displaystyle \frac{35}{8\pi }}} R_0^2 
\sum_{\mu_1 \mu_2} 
(-1)^{\mu_1} \langle 2-\mu_1 2\mu_2 |2\mu \rangle
\phi_{2\mu_1} 
{\displaystyle \frac{1}{2}}
\left(
[\Pi_{2\mu'},~R^{-2}]\Pi_{2\mu_2} + \Pi_{2\mu_2}[\Pi_{2\mu'},~R^{-2}]
\right) \\
\\[-8pt]
\!\!\!\!&+&\!\!\!\!
[\Pi_{2\mu },~\Pi_{2\mu'}] \\
\\[-8pt]
\!\!\!\!&+&\!\!\!\!
\sqrt{{\displaystyle \frac{35}{8\pi }}} R_0^2 
\sum_{\mu_1 \mu_2} 
(-1)^{\mu_1} \langle 2-\mu_1 2\mu_2 |2\mu \rangle
\phi_{2\mu_1} 
{\displaystyle \frac{1}{2}}
\left(
R^{-2}[\Pi_{2\mu'},~\Pi_{2\mu_2}] + [\Pi_{2\mu'},~\Pi_{2\mu_2}]R^{-2}
\right) .
\end{array}
\label{commurelpiPai}
\end{eqnarray}
Using
(\ref{commuphi}), (\ref{exactpi})
and 
(\ref{exactcommurelpiphi}), 
the commutation relation for 
$[\Pi_{2\mu },~\Pi_{2\mu'}]$
is also computed straightforwardly as
\begin{eqnarray}
\!\!\!\!\!\!\!\!\!\!\!\!
\begin{array}{lll}
&&
[\Pi_{2\mu },~\Pi_{2\mu'}] 
=
[\pi_{2\mu },~\pi_{2\mu'}] \\
\\[-8pt]
&&
-
\sqrt{{\displaystyle \frac{35}{8\pi }}} R_0^2 
\sum_{\mu_1 \mu_2}
(-1)^{\mu_1} \langle 2-\mu_1 2\mu_2 |2\mu \rangle 
\phi_{2\mu_1}
{\displaystyle \frac{1}{2}} \!
\left(
R^{-2}[\pi_{2\mu' },\Pi_{2\mu_2}]
\!+\!
 [\pi_{2\mu' },\Pi_{2\mu_2}]R^{-2}
\right) \\
\\[-8pt]
&&
\!+\!
(\mu \leftrightarrow \mu') \\
\\[-8pt]
&&
-
i\hbar \sqrt{{\displaystyle \frac{35}{8\pi }}} 
{\displaystyle \frac{2 \!\cdot\! 5}{4\pi }} \!
R_0^6 \!
\sum_{\mu_1 \mu_2}
(-1)^{\mu_1} \!
\langle 2-\mu_1 2\mu_2 |2\mu \rangle 
\phi_{2\mu_1}
{\displaystyle \frac{1}{2}} \!
\left(
R^{-6}\phi_{2\mu' }^* \Pi_{2\mu_2} 
\!+\!
\Pi_{2\mu_2}R^{-6}\phi_{2\mu' }^*
\right) \\
\\[-8pt]
&&
\!+\!
(\mu \leftrightarrow \mu') \\
\\[-4pt]
&&
+
\left( \!
\sqrt{{\displaystyle \frac{35}{8\pi }}} R_0^2 \!
\right)^{\!\!2} \!
\sum_{\mu_1 \mu_2} \sum_{\mu'_1 \mu'_2} 
(-1)^{\mu_1} \langle 2-\mu_1 2\mu_2 |2\mu \rangle 
(-1)^{\mu'_1} \langle 2-\mu'_1 2\mu'_2 |2\mu' \rangle \\
\\[-12pt]
&&
\hspace{100pt}
\times 
\phi_{2\mu_1} \phi_{2\mu'_1}
{\displaystyle \frac{1}{2}}
\left\{
\Pi_{2\mu' _2}
{\displaystyle \frac{1}{2}}
\left(
R^{-2}[\Pi_{2\mu_2 },~R^{-2}] + [\Pi_{2\mu_2 },~R^{-2}]R^{-2}
\right) 
\right. \\
\\[-12pt]
&&
\left.
\hspace{185pt}
+
{\displaystyle \frac{1}{2}}
\left(
R^{-2}[\Pi_{2\mu_2 },~R^{-2}] + [\Pi_{2\mu_2 },~R^{-2}]R^{-2}
\right)
\Pi_{2\mu' _2}
\right\}\\
\\[-10pt]
&&
- (\mu \leftrightarrow \mu') \\
\\
&&
+
{\displaystyle \frac{1}{2}} \!
\left( \!
\sqrt{{\displaystyle \frac{35}{8\pi }}} R_0^2 \!
\right)^{\!\!2} \!
\sum_{\mu_1 \mu_2} \sum_{\mu'_1 \mu'_2} 
(-1)^{\mu_1} \langle 2-\mu_1 2\mu_2 |2\mu \rangle 
(-1)^{\mu'_1} \langle 2-\mu'_1 2\mu'_2 |2\mu' \rangle \\
\\[-12pt]
&&
\hspace{100pt}
\times 
\phi_{2\mu_1} \phi_{2\mu'_1}
{\displaystyle \frac{1}{2}}
\left\{
R^{-2}
{\displaystyle \frac{1}{2}}
\left(
R^{-2}[\Pi_{2\mu_2 },~\Pi_{2\mu' _2}] + [\Pi_{2\mu_2 },~\Pi_{2\mu' _2}]R^{-2}
\right)
\right. \\
\\[-10pt]
&&
\left.
\hspace{182pt}
+
{\displaystyle \frac{1}{2}}
\left(
R^{-2}[\Pi_{2\mu_2 },~\Pi_{2\mu' _2}] + [\Pi_{2\mu_2 },~\Pi_{2\mu' _2}]R^{-2}
\right)
R^{-2} 
\right\}\\
\\[-12pt]
&&
- (\mu \leftrightarrow \mu') .
\end{array}
\label{commurelpaipai}
\end{eqnarray}
The term in the forth line in 
(\ref{commurelpaipai})
is transformed to
\begin{eqnarray}
\begin{array}{rcl}
&&\!\!\!\!
-
i\hbar \sqrt{{\displaystyle \frac{35}{8\pi }}} 
{\displaystyle \frac{2 \cdot 5}{4\pi }}R_0^6 
\sum_{\mu_1 \mu_2}
(-1)^{\mu_1} \langle 2-\mu_1 2\mu_2 |2\mu \rangle 
\phi_{2\mu_1}
{\displaystyle \frac{1}{2}} \!
\left(
R^{-6}\phi_{2\mu' }^* \Pi_{2\mu_2}
\!+\! 
\Pi_{2\mu_2}R^{-6}\phi_{2\mu' }^*
\right) \\
\\[-8pt]
&=&\!\!\!\!
{\displaystyle \frac{1}{2}}
(i\hbar)^2 \sqrt{{\displaystyle \frac{35}{8\pi }}} 
{\displaystyle \frac{2 \cdot 5}{4\pi }}R_0^6 R^{-6}
\sum_{\mu_1}
\langle 2\mu 2\mu' |2\mu_1 \rangle 
\phi_{2\mu_1}^* \\
\\[-6pt]
&&\!\!\!\!
-
i\hbar \sqrt{{\displaystyle \frac{35}{8\pi }}} 
{\displaystyle \frac{2 \cdot 5}{4\pi }}R_0^6 R^{-4}
\phi_{2\mu' }^*
\sum_{\mu_1 \mu_2}
(-1)^{\mu_1} \langle 2-\mu_1 2\mu_2 |2\mu \rangle 
\phi_{2\mu_1}[\Pi_{2\mu_2 },~R^{-2}] \\
\\[-6pt]
&&\!\!\!\!
-
i\hbar {\displaystyle \frac{2 \cdot 5}{4\pi }}R_0^4 R^{-4}
\phi_{2\mu' }^*
\left(
\Pi_{2\mu } - \pi_{2\mu } 
\right) .
\end{array}
\label{commurelpaipai3}
\end{eqnarray}
In
(\ref{commurelpaipai3})
we have used 
(\ref{exactpi})
again and have derived the term in the last line.

We here introduce the quantity $\widehat{\Jb}_{L\Lambda }$ defined by
\begin{eqnarray}
\!\!\!\!
\begin{array}{rl}
\widehat{\Jb}_{L\Lambda }
&\!\!\!\!
\equiv\!
\sum_{\mu,\mu' }
\langle 2\mu 2\mu'|L\Lambda \rangle 
\phi_{2\mu }^*
{\displaystyle \frac{1}{2}} \!
\left\{ \!
R^{-2}
{\displaystyle \frac{1}{2}} \!
\left( \!
R^{-2}\Pi_{2\mu' } \!+\! \Pi_{2\mu' } R^{-2}
\right)
\!+\!
{\displaystyle \frac{1}{2}} \!
\left( \!
R^{-2}\Pi_{2\mu' } \!+\! \Pi_{2\mu' } R^{-2}
\right) \!
R^{-2} \!
\right\} \\
\\[-6pt]
&\!\!\!\!
=\!
\sum_{\mu,\mu' }
\langle 2\mu 2\mu'|L\Lambda \rangle 
\phi_{2\mu }^*
{\displaystyle \frac{1}{2}} \!
\left( \!
R^{-2} \widehat{\Pi}_{2\mu' } 
\!+\! 
\widehat{\Pi}_{2\mu' } R^{-2}
\right) .
\end{array}
\label{defjl}
\end{eqnarray}
With the aid of the quantity $\widehat{\Jb}_{L\Lambda }$,
we can obtain the following two important relations: The first relation is given as\\[-14pt]
\begin{eqnarray}
\!\!\!\!\!\!\!\!
\begin{array}{rcl}
&&\!\!\!\!
i\hbar {\displaystyle \frac{2 \cdot 5}{4\pi }}R_0^4 R^{-4}\!
\left(
\phi_{2\mu }^*\Pi_{2\mu' } \!-\! \phi_{2\mu' }^*\Pi_{2\mu } 
\right) \\
\\[-4pt]
&=&\!\!\!\!
i\hbar {\displaystyle \frac{2 \cdot 5}{4\pi }}R_0^4 \! \left\{ \!
\sum_{L\Lambda } \!\!
\left( \!
\langle 2\mu 2\mu'|L\Lambda \rangle 
\!\!-\!\!
\langle 2\mu' 2\mu|L\Lambda \rangle \!
\right) \!
\widehat{\Jb}_{L\Lambda }  
\!\!-\!\!
R^{-2} \!
\left(
\phi_{2\mu }^*[\Pi_{2\mu' },~R^{-2}]
\!-\!
\phi_{2\mu' }^*[\Pi_{2\mu },~R^{-2}]
\right) \!
\right\} \\
\\[-4pt]
&=&\!\!\!\!
i\hbar {\displaystyle \frac{2 \cdot 5}{4\pi }}R_0^4 \! 
\left\{ \!
\sum_{L\Lambda } \!
\left(
1 \!-\! (-1)^L
\right) \!
\langle 2\mu 2\mu'|L\Lambda \rangle \widehat{\Jb}_{L\Lambda } 
\!-\!
R^{-2} \!
\left(
\phi_{2\mu }^*[\Pi_{2\mu' },~R^{-2}]
\!-\!
\phi_{2\mu' }^*[\Pi_{2\mu },~R^{-2}]
\right) \!
\right\} ,
\end{array}
\label{relationphaipai}
\end{eqnarray}
and the second one related to the collective angular momentum in Section 4
is expressed as\\[-14pt]
\begin{eqnarray}
\begin{array}{rcl}
&&\!\!\!\!
\sum_{L}
\left(
1 \!-\! (-1)^L
\right) \!
\left\{ \!
i\hbar {\displaystyle \frac{2 \cdot 5}{4\pi }}R_0^4 
-
i\hbar 
\left( \!
\sqrt{{\displaystyle \frac{35}{8\pi }}} R_0^2 \!
\right)^{\!2} \!\!
5W(2222;2L) \!
\right\} \!
\sum_{\Lambda }
\langle 2\mu 2\mu'|L\Lambda \rangle 
\widehat{\Jb}_{L\Lambda } \\
\\[-8pt]
&&\!\!\!\!
=
i\hbar R_0^4
\cdot 2 \!
\left( \!
{\displaystyle \frac{2 \cdot 5}{4\pi }}
-
{\displaystyle \frac{35}{8\pi }}
\cdot 5 \cdot (-1)
{\displaystyle \frac{1}{2 \cdot 5 }} \!
\right) \!
\sum_{\Lambda }
\langle 2\mu 2\mu'|1\Lambda \rangle
\widehat{\Jb}_{1\Lambda } \\
\\[-6pt]
&&\!\!\!\!
+
i\hbar R_0^4
\cdot 2 \!
\left( \!
{\displaystyle \frac{2 \cdot 5}{4\pi }}
-
{\displaystyle \frac{35}{8\pi }}
\cdot 5 \cdot 
{\displaystyle \frac{4}{5 \cdot 7 }} \!
\right) \!
\sum_{\Lambda }
\langle 2\mu 2\mu'|3\Lambda \rangle
\widehat{\Jb}_{3\Lambda } \\
\\[-6pt]
&&\!\!\!\!
=
i\hbar 
{\displaystyle \frac{15\sqrt{5}}{8\sqrt{2}\pi }}
R_0^4
\sqrt{10}
\sum_{\Lambda }
\langle 2\mu 2\mu'|1\Lambda \rangle
\widehat{\Jb}_{1\Lambda } \\
\\[-6pt]
&&\!\!\!\!
\equiv
\hbar 
{\displaystyle \frac{15\sqrt{5}}{8\sqrt{2}\pi }}
R_0^4
\sum_{\Lambda }
\langle 2\mu 2\mu'|1\Lambda \rangle
\Lb_{1\Lambda } , ~
\left(
\Lb_{1\Lambda }
=
(-1)^{1+\Lambda} \Lb_{1-\Lambda }^*
\right) .
\end{array}
\label{relationjl}
\end{eqnarray}
It is surprising to see that
in
(\ref{relationjl})
the term related to
$\widehat{\Jb}_{3\Lambda }$
vanishes.  
Substituting
(\ref{commurelpiPai})
into
(\ref{commurelpaipai})
and using
(\ref{commurelation2p})
and
(\ref{relationjl}),
finally we can get
(\ref{commurelpipi})
with the aid of the relations
(\ref{commurelpaipai3})
and 
(\ref{relationphaipai}).

\newpage


\def\thesection{\Alph{section}}
\setcounter{equation}{0}
\renewcommand{\theequation}{\Alph{section}.\arabic{equation}}
\section{\!\!\!Calculations of commutators $[\pi_{2\mu },~T]$ and
$[[{\pi }_{2\mu },~T],~\phi_{2\mu'}]$}


~~~First, 
the commutation relations for 
$[T,~R^{-2}]$
and
$[\eta_{2\mu },~T]$
are calculated, respectively, as\\[-14pt]
\begin{eqnarray}
\begin{array}{rcl}
\!\!& &\!\!
[T,~R^{-2}] 
= 
-
{\displaystyle \frac{\hbar^2}{2m}}
\sum_{\nu } \sum_{n=1}^A
(-1)^{\nu } [\vnabla _{\nu }^n \vnabla_{-\nu }^n ,~R^{-2}] \\
\\[-8pt]
\!\!&=&\!\! 
{\displaystyle \frac{\hbar^2}{m } \frac{2 \cdot 5}{3A}}
 R^{-4} \sqrt{{\displaystyle \frac{4\pi }{3}}}
\sum_{\nu } \sum_{n=1}^A
r_n Y_{1\nu }(\theta_n ,\varphi_n )  
(-1)^{\nu } \vnabla _{-\nu }^n  
+ {\displaystyle \frac{\hbar^2}{m }}5 \! 
\left( \!
1 - {\displaystyle \frac{4}{3A}} \!
\right) \!
R^{-4} ,
\end{array}
\label{comTR2} 
\end{eqnarray}\\[-26pt]
\begin{eqnarray}
\begin{array}{rcl}
[\eta_{2\mu },~T] 
= 
-
{\displaystyle \frac{\hbar^2}{m} i\hbar \frac{\sqrt{2\cdot 5}}{2}} 
\sqrt{{\displaystyle \frac{3}{4\pi }}}
\sum_{\nu \nu'}(-1)^{\mu } 
\left< 1\nu 1\nu'|2-\mu \right> 
\sum_{n=1}^A \!
\vnabla_\nu^n \vnabla_{\nu'}^n ,
\end{array}
\label{cometaT} 
\end{eqnarray}
where we have used the relations
\begin{eqnarray}
\left.
\begin{array}{c}
[\vnabla_\nu^n,~R^{-2}] 
\!=\!
-
{\displaystyle \frac{2 \cdot 5}{3A}}
\sqrt{{\displaystyle \frac{4\pi }{3}}}
R^{-4} 
r_n Y_{1\nu }(\theta_n ,\varphi_n) ,~
\vnabla_\nu^nR^{-4}
\!=\!
-
{\displaystyle \frac{4 \cdot 5}{3A}}
\sqrt{{\displaystyle \frac{4\pi }{3}}}
R^{-6} 
r_n Y_{1\nu } 
(\theta_n ,\varphi_n) ,\!\! \\
\\[-10pt] 
(-1)^{\nu } \vnabla _{-\nu }^n
r_n Y_{1\nu' }(\theta_n ,\varphi_n )
\!=\!
\sqrt{{\displaystyle \frac{3}{4\pi }}}
\delta_{\nu\nu'} .
\end{array}
\right\}
\label{nablar2r4}
\end{eqnarray}\\[-8pt]
The scalar operator
$
\sqrt{{\displaystyle \frac{4\pi }{3}}}
\sum_{\nu } \sum_{n=1}^A
r_n Y_{1\nu }(\theta_n ,\varphi_n ) 
(-1)^{\nu } \vnabla _{-\nu }^n
$
in
(\ref{comTR2})
plays an important role which is discussed in detail later.
Using the definition of
${\pi }_{2\mu }$
(\ref{definitionpi}),
the commutator
$[{\pi }_{2\mu },T]$
is computed as\\[-14pt]
\begin{eqnarray}
\begin{array}{rcl}
& & 
\hspace{-30pt} 
[{\pi }_{2\mu },~T] 
\!=\! 
-
{\displaystyle \frac{\hbar^2 R_0^2}{2m}} \!
\left\{ \!
{\displaystyle i\hbar \frac{\sqrt{2\cdot 5}}{2}} \!
\sqrt{{\displaystyle \frac{3}{4\pi }}} \!
\sum_{\nu \nu'}(-1)^{\mu } \!
\left<1\nu 1\nu'|2 \!-\! \mu \right> \!\!
\sum_{n=1}^A \!
\left( 
R^{-2} \vnabla_\nu^n \vnabla_{\nu'}^n 
\!+\!
\vnabla_\nu^n \vnabla_{\nu'}^n R^{-2}
\right)
\right. \\
\\[-10pt]
& &
\hspace{-30pt} 
+
{\displaystyle \frac{2\cdot 5}{3A}} \!
\sqrt{{\displaystyle \frac{4\pi }{3}}} \!
\sum_\nu \! \sum_{n=1}^A \!\!
\left(
r_n Y_{1\nu }(\theta_n ,\varphi_n ) R^{-4}
(-1)^\nu \vnabla_{-\nu }^n \!\cdot\! \eta_{2\mu }
\!+\!
\eta_{2\mu } \!\cdot\! 
r_n Y_{1\nu }(\theta_n ,\varphi_n ) R^{-4}
(-1)^\nu \vnabla_{-\nu }^n
\right) \\
\\[-10pt]
& & 
\left. 
\qquad 
+
5 \!
\left( \!
1-{\displaystyle \frac{4}{3A}} \!
\right) \!
\left(
R^{-4} \!\cdot\! \eta_{2\mu }
\!+\! 
\eta_{2\mu } \!\cdot\! R^{-4}
\right) 
\right\} .
\end{array}
\label{compiT2} 
\end{eqnarray}
With the use of the relations
\begin{eqnarray}
\begin{array}{rcl}
& &
-
{\displaystyle \frac{\hbar^2 R_0^2}{2m}}
{\displaystyle i\hbar \frac{\sqrt{2\cdot 5}}{2}} 
\sqrt{{\displaystyle \frac{3}{4\pi }}} 
\sum_{\nu \nu'}(-1)^{\mu } 
\left<1\nu 1\nu'|2-\mu\right> \!
\sum_{n=1}^A \!
\vnabla_\nu^n \vnabla_{\nu'}^n R^{-2} \\
\\[-6pt]
\!\!&=&\!\!
-
{\displaystyle \frac{\hbar^2 R_0^2}{2m}}
{\displaystyle i\hbar \frac{\sqrt{2\cdot 5}}{2}} 
\sqrt{{\displaystyle \frac{3}{4\pi }}} R^{-2}
\sum_{\nu \nu'}(-1)^{\mu } 
\left<1\nu 1\nu'|2-\mu\right> \!
\sum_{n=1}^A \!
\vnabla_\nu^n \vnabla_{\nu'}^n \\
\\[-6pt]
& &
-
{\displaystyle \frac{\hbar^2 R_0^2}{2m}}
{\displaystyle \frac{4 \cdot 5}{3A}} R^{-4}
\eta_{2\mu }
-
{\displaystyle \frac{\hbar^2 R_0^2}{2m}}
{\displaystyle i\hbar \frac{50}{3\pi A}} R_0^2 R^{-6}
\phi_{2\mu }^* ,
\end{array}
\label{nablanablar2} 
\end{eqnarray}
and
\begin{eqnarray}
\begin{array}{rcl}
& &
-
{\displaystyle \frac{\hbar^2 R_0^2}{2m}}
{\displaystyle \frac{2\cdot 5}{3A}}
\eta_{2\mu } \cdot R^{-4}
\sqrt{{\displaystyle \frac{4\pi }{3}}}
\sum_\nu \sum_{n=1}^A 
r_n Y_{1\nu }(\theta_n ,\varphi_n ) 
(-1)^\nu \vnabla_{-\nu }^n \\
\\[-4pt]
\!\!&=&\!\!
-
{\displaystyle \frac{\hbar^2 R_0^2}{2m}}
{\displaystyle \frac{2\cdot 5}{3A}}
R^{-4} \sqrt{{\displaystyle \frac{4\pi }{3}}}
\sum_\nu \sum_{n=1}^A 
r_n Y_{1\nu }(\theta_n ,\varphi_n ) 
(-1)^\nu \vnabla_{-\nu }^n 
\cdot \eta_{2\mu } \\
\\[-4pt]
& &
-
{\displaystyle \frac{\hbar^2 R_0^2}{2m}}
{\displaystyle i\hbar \frac{50}{3\pi A}}
R_0^2 R^{-6}
\phi_{2\mu }^*
\cdot
\sqrt{{\displaystyle \frac{4\pi }{3}}}
\sum_\nu \sum_{n=1}^A  
r_n Y_{1\nu }(\theta_n ,\varphi_n ) 
(-1)^\nu \vnabla_{-\nu }^n ,
\end{array}
\label{etar4nablar} 
\end{eqnarray}
and further using the following three relations:\\[-16pt]
\begin{eqnarray}
\left.
\begin{array}{c}
\eta_{2\mu } 
\!=\! 
R_0^{-2} R^2 {\pi }_{2\mu }
\!-\!
i\hbar
{\displaystyle \frac{5}{4\pi }}
R_0^2 R^{-2} \phi_{2\mu }^* , \\
\\[-12pt]
{\displaystyle \frac{4 \cdot 5}{3A}}
R_0^{-2} R^{-4}
\sqrt{{\displaystyle \frac{4\pi }{3}}}
\sum_\nu \sum_{n=1}^A 
r_n Y_{1\nu }(\theta_n ,\varphi_n )
\!\cdot\!
[(-1)^\nu \vnabla_{-\nu }^n ,~R^{2}]
\pi_{2\mu }
\!=\!
10 {\displaystyle \frac{4}{3A}}
R_0^{-2} R^{-2} \pi_{2\mu } , \\
\\[-6pt]
\sum_\nu \! \sum_{n=1}^A  
\left\{
r_n \! Y_{1\nu }(\theta_n ,\varphi_n ) 
(-1)^\nu \vnabla_{-\nu }^n
\!\cdot\!
R^{-2} \phi_{2\mu }^*
\!-\!
R^{-2} \phi_{2\mu }^*
\!\cdot\! 
r_n \! Y_{1\nu }(\theta_n ,\varphi_n ) 
(-1)^\nu \vnabla_{-\nu }^n 
\right\}
\!=\! 0 ,
\end{array} 
\right\}
\label{eta2m} 
\end{eqnarray}
Eq.
(\ref{compiT2})
is rewritten as\\[-20pt]
\begin{eqnarray}
\!\!\!\!\!\!\!\!\!\!\!\!\!\!
\begin{array}{rcl}
& &
[{\pi }_{2\mu },~T]
= 
-
{\displaystyle \frac{\hbar^2 R_0^2}{2m}}
\left\{ 
{\displaystyle i\hbar \sqrt{2 \cdot 5}} 
\sqrt{{\displaystyle \frac{3}{4\pi }}} R^{-2} 
\sum_{\nu \nu'}(-1)^{\mu } 
\left<1\nu 1\nu'|2-\mu\right> 
\sum_{n=1}^A \!
\vnabla_\nu^n \vnabla_{\nu'}^n 
\right. \\
\\[-14pt]
& &  
+
{\displaystyle \frac{4 \cdot 5}{3A}}
 R_0^{-2} \sqrt{{\displaystyle \frac{4\pi }{3}}}
\sum_\nu \sum_{n=1}^A 
r_n Y_{1\nu }(\theta_n ,\varphi_n ) R^{-2}
(-1)^\nu \vnabla_{-\nu }^n \cdot \pi_{2\mu }
\!\!+\!\!
{\displaystyle 
10 \!
\left( \!
1 \!+\! \frac{2}{3A} \!
\right) \!
}
R_0^{-2} R^{-2}
\!\cdot\! 
\pi_{2\mu } \\
\\[-12pt]
& &
\left.
+
{\displaystyle i\hbar \frac{25}{3\pi A}}
R_0^2 R^{-6}
\phi_{2\mu }^*
\!\cdot\!
\sqrt{{\displaystyle \frac{4\pi }{3}}}
\sum_\nu \sum_{n=1}^A  
r_n Y_{1\nu }(\theta_n ,\varphi_n ) 
(-1)^\nu \vnabla_{-\nu }^n
\!\!+\!\! 
{\displaystyle 
i\hbar \frac{25}{2\pi } \!
\left( \!
1 \!-\! \frac{2}{3A} \!
\right) \!
}
R_0^2 R^{-6}
\phi_{2\mu }^* \!
\right\} \! ,
\end{array}
\label{compiT3} 
\end{eqnarray}\\[-14pt]
where\\[-24pt]
\begin{eqnarray}
\begin{array}{c}
\sqrt{{\displaystyle \frac{4\pi }{3}}}
\sum_{\nu } \sum_{n=1}^A
r_n Y_{1\nu }(\theta_n ,\varphi_n ) 
(-1)^{\nu } \vnabla _{-\nu }^n
\!=\!
 \sum_{n=1}^A \!
 \left( \!
x_n {\displaystyle \frac{\partial }{\partial x_n}}
\!+\!
y_n {\displaystyle \frac{\partial }{\partial y_n}}
\!+\!
z_n {\displaystyle \frac{\partial }{\partial z_n}} \!
\right) .
\end{array}
\label{scalerop}
\end{eqnarray}\\[-12pt]
The scalar operator
$
\sqrt{{\displaystyle \frac{4\pi }{3}}}
\sum_{\nu } \sum_{n=1}^A
r_n Y_{1\nu }(\theta_n ,\varphi_n ) 
(-1)^{\nu } \vnabla _{-\nu }^n
$
takes a very simple form
(\ref{scalerop}).
It should be emphasized again that
such a scalar operator in the last line of
(\ref{compiT3})
plays a crucial role
in Section 6.
In deriving the relations
(\ref{nablar2r4}), (\ref{etar4nablar})
and
(\ref{eta2m}),
we have used the famous gradient formulas
for spherical harmonics\\[-16pt]
\begin{eqnarray}
\left.
\begin{array}{rcl}
\vnabla_\nu \Phi(r) Y_{l\mu }(\theta ,\varphi ) 
\!\!&=&\!\!\!\!  
-
\sqrt{{\displaystyle \frac{l \!+\! 1}{2l \!+\! 1}}}
\sum_{\mu'}(\!-\!1)^{\mu } 
\left<l\!+\!1\mu' 1\!-\!\nu|l\mu \right> \!
\left( \!
{\displaystyle \frac{d}{dr} \!-\! \frac{l}{r}} \!
\right) \!  
\Phi(r) Y_{l+1\mu' }(\theta ,\varphi ) \\
\\[-8pt]
+& &\!\!\!\!\!\!\!\!\!\!\!\!\!\! 
\sqrt{{\displaystyle \frac{l}{2l \!+\! 1}}}
\sum_{\mu'}(\!-\!1)^{\mu } 
\left<l\!-\!1\mu' 1\!-\!\nu|l\mu \right> \!
\left( \!
{\displaystyle \frac{d}{dr} \!+\! \frac{l \!+\! 1}{r}} \!
\right) \!  
\Phi(r) Y_{l-1\mu' }(\theta ,\varphi ) , \\
\\[-6pt]
\vnabla_\nu r^l Y_{l\mu }(\theta ,\varphi ) 
\!\!&=&\!\!\!\!  
\sqrt{l(2l \!+\! 1)}
\sum_{\mu'}(\!-\!1)^{\mu } 
\left<l\!-\!1\mu' 1\!-\!\nu|l\mu \right>  
r^{l-1} Y_{l-1\mu' }(\theta ,\varphi ) .
\end{array}
\right\}
\label{gradientformula} 
\end{eqnarray}
The next task is to calculate the commutator 
$[[{\pi }_{2\mu },T],\phi_{2\mu'}]$.
It is made straightforwardly as\\[-16pt]
\begin{eqnarray}
\begin{array}{rcl}
& &[[{\pi }_{2\mu },~T],~\phi_{2\mu'}] \\
\\[-8pt]
& &=\!
- {\displaystyle \frac{\hbar^2 R_0^2}{2m}} \!
\left[ 
i\hbar \sqrt{2 \cdot 5}
\sqrt{{\displaystyle \frac{3}{4\pi }}} R^{-2} 
\sum_{\nu \nu'}(-1)^\mu 
\left< 1\nu 1\nu'|2-\mu \right> 
\sum_{n=1}^A [\vnabla_\nu^n \vnabla_{\nu'}^n ,~\phi_{2\mu'}] 
\right. \\
\\[-8pt]
& & 
+ 
{\displaystyle \frac{4 \cdot 5}{3A}}
R_0^{-2}R^{-2}
\sqrt{{\displaystyle \frac{4\pi }{3}}} \!
\sum_\nu
\sum_{n=1}^A r_n Y_{1\nu }(\theta_n ,\varphi_n )(-1)^\nu 
[\vnabla_{-\nu }^n \cdot {\pi }_{2\mu },~\phi_{2\mu'}] \\
\\[-8pt]
& &
+ 
i\hbar {\displaystyle \frac{25}{3\pi A}}
R_0^2 R^{-6}  \phi_{2\mu }^*
\sqrt{{\displaystyle \frac{4\pi }{3}}} \!
\sum_\nu
\sum_{n=1}^A r_n Y_{1\nu }(\theta_n ,\varphi_n )(- 1)^\nu 
[\vnabla_{-\nu }^n ,~\phi_{2\mu'}] \\
\\[-8pt]
& &
\left. 
+
10 \!
\left( \!
1 \!+\! {\displaystyle \frac{2}{3A}} \!
\right) \!
R_0^{-2}R^{-2}
[{\pi }_{2\mu },~\phi_{2\mu'}] 
\right] .
\end{array}
\label{compiTphi}
\end{eqnarray}
Let us substitute the following three relations:
\vspace{-0.5cm}
\begin{eqnarray}
\!\!\!\!\!\!\!\!
\begin{array}{rcl}
& &
\sum_{\nu,\nu'}(-1)^\mu 
\left< 1\nu 1\nu'|2 \!-\! \mu \right> 
\sum_{n=1}^A [\vnabla_\nu^n \vnabla_{\nu'}^n ,\phi_{2\mu'}] \\
\\[-8pt]
\!\!&=&\!\! 
{\displaystyle \frac{4\pi }{3AR_0^2}}
\sqrt{2 \!\cdot\! 5}\sqrt{{\displaystyle \frac{3}{4\pi }}}
A \!\cdot\! \delta_{\mu\mu'} 
\!+\!
2 {\displaystyle \frac{4\pi }{3AR_0^2} \frac{1}{3}}\sqrt{2 \!\cdot\! 5}
\sum_\nu \! \sum_{n=1}^A r_n Y_{1\nu }(\theta_n ,\varphi_n )
(-1)^\nu \vnabla_{-\nu }^n 
\!\cdot\! 
\delta_{\mu\mu'} \\
\\[-8pt]
& &
- 
{\displaystyle \frac{10}{3AR_0^2}}
\sqrt{{\displaystyle \frac{7}{3}}} R_0^2 R^{-2} \!
\sum_{\Lambda } \!
\left< 2\Lambda 2\mu'|2\mu \right> \! \phi_{2\Lambda }^* 
\!-\!
2{\displaystyle \frac{4\pi }{3AR_0^2} \frac{i}{\hbar }}
\sqrt{{\displaystyle \frac{7}{3}}} R_0^{-2} \! R^2 \!
\sum_{\Lambda } \!
\left< 2\Lambda 2\mu'|2\mu \right> \! \pi_{2\Lambda } \\
\\[-8pt]
& &
- 
{\displaystyle \frac{4\pi }{3AR_0^2} \frac{1}{\hbar }}
\sqrt{3 \!\cdot\! 5}\sqrt{{\displaystyle \frac{3}{4\pi }}}
\sum_{\Lambda } \!
\left< 1\Lambda 2\mu'|2\mu \right> \lb_{1\Lambda } ,
\end{array}
\end{eqnarray}
\vspace{-0.1cm}
\begin{eqnarray}
\begin{array}{rcl}
& &
\sum_\nu \! \sum_{n=1}^A r_n Y_{1\nu }(\theta_n ,\varphi_n )(-1)^\nu 
[\vnabla_{-\nu }^n \!\cdot\! {\pi }_{2\mu },\phi_{2\mu'}] \\
\\[-10pt]
\!\!&=&\!\! 
-
i\hbar 
\sum_\nu \! \sum_{n=1}^A r_n Y_{1\nu }(\theta_n ,\varphi_n )
(-1)^\nu \vnabla_{-\nu }^n 
\!\cdot\! 
\delta_{\mu\mu'}
\!+\!
2\sqrt{{\displaystyle \frac{3}{4\pi }}}
\phi_{2\mu'}\pi_{2\mu } \\
\\[-12pt]
& &\!\! 
+ 
i\hbar \sqrt{{\displaystyle \frac{35}{8\pi }}} R_0^2 R^{-2} \!
\sum_{\Lambda } \!
\left< 2\Lambda 2\mu'|2\mu \right> \! \phi_{2\Lambda }^* 
\sum_\nu \! \sum_{n=1}^A r_n Y_{1\nu }(\theta_n ,\varphi_n )
\!\cdot\! 
(-1)^\nu \vnabla_{-\nu }^n , 
\end{array}
\end{eqnarray}
and
\vspace{-0.65cm}
\begin{eqnarray}
\begin{array}{c}
\sum_\nu \sum_{n=1}^A r_n Y_{1\nu }(\theta_n ,\varphi_n )(-1)^\nu 
[\vnabla_{-\nu }^n ,\phi_{2\mu'}] 
=
2\sqrt{{\displaystyle \frac{3}{4\pi }}}
\phi_{2\mu'} ,
\end{array}
\end{eqnarray}\\[-10pt]
and further substitute
$
\pi_{2\mu } 
\!=\!
\Pi_{2\mu } 
\!-\!
\sqrt{{\displaystyle \frac{35}{8\pi }}}R_0^2 \!
\left\{ \!\!
R^2 \widehat{\Jb}_{2\mu }
\!\!-\!
{\displaystyle \frac{1}{2}} \!
\sum_{\mu_1 \mu_2} \!\!
\left< 2\mu_1 2\mu_2|2\mu \right> \! \phi_{2\mu_1 }^*
[\Pi_{2\mu_2 },R^{-2} ] \!
\right\}
\!$
(\ref{exactpi})
into 
(\ref{compiTphi}).
Then we obtain\\[-12pt]
\begin{eqnarray}
\begin{array}{ll}
&
[[{\pi }_{2\mu },T],\phi_{2\mu'}] \\
\\[-6pt]
&
=
-
{\displaystyle \frac{\hbar^2 R_0^2}{2m}} \!
\left\{ \!\!
-i\hbar {\displaystyle \frac{20}{3A}}
R_0^{\!-2} \! R^{\!-2}\delta_{\mu\mu'} 
\!+\! 
i\hbar 10\sqrt{{\displaystyle \frac{35}{8\pi }}}
R^{\!-4} \!
\sum_\Lambda \!\!
\left< 2\Lambda 2 \mu'|2 \mu \right> \! \phi_{2\Lambda }^* 
\!+\! 
i\hbar {\displaystyle \frac{50}{3\pi A}}
R_0^2 R^{\!-6}\phi_{2 \mu }^* \phi_{2 \mu'} 
\right. \\
\\[-8pt]
&\!\!\!\!\!\!
+
{\displaystyle \frac{4 \cdot\! 4\pi }{3A R_0^4}} \!
\sqrt{{\displaystyle \frac{35}{8\pi }}} \!
\sum_{\Lambda } \!\!
\left< 2 \Lambda 2 \mu'|2 \mu \right> \! \pi_{2\Lambda } 
\!+\! 
{\displaystyle \frac{40}{3AR_0^{2}}} \! 
R^{-2} \! \phi_{2 \mu'}{\pi }_{2 \mu } 
\!-\! 
i{\displaystyle \frac{3}{3A R_0^{2}}} \!
\sqrt{3 \!\cdot\! 5} \! \sqrt{10} \! R^{-2} \!
\sum_{\Lambda } \!\!
\left< 1 \Lambda 2 \mu'|2 \mu \right> \! \lb_{1 \Lambda } \\
\\[-8pt]
&
+ 
\left. 
i\hbar {\displaystyle \frac{4 \!\cdot\! 5}{3A}}
\sqrt{{\displaystyle \frac{35}{8\pi }}}
R^{-4} \!
\sum_{\Lambda } \!
\left< 2 \Lambda 2 \mu'|2 \mu \right> \!
\phi_{2 \Lambda }^*
\sqrt{{\displaystyle \frac{4\pi }{3}}} \! 
\sum_\nu \! \sum_{n=1}^A r_n Y_{1 \nu }(\theta_n ,\varphi_n )
(-1)^\nu \vnabla_{-\nu }^n \!
\right\} \\
\\[-12pt]
&\!\!\!\!\!\!\!\!\!\!\!\!\!\!\!\!\!
=
- 
{\displaystyle \frac{\hbar^2 R_0^2}{2m}} \!
\left[ 
-i\hbar {\displaystyle \frac{20}{3A R_0^{2}}} \! 
R^{-2}\delta_{\mu\mu'} 
\!+\!
i\hbar 10 \! \sqrt{{\displaystyle \frac{35}{8\pi }}}
R^{-4} \!
\sum_\Lambda \!
(\!-\! 1)^{\mu'} \!\!
\left< 2\mu 2 \!\!-\!\! \mu'|2 \Lambda \right> \!\!
\phi_{2\Lambda }^*
\!+\!
i\hbar {\displaystyle \frac{50}{3\pi A}} \!
R_0^2 R^{-6}\phi_{2 \mu }^* \!
\phi_{2  \mu'}  
\right. \\
\\[-12pt]
&\!\!\!\!\!\!\!\!\!\!\!\!\!\!\!\!\!
\!+\!
\sum_{\!\Lambda } 
(\!-\! 1)^{\mu'} \!\!
\left< 2\mu 2 \!-\! \mu'|2 \Lambda \right> \!\!
\left\{ \!\!
{\displaystyle \frac{4 \!\cdot\! 4\pi }{3A R_0^4}} \!
\sqrt{{\displaystyle \frac{35}{8\pi }}} 
\Pi_{2\Lambda }
\!-\!\!
{\displaystyle \frac{70}{3AR_0^{2}}} \!\!
\left( \!\!
R^{2} \! \widehat{\Jb}_{\! 2 \Lambda }
\!-\!
{\displaystyle \frac{1}{2}} \!\!
\sum_{\mu_1 \mu_2 } \!\!
\left< 2 \mu_1 2 \mu_2|2 \Lambda \right> \!\!
\phi_{2\mu_1 }^*
\! [\Pi_{2\mu_2 },R^{\!-2}] \!
\right) \!\!
\right\} \\
\\[-10pt]
&
+ 
{\displaystyle \frac{40}{3A R_0^{2} }}
\! R^{-2} 
\phi_{2 \mu'} 
\Pi_{2\mu }
\!-\!
{\displaystyle \frac{40}{3A}} \!
\sqrt{{\displaystyle \frac{35}{8\pi }}}
\phi_{2 \mu'} \!
\left( \!\!
\widehat{\Jb}_{2\mu }
\!-\!
{\displaystyle \frac{1}{2}} \! R^{-2} \!
\sum_{\mu_1 \mu_2 } \!\!
\left< 2\mu_1 2\mu_2|2\mu \right> \! 
\phi_{2 \mu_1}^* \!
[\Pi_{2\mu_2 },R^{-2}] \!
\right) \\
\\[-8pt]
&~~~~~~~~~~~~~~~~~~~~~~~~~~~~~~~~~~~~~~~~~~~~~~
~~~~
+
i{\displaystyle \frac{15}{3A R_0^{2}}} 
\sqrt{10} R^{-2} \!
\sum_{\Lambda } 
(\!-\! 1)^{\mu'} \!\!
\left< 2\mu 2 \!-\! \mu'|1 \Lambda \right> \!
\lb_{1 \Lambda } \\
\\[-12pt]
&
+ 
\left. 
i\hbar {\displaystyle \frac{4 \!\cdot\! 5}{3A}}
\sqrt{{\displaystyle \frac{35}{8\pi }}}
R^{-4} \!
\sum_{\Lambda } 
(\!-\! 1)^{\mu'} \!\!
\left< 2\mu 2 \!-\! \mu'|2 \Lambda \right> \! 
\phi_{2 \Lambda }^*
\sqrt{{\displaystyle \frac{4\pi }{3}}}
\sum_\nu \sum_{n=1}^A r_n Y_{1\nu }(\theta_n ,\varphi_n )
(-1)^\nu \vnabla_{-\nu }^n \!
\right] \!\! .
\end{array}
\label{commutatorpiTphi} 
\end{eqnarray}
Finally,
in the forth and third line from the bottom of
(\ref{commutatorpiTphi}),
the approximate relation
$[\Pi_{2\mu },R^{-2}]
\approx
i\hbar {\displaystyle \frac{2 \cdot 5}{4\pi }}
R_0 ^4 R^{-6}
\phi_{2\mu }^*$
(\ref{apprcommuPaiR})
should be put into
$[\Pi_{2\mu_2 },R^{-2}]$.

\newpage


\def\thesection{\Alph{section}}
\setcounter{equation}{0}
\renewcommand{\theequation}{\Alph{section}.\arabic{equation}}
\section{Calculation of term
$\sum_{\Lambda } 
\phi_{2\Lambda }^*
(\!-\! 1)^{\Lambda } \Pi_{2 \!-\! \Lambda }$}


For our purpose,
using
(\ref{exactpi}), (\ref{definitionphi}),
(\ref{definitionR2}) and (\ref{definitionpi}), 
we calculate approximately
the term
$
\sum_{\Lambda } 
\phi_{2\Lambda  }^*
(\!-\! 1)^{\Lambda } \Pi_{2 \!-\! \Lambda }
$
in the following way:
\begin{eqnarray}
\!\!\!\!
\begin{array}{cc}
&
\sum_{\Lambda } 
\phi_{2\Lambda  }^*
(\!-\! 1)^{\Lambda } \Pi_{2-\Lambda }
\!\approx\!
\sum_{\Lambda } 
\phi_{2\Lambda  }
\pi_{2\Lambda }
\!=\!
R_0^2 R^{-2} \!
\sum_{\Lambda } 
\phi_{2\Lambda }
\eta_{2\Lambda }
+
{\displaystyle \frac{1}{2}} \! R_0^2 \!
\sum_{\Lambda } 
\phi_{2\Lambda }
\left[
\eta_{2\Lambda }, R^{-2}
\right] \\
\\[-10pt]
&
=
i\hbar
{\displaystyle \frac{5}{4 \pi }} \! 
R_0^4 R^{-4} \!
\sum_{\Lambda } 
\phi_{2\Lambda }^*
\phi_{2\Lambda } \\
\\[-10pt]
&\!\!\!\!\! 
-i\hbar 
{\displaystyle \frac{\sqrt{2 \!\cdot\! 5}}{2}} \!
R_0^2 R^{\!-2} \!
{\displaystyle \frac{4\pi }{3AR_0^2}} \!\!
\sum_{\Lambda } \!\!
\sum_{n=1}^A \! 
r_{\!n}^2 Y_{\! 2\Lambda } (\!\theta_n ,\!\varphi_n) \!\!
\sum_{\kappa \nu } 
(\!-\! 1)^{\Lambda } \!
\langle 1 \kappa 1 \nu|2 \!\!-\!\! \Lambda \rangle \!\!
\sum_{n'=1}^A \! 
r_{\!n'} Y_{\! 1\kappa }(\!\theta_{n'} ,\!\varphi_{n'}) 
\!\cdot\! 
\vnabla_{\nu }^{n'} \! ,
\end{array}
\label{SigmaphiPai}
\end{eqnarray}
where we have used the definition of the $\eta_{2\Lambda }$
(\ref{definitioneta})
and the commutation relation
$\left[ \eta_{2\Lambda }, R^{-2} \right]$
(\ref{commuphietar2}).
Eq.
(\ref{SigmaphiPai})
is further calculated as follows:
\begin{eqnarray}
\begin{array}{cc}
&
\sum_{\Lambda } 
\phi_{2\Lambda  }^*
(\!-\! 1)^{\Lambda } \Pi_{2-\Lambda }
\approx
i\hbar
{\displaystyle \frac{5}{4 \pi }} \!
R_0^4 R^{-4} \!
\sum_{\Lambda } 
\phi_{2\Lambda }^*
\phi_{2\Lambda } \\
\\[-10pt]
&\!\!\!\!\!\!\!\!\!\!  
-i\hbar
{\displaystyle \frac{\sqrt{2 \!\cdot\! 5}}{2 \!\cdot\! 5}} \!
4\pi \!
\sum_{n=1}^A \! 
r_{n} \!
\sum_{\Lambda } \! 
\sum_{\kappa \nu } 
(\!-\! 1)^{\Lambda } \!
\langle 1 \kappa 1 \nu|2 \!-\! \Lambda \rangle \\
\\[-10pt]
&\!\!\!\!\!\!\!\!\!\!
\!\times\!
\sum_{L K} \!
{\displaystyle \sqrt{\frac{5 \!\cdot\! 3}{4\pi (2L \!+\! 1)}}} \!
\langle 2 \Lambda 1 \kappa |L K \rangle \!
\langle 2 0 1 0|L 0 \rangle \! 
Y_{LK }(\theta_{n} ,\varphi_{n}) 
\!\cdot\! 
\vnabla_\nu^n \\
\\[-10pt]
&
=
i\hbar
{\displaystyle \frac{5}{4 \pi }} \! 
R_0^4 R^{-4} \!
\sum_{\Lambda } 
\phi_{2\Lambda }^*
\phi_{2\Lambda } \\
\\[-10pt]
&\!\!\!\!\!\!\!\!\!\!\!\!\!\!\!   
-i\hbar
{\displaystyle \frac{\sqrt{2 \!\cdot\! 5}}{2 \!\cdot\! 5}} \!
4\pi \!
\sum_{n=1}^A \! 
r_{n} \!
\sum_{L K} \!
{\displaystyle \sqrt{\frac{5 \!\cdot\! 3}{4\pi (2L \!+\! 1)}}} \!
\langle 2 0 1 0|L 0 \rangle \\
\\[-10pt]
&\!\!\!\!\!\!\!\!\!\!
\!\times\!
\sum_{\kappa \nu } \!
\sum_{\Lambda } \!  
(\!-\! 1)^{\kappa } \!
\langle 1 \kappa 1 \nu|2 \Lambda \rangle \!
\langle 2 \Lambda 1 \!-\! \kappa |L K \rangle \! 
Y_{LK }(\theta_{n} ,\varphi_{n}) 
\!\cdot\! 
(\!-\! 1)^{\nu } \! 
\vnabla_{-\nu }^n \\
\\[-10pt]
&
=
i\hbar
{\displaystyle \frac{5}{4 \pi }} \! 
R_0^4 R^{-4} \!
\sum_{\Lambda } 
\phi_{2\Lambda }^*
\phi_{2\Lambda } \\
\\[-10pt]
&\!\!\!\!\!\!\!\!\!\!\!\!\!  
-i\hbar
{\displaystyle \frac{\sqrt{2 \!\cdot\! 5}}{2 \!\cdot\! 5}} \!
4\pi \!
\sum_{n=1}^A \! 
r_{n} 
\sum_{L K} \!
{\displaystyle \sqrt{\frac{5 \!\cdot\! 3}{4\pi (2L \!+\! 1)}}} 
\langle 2 0 1 0|L 0 \rangle \!
\sum_{L'} \!
\sqrt{5 (2L' \!+\! 1)}
W(11L1;2L') \\
\\[-10pt]
&\!\!\!\!\! 
\!\times\!
\sum_{\nu } \!
\sum_{\kappa K'} 
(\!-\! 1)^{\kappa } 
(\!-\! 1)^{1 \!+\! \kappa } 
{\displaystyle \sqrt{\frac{2L' \!+\! 1}{3}}} 
\langle 1 \kappa L' K'|1 \nu \rangle 
\langle 1 \kappa L' K'|L K \rangle 
Y_{LK }(\theta_{n} ,\varphi_{n}) 
\!\cdot\! 
(\!-\! 1)^{\nu } 
\vnabla_{-\nu }^n \\
\\[-10pt]
&\!\!\!\!\!\!\!\!\!\!
\!=\!
i\hbar \!
{\displaystyle \frac{5}{4 \pi }} \! 
R_0^4 R^{\!-4} \!
\sum_{\!\Lambda } \!
\phi_{2\Lambda }^*
\phi_{2\Lambda } \\
\\[-10pt]
&\!\!\!\!\! 
\!-\!
i\hbar
\sum_{L \!=\! 0,1,2 } 
(2L \!\!+\!\! 1)W \! (1111;2L) \!
{\displaystyle \sqrt{\frac{4\pi }{3}}} \!
\sum_{\!\nu } \!
\sum_{n=1}^A \! 
r_{n} 
Y_{\! 1 \nu }(\!\theta_{n} ,\!\varphi_{n}) 
\!\cdot\!
(\!-\! 1)^{\nu } 
\vnabla_{\!-\nu }^n \\
\\[-10pt]
&\!\!\!\!\! 
=
i\hbar
{\displaystyle \frac{5}{4 \pi }} \! 
R_0^4 R^{-4} \!
\sum_{\Lambda } \!
\phi_{2\Lambda }^*
\phi_{2\Lambda } 
\!-\!
i\hbar
{\displaystyle \sqrt{\frac{4\pi }{3}}} 
\sum_{\nu } \! 
\sum_{n=1}^A \! 
r_{n} 
Y_{1 \nu }(\theta_{n} ,\varphi_{n}) 
\!\cdot\!
(\!-\! 1)^{\nu } 
\vnabla_{-\nu }^n .
\end{array}
\label{sigmaphipi} 
\end{eqnarray}
Then we obtain the approximated result for the term
$
\sum_{\Lambda }
\phi_{2\Lambda  }^*
(\!-\! 1)^{\Lambda } \Pi_{2 \!-\! \Lambda }
$
in the following form:
\begin{eqnarray}
\!\!\!\!\! 
\begin{array}{l}
\sum_{\Lambda } \!
\phi_{2\Lambda }^*
(\!-\! 1)^{\Lambda } \Pi_{2-\Lambda }
\!\approx\!
i\hbar
{\displaystyle \frac{5}{4 \pi }} \! 
R_0^4 R^{-4} \!
\sum_{\Lambda } \!
\phi_{2\Lambda }^*
\phi_{2\Lambda } 
\!-\!
i\hbar
{\displaystyle \sqrt{\frac{4\pi }{3}}} \!
\sum_{\nu } \!\!
\sum_{n=1}^A \! 
r_{n} 
Y_{1 \nu }(\theta_{n} ,\varphi_{n}) 
\!\cdot\!
(\!-\! 1)^{\nu } 
\vnabla_{-\nu }^n . 
\end{array}
\label{sigmaphipi2} 
\end{eqnarray}

\newpage


\def\thesection{\Alph{section}}
\setcounter{equation}{0}
\renewcommand{\theequation}{\Alph{section}.\arabic{equation}}
\section{Calculation of term
$
\sum_{\Lambda }
\phi_{2\Lambda }^*
R^4
(\!-\! 1)^{\Lambda }
\widehat{\Jb}_{2 \!-\! \Lambda }
$}


~~~~~Using the relation\\[-16pt]
\begin{eqnarray}
\begin{array}{c}
\Jb_{L\Lambda } 
\!=\! 
\sum_{\mu\mu'}
\left< 2 \mu 2 \mu'|L \Lambda \right> \!
\phi_{2\mu }^* \Pi_{2\mu'} 
\!=\!
R^4
\widehat{\Jb}_{L\Lambda }
\!-\!
R^2 \!
\sum_{\mu\mu'}
\left< 2 \mu 2 \mu'|L \Lambda \right>
\phi_{2\mu }^*
\left[ 
\Pi_{2\mu'},R^{-2} 
\right] ,
\end{array}
\label{defjandhatj}
\end{eqnarray}\\[-12pt]
we compute approximately
the term
$
\sum_{\Lambda } 
\phi_{2\Lambda }^*
R^4 
(\!-\! 1)^{\Lambda }
\widehat{\Jb}_{2 \!-\! \Lambda }
$
in a manner similar to 
(\ref{sigmaphipi}):\\[-14pt]
\begin{eqnarray}
\begin{array}{cc}
&\sum_{\Lambda } \!
\phi_{2\Lambda }^*
R^4 \!
(\!-\! 1)^{\Lambda }
\widehat{\Jb}_{2 \!-\! \Lambda }
\!=\!
\sum_{\Lambda } \!
\phi_{2\Lambda }^* 
(\!-\! 1)^{\Lambda } \!
\left( \!
\Jb_{2 \!-\! \Lambda }
\!+\!
R^2 \!
\sum_{\mu_1\mu_2} \!
\left< 2 \mu_1 2 \mu_2|2 \!-\! \Lambda \right> \!
\phi_{2\mu_1}^* \!
\left[
\Pi_{2\mu_2}, R^{-2}
\right] \!
\right) \\
\\[-12pt]
&
\approx
\sum_{\Lambda } \!
\phi_{2\Lambda }^* \!
\sum_{\mu_1\mu_2} \! 
(\!-\! 1)^{\Lambda } \!
\left< 2 \mu_1 2 \mu_2|2 \!-\! \Lambda \right> \!
\left( \!
i\hbar
{\displaystyle \frac{2 \!\cdot\! 5}{4 \pi }} \! 
R_0^4 R^{-4}
\phi_{2\mu_1}^* 
\phi_{2\mu_2}^* \!
\!+\!
\phi_{2\mu_1}^* 
\pi_{2\mu_2} \!
\right) \\
\\[-12pt]
&
\!=\!
-i\hbar
2
\sqrt{{\displaystyle \frac{2}{7}}} \!
\sqrt{{\displaystyle \frac{5}{4 \pi }}} \! 
R_0^2 R^{-2} \!
\sum_{\Lambda } \!
\phi_{2\Lambda }^*
\phi_{2\Lambda } \\
\\[-14pt]
&
+
R_0^2
\sum_{\Lambda } \!
\phi_{2\Lambda }^* \!
\sum_{\mu_1\mu_2} \! 
(\!-\! 1)^{\Lambda } \!
\left< 2 \mu_1 2 \mu_2|2 \!-\! \Lambda \right> \!
\phi_{2\mu_1}^* \!\!
\left( \!\!
R^{-2}
\eta_{2\mu_2}
\!+\!
{\displaystyle \frac{1}{2}} \!
\left[
\eta_{2\mu_2}, R^{-2} 
\right] \!
\right) \\
\\[-14pt]
&
\!=\!
-i\hbar
3
\sqrt{{\displaystyle \frac{2}{7}}} \!
\sqrt{{\displaystyle \frac{5}{4 \pi }}} \! 
R_0^2 R^{-2} \!
\sum_{\Lambda } \!
\phi_{2\Lambda }^*
\phi_{2\Lambda } \\
\\[-14pt]
&\!\!\!\!\!\!\!\!
\!-\!
i\hbar 
{\displaystyle \frac{\sqrt{2 \!\cdot\! 5}}{2}} 
R_0^2 R^{-2} \!
\sum_{\Lambda } \!
\phi_{2\Lambda }^* \!
\sum_{\mu_1\mu_2} \!
\left< 2 \mu_1 2 \mu_2|2 \Lambda \right> \!
\phi_{2\mu_1} \!\!
\sum_{\kappa \nu } 
\langle 1\kappa 1\nu|2 \mu_2 \rangle \!
\sum_{n=1}^A \!
r_n Y_{1\kappa }(\theta_n ,\varphi_n) 
\!\cdot\! \vnabla_\nu^n ,
\end{array}
\label{R4sigmaphijhat}
\end{eqnarray}\\[-4pt]
where we have used the relations
(\ref{definitioneta}), (\ref{definitionpi})
and
(\ref{commuphietar2})
and the approximate formula\\[-14pt]
\begin{eqnarray}
&
\sum_{\Lambda } \!
\phi_{2\Lambda }^*
\sum_{\mu_1\mu_2} \! 
(\!-\! 1)^{\Lambda } \!
\left< 2 \mu_1 2 \mu_2|2 \!-\! \Lambda \right> \!
\phi_{2\mu_1}^* 
\phi_{2\mu_2}^* 
\!\approx\!
-
\sqrt{{\displaystyle \frac{2}{7}}} \!
\sqrt{{\displaystyle \frac{4 \pi }{5}}} \!
R_0^{-2} R^2 \!
\sum_{\Lambda } \!
\phi_{2\Lambda }^*
\phi_{2\Lambda } ,
\label{sigmaphiphiphi}
\end{eqnarray}\\[-14pt]
which can be derived from
(\ref{definitionphi}) and (\ref{definitionR2}).

Using
(\ref{definitionphi}) and (\ref{definitionR2}),
the term in the last line of
(\ref{R4sigmaphijhat})
is approximated as\\[-14pt]
\begin{eqnarray}
\begin{array}{cc}
&\!\!\!\!\!\!\!\!\!\!\!\!\!\!\!\!
\!-\!
i\hbar 
{\displaystyle \frac{\sqrt{2 \!\cdot\! 5}}{2}} 
{\displaystyle \frac{4 \pi }{5}} \!
\sum_{\Lambda } 
\phi_{2\Lambda }^* \!
\sum_{\mu_1\mu_2} \!
\left< 2 \mu_1 2 \mu_2|2 \Lambda \right> \!\!
\sum_{\kappa \nu } 
\langle 1\kappa 1\nu|2 \mu_2 \rangle \!
\sum_{n=1}^A \!
r_n 
Y_{2\mu_1}(\theta_n ,\varphi_n) 
Y_{1\kappa }(\theta_n ,\varphi_n) 
\!\cdot\! \vnabla_\nu^n \\
\\[-12pt]
&\!\!\!\!\!\!\!\!\!\!\!\!\!\!\!\!\!\!\!\!
=
\!-\!
i\hbar 
{\displaystyle \frac{\sqrt{2 \!\cdot\! 5}}{2}} \!
{\displaystyle \sqrt{\frac{4 \pi }{5}}} \!
\sum_{\Lambda } \!
\phi_{2\Lambda }^* \!
\sum_{L K } \!
\left< 2 0 1 0|L 0 \right> \!\!
\sum_{\mu_1\mu_2} \!\!
\left< 2 \mu_1 2 \mu_2|2 \Lambda \right> \!\!
\sum_{\kappa \nu }
(\!-\! 1)^{\mu_1 } \!\!
\left< L K 2 \!-\! \mu_1 |1 \kappa \right> \!\!
\langle 1 \kappa 1 \nu|2 \mu_2 \rangle \! \\
\\[-14pt]
&
\times
\sum_{n=1}^A  
r_n 
Y_{L K}(\theta_n ,\varphi_n) 
\!\cdot\! \vnabla_\nu^n \\
\\[-14pt]
&
=
\!-\!
i\hbar 
{\displaystyle \frac{\sqrt{2 \!\cdot\! 5}}{2}} 
{\displaystyle \sqrt{\frac{4 \pi }{5}}} \!
\sum_{\Lambda } \!
\phi_{2\Lambda }^* \!
\sum_{L K } \!
\left< 2 0 1 0|L 0 \right> \!\!
\sum_{L' K' } \!
\sqrt{3(2L' \!+\! 1)} W(L221;1L') \\
\\[-10pt]
&
\!\!\!\!\!\!\!\!\!\!\!\!
\times
\sum_{\mu_1} \!
\sum_{\nu }
(\!-\! 1)^{\mu_1 } \!
\left< 2 \!-\! \mu_1 1 \nu |L' K' \right> \!\!
\sum_{\mu_2} \!
\langle L K L' K'|2 \mu_2 \rangle \!
\left< 2 \mu_2 2 \mu_1|2 \Lambda \right> \!\!
\sum_{n=1}^A \!
r_n 
Y_{\! L K}(\theta_n ,\varphi_n) 
\!\cdot\! \vnabla_\nu^n \\
\\[-10pt]
&\!\!\!\!\!\!\!\!\!\!\!\!\!\!\!\!\!
=
\!-\!
i\hbar 
{\displaystyle \frac{\sqrt{2 \!\cdot\! 5}}{2}} \!
{\displaystyle \sqrt{\frac{4 \pi }{5}}} \!\!
\sum_{\Lambda } \!
\phi_{2  \Lambda }^* \!
\sum_{L L'  L''  } \!
\left< 2 0 1 0|L 0 \right> \!\! 
(2L' \!\!+\!\! 1) W \! (L221;1L') \!
\sqrt{5(2L'' \!\!+\!\! 1)} W \! (LL'22;2L'') \\
\\[-8pt]
&\!\!\!\!\!\!\!\!\!\!
\times
\sum_{K  K' K'' } \!
\sum_{\mu_1 \nu} \!
\left< L' K' 2 \mu_1|1 \nu  \right> \!
\langle L' K' 2 \mu_1|L'' K'' \rangle \!
\left< L K L'' K''|2 \Lambda \right> \!\!
\sum_{n=1}^A  
r_n 
Y_{L K}(\theta_n ,\varphi_n) 
\!\cdot\! \vnabla_\nu^n \\
\\[-10pt]
&
= 
{\displaystyle \sqrt{\frac{4 \pi }{5}}} \!
\sum_{L} \!
\left< 2 0 1 0|L 0 \right> \!
\sum_{L' } 
(2L' \!+\! 1) W(L221;1L') 
\sqrt{5 \!\cdot\! 3} W(LL'22;21) \\
\\[-12pt]
&
\times
\sum_{\Lambda } \!
\phi_{2\Lambda } 
\left(\!-\! i\hbar \right) 
{\displaystyle \frac{\sqrt{2 \!\cdot\! 5}}{2}} \!
\sum_{\kappa \nu } 
(\!-\! 1)^{\Lambda } \!
\left< L \kappa 1 \nu|2 \!-\! \Lambda \right> \!\!
\sum_{n=1}^A  
r_n 
Y_{L \kappa }(\theta_n ,\varphi_n) 
\!\cdot\! \vnabla_\nu^n \\
\\[-10pt]
&\!\!\!\!\!\!\!\!\!\!\!\!\!\!\!\!\!\!\!\!
\!\approx\!
-
\sqrt{6}
{\displaystyle \sqrt{\frac{4 \pi }{5}}}
\sum_{L = 1,2,3 } 
(2L \!+\! 1) W(1221;1L) W(1L22;21) \!
\sum_{\Lambda }
\phi_{2 \Lambda } 
\eta_{2 \Lambda } \\
\\[-8pt]
&\!\!\!\!\!\!\!\!\!\!\!\!\!\!\!\!\!\!\!\!
\!=\!
-
\sqrt{6}
{\displaystyle \frac{\sqrt{4 \pi }}{5}}
{\displaystyle
\frac{\sqrt{7}( \sqrt{7} \!\!+\!\! \sqrt{3})}{60 }
} \!
\sum_{\Lambda } \!
\phi_{2 \Lambda } 
\eta_{2 \Lambda } ,~
(\mbox{Pick up only the $L\!=\!1$ term and for}~\eta_{2\Lambda }~
\mbox{see}~(\ref{definitioneta}).
).
\end{array}
\label{R4sigmaphijhat2}
\end{eqnarray}

\newpage


\def\thesection{\Alph{section}}
\setcounter{equation}{0}
\renewcommand{\theequation}{\Alph{section}.\arabic{equation}}
\section{Calculation of commutator
$[f_{2\mu },~\phi_{2\mu'}]$}


With the aid of the relations
(\ref{R4sigmaphijhat}), (\ref{sigmaphiphiphi}) and (\ref{R4sigmaphijhat2})
appeared in Appendix D,
\vspace{-0.1cm}
\begin{eqnarray}
\!\!\!\!\!\!\!\!\!\!\!\!\!\!\!\!
&
\left[f_{2\mu },~ \phi_{2\mu' }\right]
\approx \left[
\left[\pi_{2\mu },~T
\right],~ \phi_{2\mu' }
\right] \nonumber \\
\nonumber \\[-4pt]
&\!\!\!\!
-
{\displaystyle \frac{\hbar^2 R_0^2}{2m}} \!
\left[
i\hbar
{\displaystyle \frac{20}{3A R_0^2 }} R^{-2}
\delta_{\mu\mu'}
\!+\!
i\hbar 
{\displaystyle \frac{35}{3A R_0^2} } R^{-2}
\delta_{\mu \mu'}
\!+\!
{\displaystyle \frac{40}{3A R_0^2 }} R^{-2} 
\phi_{2\mu }^* \Pi_{2 \mu'}^*
\!-\!
i\hbar
{\displaystyle \frac{50}{3 \pi A }} R_0^2 R^{-6}
\phi_{2\mu }^* 
\phi_{2 \mu'} 
\right. \nonumber \\
\nonumber \\[-4pt]
&\!\!\!\!\!\!\!\!
\left.
-
{\displaystyle \frac{4 \!\cdot\! 4\pi }{3A R_0^4}}
\sqrt{{\displaystyle \frac{35}{8\pi }}} 
\sum_{\Lambda } \!
(\!-\! 1)^{\mu'} \!
\langle 2 \mu 2 \!-\! \mu'|2 \Lambda \rangle 
\Pi_{2\Lambda } 
\!-\!
i\hbar
{\displaystyle \frac{50}{3A}} \!
\sqrt{{\displaystyle \frac{35}{8\pi }}} R^{-4} \!
\sum_{\Lambda } \!
(\!-\! 1)^{\mu'} \!
\langle 2 \mu 2 \!-\! \mu'|2 \Lambda \rangle  
\phi_{2\Lambda }^*  
\right. \nonumber \\
\nonumber \\[-2pt]
&
\left.
+
{\displaystyle \frac{40}{3A }} 
\sqrt{{\displaystyle \frac{35}{8\pi }}} R^{-4}
\phi_{2 \mu'} \!
\left( \!
R^4
\widehat{\Jb}_{2 \mu }
\!-\!
i\hbar
{\displaystyle \frac{2 \!\cdot\! 5}{4\pi }}
R_0^4 R^{-4} \!
\sum_{\mu_1\mu_2} 
\langle 2 \mu_1 2 \mu_2|2 \mu \rangle 
\phi_{2\mu_1 }^*
\phi_{2\mu_2 }^* \!
\right)
\right. \nonumber \\
\nonumber \\[-4pt]
&
\left.
-
i\hbar 
{\displaystyle \frac{50}{3 \pi A }} 
\sqrt{{\displaystyle \frac{35}{8\pi }}} R_0^4 R^{-8}
\phi_{2 \mu'} \!
\sum_{\mu_1\mu_2} 
\langle 2 \mu_1 2 \mu_2|2 \mu \rangle 
\phi_{2\mu_1 }^*
\phi_{2\mu_2}^* 
\right. \nonumber \\
\nonumber \\[-2pt]
&
\left.
+
i\hbar
{\displaystyle \frac{50}{3 \pi A }} 
\sqrt{{\displaystyle \frac{35}{8\pi }}} R_0^4 R^{-8} \!
\sum_{\Lambda } 
(\!-\! 1)^{\mu'} \!
\langle 2 \mu 2 \!-\! \mu'|2 \Lambda \rangle 
\phi_{2\Lambda }^*
\!\cdot\!
\sum_{\Lambda' } \!
\phi_{2\Lambda' }^*
\phi_{2\Lambda' }  
\right. \nonumber \\
\nonumber \\[-4pt]
&
\left.
-
i\hbar
{\displaystyle \frac{40}{3A }} 
\sqrt{{\displaystyle \frac{35}{8\pi }}} R^{-4} \!
\sum_{\Lambda } 
(\!-\! 1)^{\mu'} \!
\langle 2 \mu 2 \!-\! \mu'|2 \Lambda \rangle 
\phi_{2\Lambda }^* \!
{\displaystyle \sqrt{\frac{4\pi }{3}}} \!
\sum_{\nu } \! 
\sum_{n=1}^A \! 
r_{n} 
Y_{1 \nu }(\theta_{n} ,\varphi_{n}) 
\!\cdot\!
(\!-\! 1)^{\nu } 
\vnabla_{-\nu }^n \!
\right. \nonumber \\
\nonumber \\[-4pt]
&
\left.
-
i\hbar
{\displaystyle \frac{50}{3 \pi A }} 
\sqrt{{\displaystyle \frac{35}{8\pi }}} R_0^4 R^{-8} \!
\sum_{\Lambda } 
(\!-\! 1)^{\mu'} \!
\langle 2 \mu 2 \!-\! \mu'|2 \Lambda \rangle 
\phi_{2\Lambda }^* 
\!\cdot\!
\sum_{\Lambda'} \!
\phi_{2\Lambda' }^*
\phi_{2\Lambda' } 
\right. \nonumber \\
\nonumber \\[-2pt]
&
\left.
-
i\hbar
{\displaystyle \frac{20}{3A }} 
\sqrt{{\displaystyle \frac{35}{8\pi }}} R^{-4} \!
\sum_{\Lambda } 
(\!-\! 1)^{\mu'} \!
\langle 2 \mu 2 \!-\! \mu'|2 \Lambda \rangle 
\phi_{2\Lambda }^* 
\!\cdot\!
5
\right. \nonumber \\
\nonumber \\[-2pt]
&\!\!\!\!\!\!\!\!
\left.
-
{\displaystyle \frac{70}{3A R_0^2} } R^{2} \!
\sum_{\Lambda }
(\!-\! 1)^{\mu' } \!
\langle 2 \mu 2 \!-\! \mu'|1 \Lambda \rangle 
\widehat{\Jb}_{1 \Lambda }
\!+\!
{\displaystyle \frac{80}{3A R_0^2} } R^{2} \!
\sum_{\Lambda }
(\!-\! 1)^{\mu' } \!
\langle 2 \mu 2 \!-\! \mu'|3 \Lambda \rangle 
\widehat{\Jb}_{3\Lambda }
\right. \nonumber \\
\nonumber \\
&\!\!\!\!\!\!\!\!\!\!\!\!\!\
\left.
-
{\displaystyle \frac{350}{3 \pi A}} R_0^2 R^{-2} \!
\sum_{\Lambda' \Lambda'' } 
(\!-\! 1)^{\mu' } \!
\langle 2 \Lambda' 2 \Lambda'' |2 \!-\! \mu' \rangle 
\phi_{2\Lambda' }^* 
\phi_{2\Lambda'' }^*  
\widehat{\Jb}_{2 \mu }  
\!+\!
i\hbar
{\displaystyle \frac{70}{3 \pi A } } R_0^2 R^{-6} \!
\sum_{\Lambda' } \!
\phi_{2\Lambda' }^*
\phi_{2\Lambda' }
\delta_{\mu\mu'}
\right. \nonumber \\
\nonumber \\
&\!\!\!\!\!\!\!\!
\left.
-
i\hbar
{\displaystyle \frac{75}{3 \pi A } } R_0^2 R^{-6} \!
\sum_{\Lambda } 
(\!-\! 1)^{\mu' } \!
\langle 2\mu 2 \!-\! \mu'|2 \Lambda \rangle \!
\sum_{\Lambda' \Lambda'' } \!
\langle 2 \Lambda' 2 \Lambda''|2 \Lambda \rangle
\phi_{2\Lambda' }^*
\phi_{2\Lambda'' }^*
\right. \nonumber \\
\nonumber \\
&\!\!\!\!\!\!\!\!\!\!\!\!
\left.
-
i\hbar
{\displaystyle \frac{175}{3A } } 
{\displaystyle \frac{25}{4 \pi^2 } }
R_0^6 R^{-10} \!
\sum_{\mu_1\mu_2} 
\langle 2 \mu_1 2 \mu_2|2 \mu \rangle 
\phi_{2\mu_1 }^*
\phi_{2\mu_2 }^* \!
\sum_{\mu'_1\mu'_2}
(\!-\! 1)^{\mu' } \!
\langle 2 \mu'_1 2 \mu'_2|2 \!-\! \mu' \rangle 
\phi_{2\mu'_1 }^*
\phi_{2\mu'_2 }^*
\right. \nonumber \\
\nonumber \\
&\!\!\!\!\!\!\!\!\!\!\!\!
\left.
+
{\displaystyle \frac{70}{3A R_0^2}} R^2 \!
\sum_{\Lambda } 
(\!-\! 1)^{\mu' } \!
\langle 2 \mu 2 \!-\! \mu'|2 \Lambda \rangle
\widehat{\Jb}_{2 \Lambda } 
\right. \nonumber \\
\nonumber \\
&\!\!\!\!\!\!\!\!\!\!\!\!
\!\!\!\!
\left. 
-
{\displaystyle \frac{175}{3 \pi A }} R_0^2 R^{-6} \!
\sum_{\Lambda } 
(\!-\! 1)^{\mu' } \!
\langle 2 \mu 2 \!-\! \mu'|2 \Lambda \rangle 
\phi_{2\Lambda }^* \!
\left( \!
\!-\!
i\hbar
\sqrt{{\displaystyle \frac{2}{7}}} \!
\sqrt{{\displaystyle \frac{5}{4 \pi }}} \!
R_0^2 R^{-2} \!
\sum_{\Lambda' } \!
\phi_{2\Lambda' }^*
\phi_{2\Lambda' }
\right.
\right. \nonumber \\
\nonumber \\
&\!\!\!\!\!\!\!\!\!\!
\left.
\left.
\!+\!
i\hbar
\sqrt{6}
{\displaystyle \sqrt{\frac{4 \pi }{5}}} 
{\displaystyle \frac{\sqrt{7}(\sqrt{7} \!+\! \sqrt{3})}{60}}
R_0^{-2} R^2
{\displaystyle \sqrt{\frac{4\pi }{3}}} \!
\sum_{\nu } \! 
\sum_{n=1}^A \! 
r_{n} 
Y_{1 \nu }(\theta_{n} ,\varphi_{n}) 
\!\cdot\!
(\!-\! 1)^{\nu } 
\vnabla_{-\nu }^n \!
\right)
\right] .
\label{fcommutatorphi10}
\end{eqnarray}
Substituting
$\left[
\left[\pi_{2\mu },T
\right], \phi_{2\mu' }
\right]\!,$~i.e.,
(\ref{commutatorpiTphi}),
into
(\ref{fcommutatorphi10}),
then we obtain
the following result:

\begin{eqnarray}
\!\!\!\!\!\!
&
\left[f_{2\mu },\phi_{2\mu' }\right]
\!\approx\!
-
{\displaystyle \frac{\hbar^2 R_0^2}{2m}} \!
\left[ \!
-
{\displaystyle \frac{40}{3A R_0^2 }} \! R^{-2}
\left\{ \!
\phi_{2\mu }^*  (\!-\! 1)^{\mu' } \Pi_{2 -\mu'}
\!-\!
(\!-\! 1)^{\mu' } \phi_{2 -\mu'}^*
\Pi_{2\mu } \!
\right\}
\right. \nonumber \\
\nonumber \\[-4pt]
&
\left.
+
i\hbar
10 \!
\left( \!
1
\!-\!
{\displaystyle \frac{15}{3A}} \!\!
\right) \!
\sqrt{{\displaystyle \frac{35}{8\pi }}} \! R^{-4} \!
\sum_{\Lambda }
(\!-\! 1)^{\mu'}
\langle 2 \mu 2 \!\!-\!\! \mu'|2 \Lambda \rangle 
\phi_{2\Lambda }^* 
\right. \nonumber \\
\nonumber \\[-4pt]
&
\left.
+
i\hbar 
{\displaystyle \frac{50}{3 \pi A }} 
\sqrt{{\displaystyle \frac{35}{8\pi }}} R_0^4 R^{-8}
\phi_{2 \mu'} \!
\sum_{\mu_1\mu_2} 
\langle 2 \mu_1 2 \mu_2|2 \mu \rangle 
\phi_{2\mu_1 }^*
\phi_{2\mu_2}^* 
\right. \nonumber \\
\nonumber \\[-4pt]
&
\left.
+
i\hbar \!
\left( \! 
{\displaystyle \frac{35}{3A R_0^2} } R^{-2}
\!+\!
{\displaystyle \frac{70}{3 \pi A } } R_0^2 R^{-6} \!
\sum_{\Lambda' } \!
\phi_{2\Lambda' }^*
\phi_{2\Lambda' } \!
\right) \!
\delta_{\mu\mu'}
\right. \nonumber \\
\nonumber \\[-4pt]
&
\left. 
+
i\hbar
{\displaystyle \frac{100}{3 \pi A }} 
\sqrt{{\displaystyle \frac{35}{8 \pi }}} \!
R_0^4 R^{-8} \!
\sum_{\Lambda }
(\!-\! 1)^{\mu' } \!
\langle 2 \mu 2 \!-\! \mu'|2 \Lambda \rangle 
\phi_{2\Lambda }^* \!
\sum_{\Lambda' } \!
\phi_{2\Lambda' }^*
\phi_{2\Lambda' }
\right. \nonumber \\
\nonumber \\[-4pt]
&\!\!\!\!\!
\left.
+
i{\displaystyle \frac{15}{3A R_0^{2}}} 
\sqrt{10} R^{-2} \!
\sum_{\Lambda } 
(\!-\! 1)^{\mu'} \!\!
\left< 2\mu 2 \!-\! \mu'|1 \Lambda \right> \!
\lb_{1 \Lambda } 
\right. \nonumber \\
\nonumber \\[-4pt]
&\!\!\!\!\!\!\!\!\!\!\!\!\!\!
\left.
-
{\displaystyle \frac{70}{3A R_0^2} } R^{2} \!
\sum_{\Lambda }
(\!-\! 1)^{\mu' } \!
\langle 2 \mu 2 \!-\! \mu'|1 \Lambda \rangle 
\widehat{\Jb}_{1 \Lambda }
\!\!-\!
{\displaystyle \frac{350}{3 \pi A}} R_0^2 R^{-2} \!
\sum_{\Lambda' \Lambda'' } 
(\!-\! 1)^{\mu' } \!
\langle 2 \Lambda' 2 \Lambda'' |2 \!-\! \mu' \rangle 
\phi_{2\Lambda' }^* 
\phi_{2\Lambda'' }^*  
\widehat{\Jb}_{2 \mu }
\right. \nonumber \\
\nonumber \\[-6pt]
&\!\!\!\!\!\!\!\!
\left.
\!+\!
{\displaystyle \frac{80}{3A R_0^2} } R^{2} \!
\sum_{\Lambda }
(\!-\! 1)^{\mu' } \!
\langle 2 \mu 2 \!-\! \mu'|3 \Lambda \rangle 
\widehat{\Jb}_{3\Lambda }
\right. \nonumber \\
\nonumber \\[-6pt]
&\!\!\!\!\!\!\!\!
\left.
+
i\hbar
{\displaystyle \frac{275}{3 \pi A } } R_0^2 R^{-6} \!
\sum_{\Lambda } 
(\!-\! 1)^{\mu' } \!
\langle 2\mu 2 \!-\! \mu'|2 \Lambda \rangle \!
\sum_{\Lambda' \Lambda'' } \!
\langle 2 \Lambda' 2 \Lambda''|2 \Lambda \rangle
\phi_{2\Lambda' }^*
\phi_{2\Lambda'' }^*
\right. \nonumber \\
\nonumber \\[-4pt]
&\!\!\!\!\!\!\!\!\!\!\!\!\!\!\!\!\!\!\!\!
\left.
-
i\hbar
{\displaystyle \frac{175}{3A } } 
{\displaystyle \frac{25}{4 \pi^2 } }
R_0^6 R^{-10} \!
\sum_{\mu_1\mu_2} 
\langle 2 \mu_1 2 \mu_2|2 \mu \rangle 
\phi_{2\mu_1 }^*
\phi_{2\mu_2 }^*
\sum_{\mu'_1\mu'_2}
(\!-\! 1)^{\mu' } \!
\langle 2 \mu'_1 2 \mu'_2|2 \!-\! \mu' \rangle 
\phi_{2\mu'_1 }^*
\phi_{2\mu'_2 }^*
\right. \nonumber \\
\nonumber \\[-4pt]
&\!\!\!\!\!\!\!\!
\left.
\!-\!
i\hbar
10 \!
\left( \!
5
\!\!+\!\!
{\displaystyle \frac{7 \sqrt{21}}{3}} \!
\right) \!\!
{\displaystyle \frac{1}{3A }} \! 
\sqrt{{\displaystyle \frac{35}{8\pi }}} \!
R^{-4} \!
\sum_{\Lambda } 
(\!-\! 1)^{\mu'} \!
\langle 2 \mu 2 \!-\! \mu'|2 \Lambda \rangle 
\phi_{2\Lambda }^* \!
\right. \nonumber \\
\nonumber \\[-4pt]
&\!\!\!\!\!\!\!\!\!\!\!\!\!\!\!\!\!\!\!\!\!
\left.
\times
{\displaystyle \sqrt{\frac{4\pi }{3}}} \!
\sum_{\nu } \!\!
\sum_{n=1}^A \!
r_{\! n} 
Y_{\! 1 \nu }(\!\theta_{n} ,\!\varphi_{n}) 
\!\cdot\!
(\!-\! 1)^{\nu }
\vnabla_{\!-\nu }^n \!
\right] ,
\label{fcommutatorphi11}
\end{eqnarray}
in Eq.
(\ref{fcommutatorphi11}), 
as shown in
(\ref{relationphaipai}),
the term in curly bracket of  the first line
is transformed to the quantities
$\widehat{\Jb}_{1\Lambda }$ and $\widehat{\Jb}_{3\Lambda }$
as
\begin{eqnarray}
&
-
{\displaystyle \frac{40}{3A R_0^2 }} R^{2} R^{-4}
\left\{
\phi_{2\mu }^*  (\!-\! 1)^{\mu' } \Pi_{2 -\mu'}
\!-\!
(\!-\! 1)^{\mu' } \phi_{2 -\mu'}^*
\Pi_{2\mu }
\right\} \nonumber \\
\nonumber \\[-4pt]
&
\!=\!
-
{\displaystyle \frac{40}{3A R_0^2 }} R^{2}
\sum_{L} \!
\left\{
1
\!-\! 
(\!-\! 1)^L
\right\} \!
\sum_{\Lambda } 
(\!-\! 1)^{\mu'} \!
\langle 2 \mu 2 \!-\! \mu'|L \Lambda \rangle
\widehat{\Jb}_{L\Lambda } \nonumber \\
\nonumber \\[-4pt]
&
+
{\displaystyle \frac{40}{3A R_0^2 }} R^{2} R^{-2}
\left(
\phi_{2\mu }^* \left[(\!-\! 1)^{\mu'} \! \Pi_{2-\mu'}, R^{-2}\right]
-
(\!-\! 1)^{\mu'} \! \phi_{2-\mu'}^* \left[\Pi_{2-\mu}, R^{-2}\right] 
\right) \nonumber \\
\nonumber \\[-4pt]
&
=
-
{\displaystyle \frac{80}{3A R_0^2 }} R^{2} \!
\sum_{\Lambda } 
(\!-\! 1)^{\mu'} \!
\langle 2 \mu 2 \!-\! \mu'|1 \Lambda \rangle
\widehat{\Jb}_{1\Lambda }
\!-\!
{\displaystyle \frac{80}{3A R_0^2 }} R^{2} \!
\sum_{\Lambda } 
(\!-\! 1)^{\mu'} \!
\langle 2 \mu 2 \!-\! \mu'|3 \Lambda \rangle
\widehat{\Jb}_{3\Lambda } .
\label{fcommutatorphi12}
\end{eqnarray}
Notice that the term $\widehat{\Jb}_{3\Lambda }$ should vanish.
Put the relation
(\ref{fcommutatorphi12})
into the above result
(\ref{fcommutatorphi11}),
we can reach
the approximate expression for the commutation relation
$\left[f_{2\mu },\phi_{2\mu' }\right]$
given by
(\ref{fcommutatorphi13})
in Section 5.

\newpage


\def\thesection{\Alph{section}}
\setcounter{equation}{0}
\renewcommand{\theequation}{\Alph{section}.\arabic{equation}}
\section{Expression for $C_0(R^2)$ in terms of 
$\phi_{2\mu }$ and $\eta_{2\mu }$}


\begin{eqnarray}
\!\!\!\!\!\!\!\!\!\!\!\!\!\!\!\!\!\!\!\!\!\!\!\!
\begin{array}{lll}
~~
C_0 (R^2) 
\!=\!\! 
& &
-\!\!\!\!\!\!\!\!\!\!\!\!
{ \displaystyle  \frac{\hbar^2 R_0^2}{2m}\frac{10}{3AR_0^4} }
R_0^2 R^{-2} \!
\sum_{\mu} \!\! 1
\!+\!
T
\!-\!
{\displaystyle \frac{\hbar^2 R_0^2}{2m}}
\!\cdot\!
{\displaystyle \frac{25}{4 \pi }} \!\!
\left( \!\!
1 \!-\! 4 
\!\cdot\! 
{\displaystyle \frac{1}{3A}}
\!+\!
{\displaystyle \frac{60}{7}}
\!\cdot\!
{\displaystyle \frac{1}{3A}}
\!\cdot\!
{\displaystyle
\frac{21A \!\!-\!\! 93}
{15 \!\!+\!\! 7 \sqrt{21}} \!
} \!
\right) \!\! 
R_0^2 R^{-6} \!
\sum_{\mu} \!\! 
\phi_{2\mu }^* \! \phi_{2\mu} \\
\\[-8pt]
& &
\!\!\!\!\!\!\!\!\!\!
\!-
i\hbar 
{\displaystyle \frac{5}{4 \pi}}
{\displaystyle \frac{4\pi }{3AR_0^4} \frac{1}{m} } \!
R_0^4 R^{-2} \!
\sum_{\mu } \!\!
\left( 
\phi_{2\mu } 
\pi_{2\mu }
\!+\!
\phi_{2\mu }^* 
(\!-\!1)^\mu \!
\pi_{2 \!-\! \mu } 
\right) \\
\\[-12pt]
& &
\!\!\!\!\!\!\!\!\!\!\!\!\!\!\!\!\!\!\!\!\!\!\!\!\!\!\!\!\!\!\!
+
{\displaystyle \frac{4\pi }{3AR_0^4} \frac{1}{m} } \!
\left[
i\hbar 
\left( \!\!
\sqrt{{\displaystyle \frac{35}{8\pi }}} R_0^2 \!\!
\right)^{\!\!2} \!\!
R^{-2} \!
\sum_{\mu' } \!\!
\sum_{\mu \mu_1 \mu_2} \!
(\!-\!1)^\mu \!
\langle 2 \mu_1 2\mu_2 |2 \!-\! \mu \rangle \!
(\!-\!1)^\mu \!  (\!-\!1)^{\mu_1} \!
\langle 2 \!-\! \mu 2 \!-\! \mu_1 |2 \mu' \rangle
\phi_{2 \mu' } 
\pi_{2 \mu_2 }
\right. \\
\\[-14pt]
& &
\!\!\!\!
\left.
-
R^2 \!
\sum_\mu \!
\pi_{2\mu }
(\!-\!1)^\mu
\pi_{2 \!-\! \mu }
\!-\!
\sqrt{{\displaystyle \frac{35}{8\pi }}} R_0^2 \!
\sum_{\mu } \!
\sum_{\mu_1 \mu_2} \!
\langle 2 \mu_1 2\mu_2 |2\mu \rangle
\phi_{2 \mu_1}^* \!
\pi_{2 \mu_2 }
(\!-\!1)^\mu
\pi_{2 \!-\! \mu } \!\! .
\right] \\
\\[-14pt]
& &
\!\!\!\!\!\!\!\!\!\!\!\!\!\!\!\!\!\!\!\!\!\!\!\!\!\!\!\!\!\!\!
+{\displaystyle \frac{4\pi }{3AR_0^2}
\frac{1}{m}} \!
\sqrt{{\displaystyle \frac{35}{8\pi }}} \!
\sum_{\mu \mu' \mu''}
(\!-\!1)^{\mu }
\langle 2 \mu' 2 \mu'' |2 \!-\! \mu \rangle \!\!\!
\left[ \!
\phi_{2 \mu'' }^* \!\!
\pi_{2\mu }
\pi_{2\mu' }
\!\!+\!\!
\sqrt{{\displaystyle \frac{35}{8\pi }}} \!
R_0^2 R^{-2} \!
\phi_{2 \mu'' }^* \!\!
\sum_{\mu_2} \!
(\!-\!1)^{\mu} \!
\langle 2 \!-\! \mu 2\mu_2 |2\mu' \rangle \!
\pi_{2 \mu_2 } 
\right. \\
\\[-14pt]
& &
\!\!\!\!
\left.
\!+\!
\left( i\hbar \right)^2 \! 
{\displaystyle \frac{5}{4 \pi}} \!
\sqrt{{\displaystyle \frac{35}{8\pi }}} 
R_0^6 R^{-6} 
(\!-\!1)^{\mu}  \! (\!-\!1)^{\mu'} 
\langle 2 \!-\! \mu 2 \!-\! \mu' | 2 \mu'' \rangle
\phi_{2 \mu'' }^*
\phi_{2 \mu''} 
\right] \\
\\[-14pt]
& &
\!\!\!\!\!\!\!\!\!\!\!\!
\!\approx\! 
-
{ \displaystyle  \frac{\hbar^2 R_0^2}{2m}\frac{10}{3AR_0^4} }
R_0^2 R^{-2} \!
\sum_{\mu} \! 1
\!+\!
{\displaystyle \frac{4 \pi}{5 A}}
{\displaystyle \frac{1}{m}}
R^{-2} \!
\sum_{\mu } \!
\eta_{2 \mu }
(\!-\!1)^{\mu}
\eta_{2 - \mu }  \\
\\[-6pt]
& & 
\!\!\!\!\!\!
-
{\displaystyle \frac{\hbar^2 R_0^2}{2m}}
\!\cdot\!
{\displaystyle \frac{25}{4 \pi }} \!
\left( \!
1 \!-\! {\displaystyle \frac{4}{3A}}
\!+\!
{\displaystyle \frac{20}{7A}}
\!\cdot\!
{\displaystyle
\frac{21A \!-\! 93}
{15 \!+\! 7 \sqrt{21}} \!
}
\right) \!\! 
R_0^2 R^{-6} \!
\sum_{\mu} \! 
\phi_{2\mu }^* \phi_{2\mu} \\
\\[-10pt]
& &
\!\!\!\!\!\!
-
{\displaystyle \frac{4\pi }{3AR_0^4}
\frac{1}{m} }
R^2 \! 
\sum_\mu \!
\pi_{2\mu }
(\!-\!1)^\mu
\pi_{2 \!-\! \mu }
\!-\!
{\displaystyle \frac{4\pi }{3AR_0^4}
\frac{1}{m} } 
\sqrt{{\displaystyle \frac{35}{8\pi }}}
R_0^2 
\phi_{2 0}^*
R^{-2} R^2 \!
\sum_{\mu }
\pi_{2 \mu }
(\!-\!1)^\mu
\pi_{2 \!-\! \mu } \\
\\[-8pt]
& &
\!\!\!\!\!\!\!\!\!\!\!\!
-
i\hbar \!
{\displaystyle \frac{4\pi }{3AR_0^4} \!
\frac{1}{m} } \!
{\displaystyle \frac{5}{4 \pi}} \!
R_0^4 R^{-2} \!
\sum_{\mu } \!\!
\left( \!
\phi_{2\mu } 
\pi_{2\mu }
\!+\!
\phi_{2\mu }^* 
(\!-\!1)^\mu 
\pi_{2 \!-\! \mu } \!
\right)
\!\!+\!\!
{\displaystyle \frac{4\pi }{3AR_0^2} \!
\frac{1}{m}} \!
{\displaystyle \frac{1}{5}} \!
\sqrt{{\displaystyle \frac{35}{8\pi }}} \!
\phi_{2 0 }^* \!
R^{-2} R^2 \!
\sum_{\mu } \!\! 
\pi_{2\mu }
(\!-\!1)^{\mu }
\pi_{2 \!-\! \mu } \\
\\[-4pt]
& &
\!\!\!\!\!\!
\!-\!
i\hbar
{\displaystyle \frac{4\pi }{3AR_0^4} \!
\frac{1}{m}} \!
5
{\displaystyle \frac{35}{8\pi }} \!
R_0^4 R^{-2} \!
\sum_{\mu } \!
\left( \!
\phi_{2 \mu }
\pi_{2\mu }
\!+\!
\phi_{2 \mu }^* 
(\!-\!1)^{\mu }
\pi_{2 \!-\! \mu } \!
\right)
\!+\!
\left( i\hbar \right)^2 \!\! 
{\displaystyle \frac{5}{3AR_0^2} \!
\frac{1}{m}} \!
5
{\displaystyle \frac{35}{8\pi }} \!
R_0^6 R^{-6}  \!
\sum_{\mu } \!
\phi_{2 \mu }^*
\phi_{2 \mu } .
\end{array}
\label{expansionC02}
\end{eqnarray}\\[-4pt]
Thus we can get the approximate expression for
$C_0 (R^2)$
as\\[-12pt]
\begin{eqnarray}
\begin{array}{lll}
& &
\!\!\!\!\!\!\!\!\!\!\!\!\! 
C_0 (R^2)
\!\approx\!
-
{ \displaystyle  \frac{\hbar^2 R_0^2}{2m}
\frac{10}{3AR_0^4} } \! 
\left\{ \!
1
\!+\!
\left( \!
1
\!+\!
{\displaystyle \frac{4}{5 }}
\sqrt{{\displaystyle \frac{7}{2 }}} \!
\right) \!
\right\} \!
R_0^2 R^{-2} \!
\sum_{\mu} \! 1 \\
\\[-10pt]
& &
\!\!\!\!\!\!\!\!\!\!\!\!\! 
\!+\!
{\displaystyle \frac{4 \pi}{5 A}}
{\displaystyle \frac{1}{m}} \!
\left\{ \!
1
\!-\!
\left( \!
5
\!+\!
4
\sqrt{{\displaystyle \frac{7}{2 }}} \!
\right) \!
\right\} \!
R^{-2} \!
\sum_{\mu } \!
\eta_{2 \mu }
(\!-\!1)^{\mu}
\eta_{2 - \mu } \\
\\[-10pt]
& &
\!\!\!\!\!\!\!\!\!\!\!\!\!
-
i\hbar
{\displaystyle \frac{4\pi }{3AR_0^4}
\frac{1}{m} } \!
\left\{ \!
\left( \!
1
\!+\!
{\displaystyle \frac{4}{5 }}
\sqrt{{\displaystyle \frac{7}{2 }}} \!
\right) 
\!+\!
\left( \! 1 \!+\! {\displaystyle \frac{35}{2}} \! \right) \! \!
\right\} \!
{\displaystyle \frac{5}{4 \pi}} 
R_0^6 R^{-4} \!
\sum_{\mu } \!\!
\left(
\phi_{2\mu }  
\eta_{2\mu }
\!+\!
\phi_{2\mu }^* 
(\!-\!1)^\mu 
\eta_{2 \!-\! \mu } 
\right)   \\
\\[-8pt]
& &
\!\!\!\!\!\!\!\!\!\!\!\!\!
-
{\displaystyle \frac{\hbar^2 R_0^2}{2m}}
\!\cdot\!
{\displaystyle \frac{25}{4 \pi }} \!
\left\{ \!
1 \!-\! {\displaystyle \frac{4}{3A}}
\!+\!
{\displaystyle \frac{20}{7A}}
\!\cdot\!
{\displaystyle
\frac{21A \!\!-\!\! 93}
{15 \!\!+\!\! 7 \sqrt{21}}
}
\!-\!
\left( \! 1 \!\!+\!\! {\displaystyle \frac{35}{2}} \! \right) \! 
{\displaystyle \frac{4 }{3A}}
\!+\!
{\displaystyle \frac{35}{3A}}
\!-\!
\left( \!
1
\!\!+\!\!
{\displaystyle \frac{4}{5 }}
\sqrt{{\displaystyle \frac{7}{2 }}} \!
\right)  \!\!
{\displaystyle \frac{10}{3A}} \!
\right\} \!\! 
R_0^2 R^{-6} \!
\sum_{\mu} \! 
\phi_{2\mu }^* \phi_{2\mu} ,
\end{array}
\label{expansionC03}
\end{eqnarray}\\[-6pt] 
where we have used the important relation for $T\!\!$
(\ref{Tetaeta}),
approximate relations for
$\pi_{2 \mu}\!\!$
(\ref{paiapproxaeta})
and
$R^2\!  \sum_{\mu} \pi_{2 \mu} (\!-\!1)^\mu \pi_{2 \!-\! \mu}\!\!$
(\ref{pieta})
and lastly approximate one for
$\phi_{2 0}$
(\ref{approxphi20}).
The first term of the R.H.S of
(\ref{expansionC03})
is the the important result.
The others are expected to be vanished.


\end{document}